\documentclass[5p,times]{elsarticle}

\usepackage{amsmath}
\usepackage{amssymb}
\usepackage{float}
\usepackage[final]{listings}
\usepackage{subcaption}
\usepackage{makecell}
\usepackage{multicol}
\usepackage{multirow}
\usepackage{setspace}
\usepackage[table]{xcolor}
\usepackage{graphicx}

\usepackage{newfloat}
\DeclareFloatingEnvironment[name={Figure}]{suppfigure}

\DeclareFloatingEnvironment[name={Table}]{supptable}

\newfloat{lstfloat}{b!}{lop}
\floatname{lstfloat}{Listing}

\lstdefinelanguage
 [x64]{Assembler}     
 [x86masm]{Assembler} 
 {morekeywords={cmpps, pand, paddd, ptest, cmovz, CDQE,CQO,CMPSQ,CMPXCHG16B,JRCZ,LODSQ,MOVSXD, 
 POPFQ,PUSHFQ,SCASQ,STOSQ,IRETQ,RDTSCP,SWAPGS, 
 ecx,edx,xmm0,xmm1,xmm2,xmm3,xmm4,xmm5,xmm6,
 xmm7,xmm8,xmm9,xmm10,xmm11,xmm12,xmm13,xmm14,xmm15,
 rax,rdx,rcx,rbx,rsi,rdi,rsp,rbp, 
 r8,r8d,r8w,r8b,r9,r9d,r9w,r9b, 
 r10,r10d,r10w,r10b,r11,r11d,r11w,r11b, 
 r12,r12d,r12w,r12b,r13,r13d,r13w,r13b, 
 r14,r14d,r14w,r14b,r15,r15d,r15w,r15b}} 

\journal{}
\makeatletter
\def\ps@pprintTitle{%
 \let\@oddhead\@empty
 \let\@evenhead\@empty
 \def\@oddfoot{}%
 \let\@evenfoot\@oddfoot}
\makeatother
\begin{document}

\begin{frontmatter}

\title{High performance computing on Android  devices - a case study.}

%\texttrademark {}

\author[afil1]{Robert Fritze\corref{mycorrespondingauthor}\fnref{orcid1}}
\cortext[mycorrespondingauthor]{Corresponding author}
\ead{Robert.Fritze99@gmail.com}
\fntext[orcid1]{ORCID: 0000-0001-7061-9587}

\author[afil1]{Claudia Plant}

\address[afil1]{University of Vienna, Faculty of Computer Science, Research Group Data Mining, W\"{a}hringer Straße 29, 1090, Vienna, Austria}

\begin{abstract}

High performance computing for low power devices can be useful to speed up calculations on processors that use a lower clock rate than computers for which energy efficiency is not an issue. In this trial,  different high performance techniques for Android devices have been compared, with a special focus on the use of the GPU. Although not officially supported, the OpenCL framework can be used on Android tablets.

For the comparison of the different parallel programming paradigms, a benchmark was chosen that could be implemented easily with all frameworks. The Mandelbrot algorithm is computationally intensive and has very few input and output operations.
The algorithm has been implemented in Java, C, C with assembler, C with SIMD assembler, C with OpenCL and scalar instructions and C with OpenCL and vector instructions. The implementations have been tested for all architectures currently supported by Android.

High speedups can be achieved using SIMD and OpenCL,
although the implementation is not straightforward for either one.
Apps that use the GPU must account for the fact that they can be suspended by the user at any moment.
In using the OpenCL framework on the GPU of Android devices, a computational power comparable to those of modern high speed CPUs can be made available to the software developer. 

\end{abstract}


\end{frontmatter}

%\linenumbers

\section{Introduction}

Low-power CPUs (central processing units) have become increasingly popular in recent years, especially for mobile computing where energy efficiency is an issue. These CPUs often have a lower clock rate and, therefore, efficient parallelization techniques may be important to optimize runtime for more complex workloads \cite{MontBlanc}.

Android is a widely used Linux-based open-source operating system for mobile devices and allows an easy user interaction. % with users that do not need to be computer experts in order to use such a device.
Android Studio provides a free and easy to use IDE (integrated development environment) for the creation of apps.  
Android supports only a limited set of processor architectures (currently four), which makes it easier to produce different flavors of assembler code.

The Mandelbrot set \cite{Douady1984} is the set of complex numbers $c$ for which the sequence of complex numbers $(z_n)_{n\in \mathbb{N}}$ defined by the iteration 

\begin{align*} 
z_0 &= 0 \\ 
z_{n+1} &=  z_n^2 + c
\end{align*}

is limited. 
In  practice,  a  number $N\in\mathbb{N}$ is  selected.  If  the  absolute value  of $z_N$ is  still  below 2,  the  sequence  is  regarded  as limited.  %Conversely, the  number  of  iterations  needed  to  show  that the  sequence  is  not  limited  depends  on  the  choice  of c.
The calculation of the iteration is computationally intensive and has few input/output operations compared to the number of arithmetic operations.
Moreover, this algorithm can be implemented rather easily with different parallel programming paradigms (multithreading, vector parallelism, OpenCL) 
and  assembler code can be produced with moderate effort.
Used as a benchmark, this algorithm measures the CPU performance for floating point operations executed serially or in parallel. 
Implemented with different parallelization paradigms, the performance of the algorithm  can be compared in a  repeatable and transparent manner.
The calculation of the Mandelbrot  set  has been used several times before as a benchmark \cite{Brodtkorb2008, Wang2016}.

%It allows to use various kinds of parallelism (multithreading, vector parallelism) and it can be calculated by a GPU (graphical processing unit). The algorithm is computationally intensive and does not have much data transfer.

%The aim of this trial was to optimize the code for the calculation of three different windows of the Mandelbrot set and test different parallel programming paradigms on Android devices. 

\subsection{Contributions}

\begin{itemize}

  \item A survey of how the OpenCL framework can be used on Android mobile devices
  \item A wrapper library for the use of OpenCL libraries on Android devices
  \item A locking mechanism to allow the unloading of the wrapper library and memory optimization while the app is not in use
  \item A locking mechanism to allow calculations performed on the GPU to be safely aborted when the user has stopped the app
  \item A comparison of parallelization techniques (e.g. SIMD, multithreading) for all currently-supported Android architectures
  \item A comparison of parallel programming paradigms performed on an ARMv7 tablet GPU (Mali G-71 MP 2)
  
%  \item -------------
%    \item A survey of how the OpenCL framework can be used on Android mobile devices
%    \item A wrapper library for the flawless use of OpenCL libraries on Android devices has been written
%    \item A locking mechanism has been developed, that allows to unload the wrapper library and save memory when the app is not used
%    \item A locking mechanism has been developed, that allows to abort in a save manner calculations on the GPU when the app is stopped by the user
%    \item For all architectures currently supported by Android different parallelization techniques (like SIMD and multithreading) have been compared to each other
%    \item The performance of a GPU (Mali G-71 MP 2) on an Arm-v7 tablet has been compared to other parallel programming paradigms on the same Android tablet 

\end{itemize}

\subsection{Previous literature}

Many research projects currently focus on the ability to use lower power processors (especially ARM) for high performance computing but few trials investigated the use of OpenCL on mobile devices and compared it to other parallel programming paradigms. 
Acosta et al. \cite{Acosta2018} gave a nice overview of the availability of OpenCL capable GPUs on mobile devices.
Ross et al. \cite{Ross2014} published a paper in 2014 where they compare different types of graphic cards. One of the devices was a GPU on a mobile device.
The Mont Blanc project \cite{MontBlanc} aims to create a basis for high performance computing on low power CPUs. For this project  Pérez et al. \cite{Perez2019}, \cite{Perez2017} tried to create an OpenCL environment, that can be executed on multiple low power devices concurrently. 

Wang et al. \cite{Wang2016} used the Mandelbrot set to evaluate the performance of an Android Adreno-GPU. They only compared OpenCL and Java. They used a slightly different setting and the speedups were lower compared to those achieved in this trial. 
In 2013, Wang et al. \cite{Wang2013} implemented an image processing application on Android mobile devices with GPUs using OpenCL, that allowed the removal of objects from images.
Yokoyama et al. \cite{Yokoyama2019} summarize in a 2019 review the efforts that have been made to use ARM processors for high performance computing.

%Besides of shedding light on the performance of different parallelization techniques, difficulties of addressing the GPU on Android devices have been investigated and solutions developed.
%For example apps can be stopped or terminated at any moment (by the user or the operating system) for which an application that uses the GPU has to account.

%This trial should guide future research and development. 

%help to decide how efficient each of the parallel programming types on Android devices can be and 

%Not for all low power CPUs with a GPU there are already OpenCL libraries available (like Raspberry Pi).

%The aim of this trial the was to measure wall clock times for the calculation for three different windows of the Mandelbrot set applying different parallel programming paradigms. 

\section{Calculations}%Materials and Methods}

\subsection{Test environment}

\begin{figure*}[bt]
\centering
\begin{subfigure}[b]{0.475\textwidth}
  \centering
  \includegraphics[width=\textwidth]{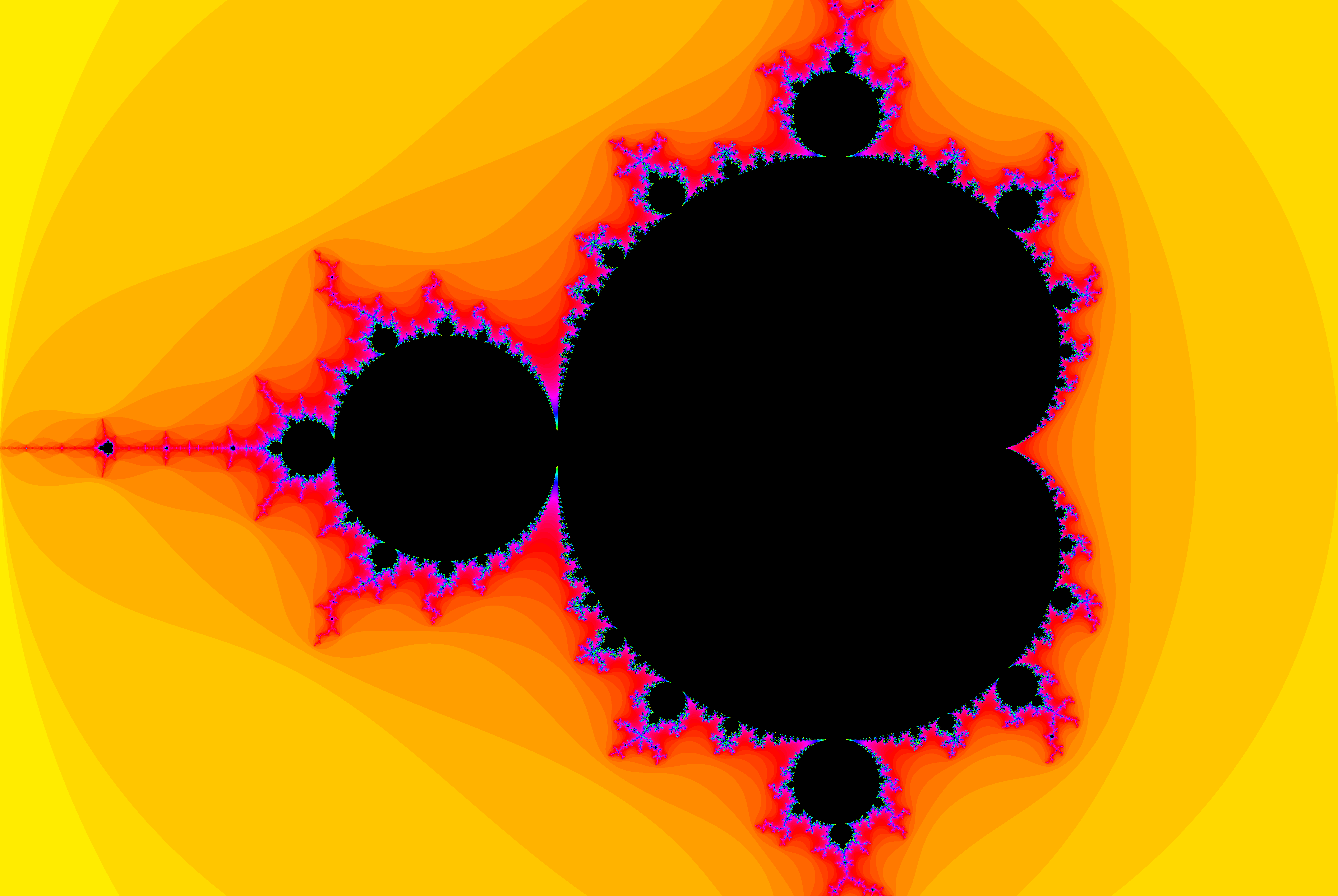}
  \caption{Window 1}
  \label{fig:win1}
\end{subfigure}
\hfill
\begin{subfigure}[b]{0.475\textwidth}  
  \centering 
  \includegraphics[width=\textwidth]{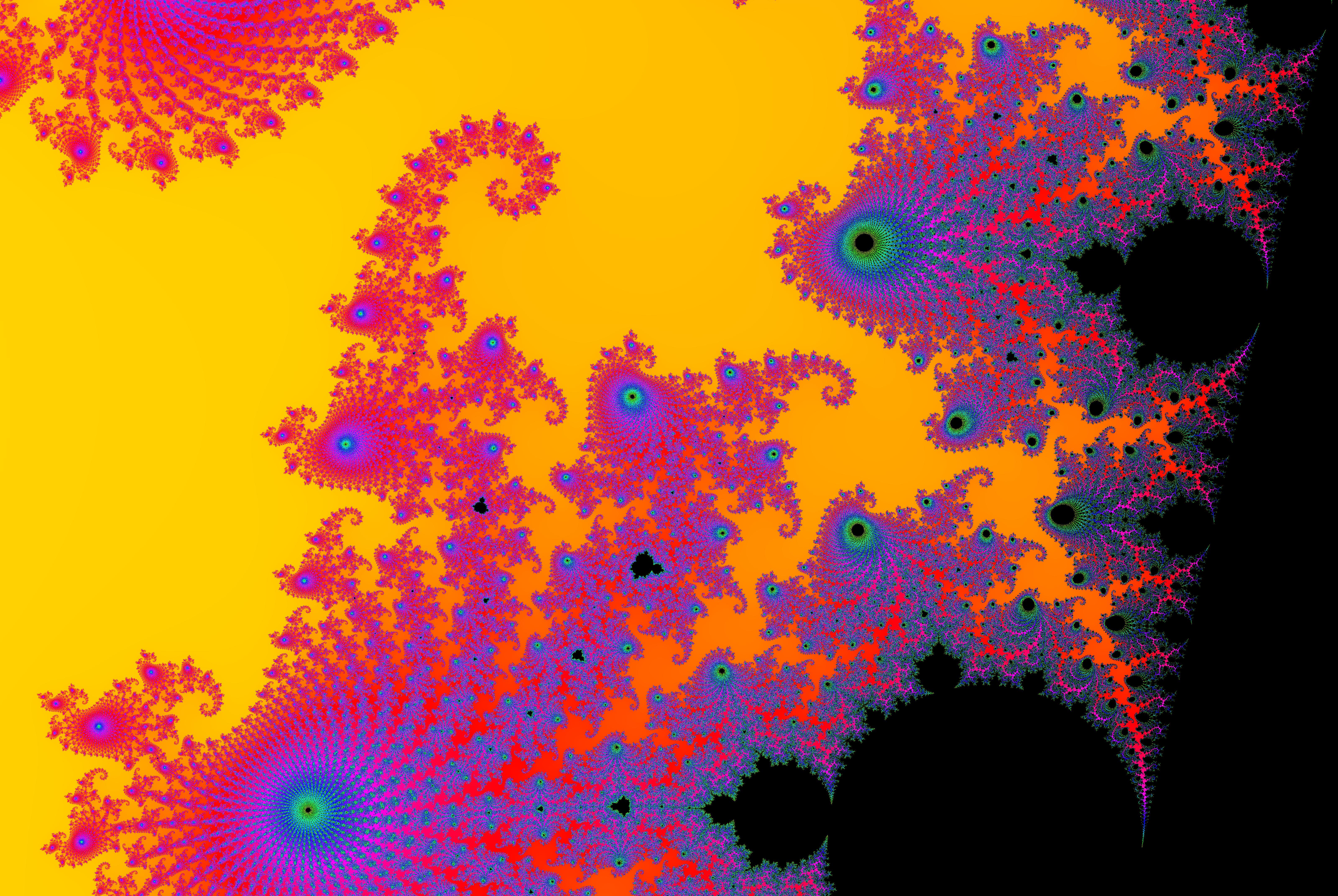}
  \caption{Window 2}    
  \label{fig:win2}
\end{subfigure}
\vskip\baselineskip
\begin{subfigure}[b]{0.475\textwidth}   
  \centering 
  \includegraphics[width=\textwidth]{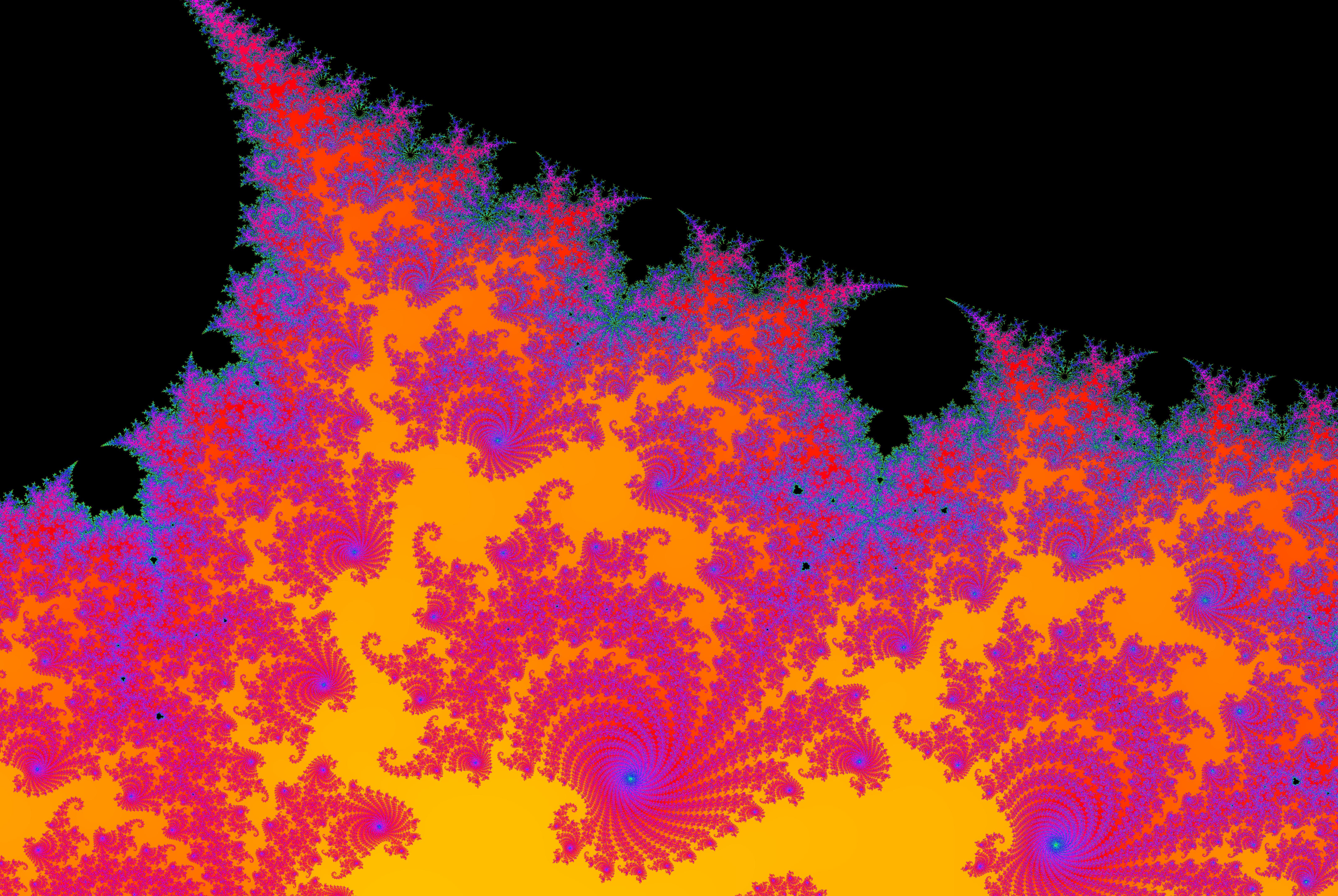}
  \caption{Window 3}    
  \label{fig:win3}
\end{subfigure}
\hfill
\begin{subfigure}[b]{0.475\textwidth}   
  \centering 
  \includegraphics[width=\textwidth]{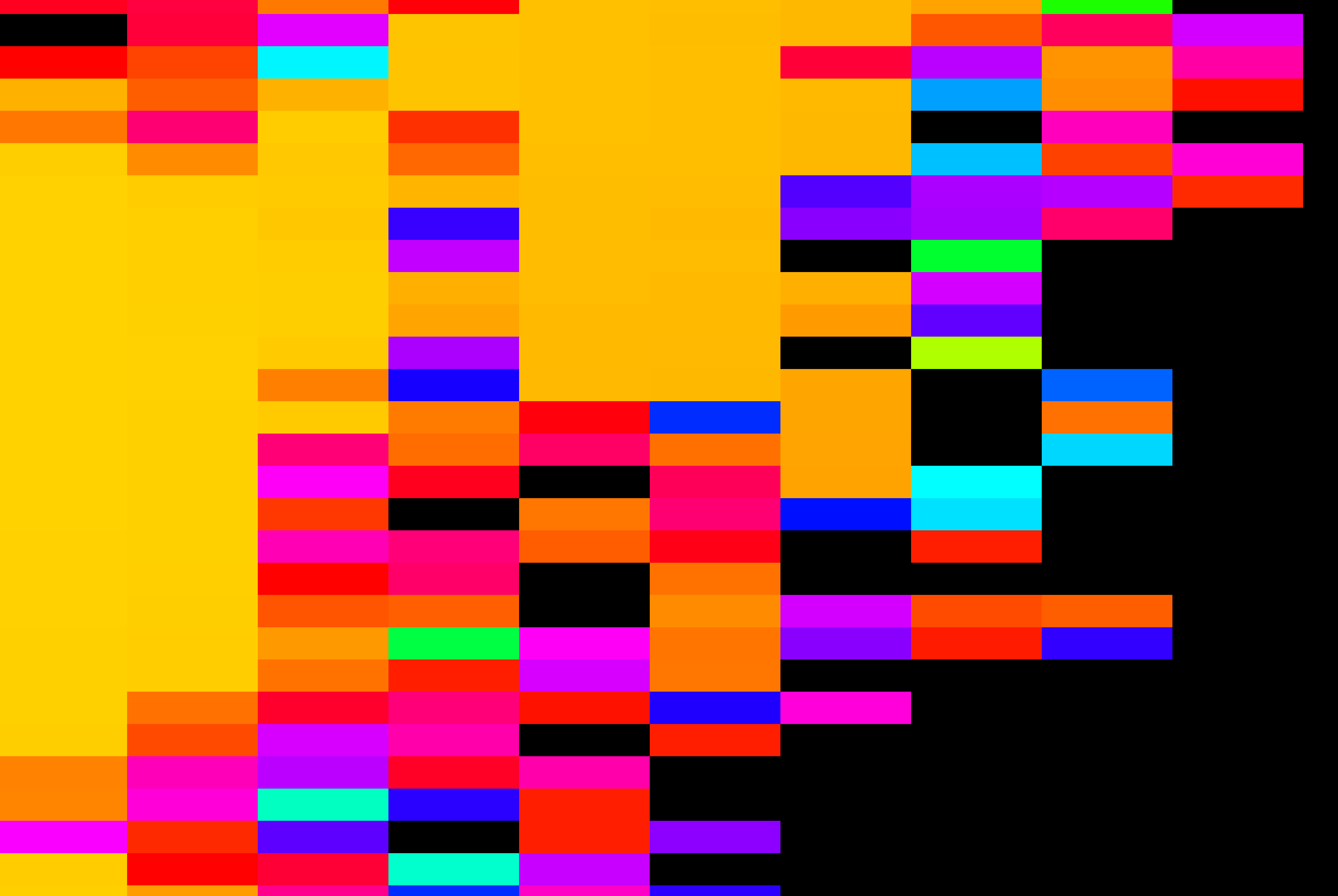}
  \caption{Window 2 calculated with half precision}    
  \label{fig:win2half}
\end{subfigure}
\caption{(a)-(c) The three regions of the Mandelbrot set selected for the runtime evaluations. (d) Window 2 calculated with half precision.} 
\label{fig:wins}
\end{figure*}

For this trial three distinct windows of the Mandelbrot set were arbitrarily selected (see Table \ref{tab:wins} and Figures \ref{fig:wins}a-c), representing modest, intermediate and high workload. All three windows had a size of 1600x1072 pixels.
For each of the architectures 
%that Android currently supports
(Intel-x86, Intel-x86\_64, Arm-v7, Arm-v8, GPU), each possible number of threads available on the device (except GPU) and each precision available, the wall clock time of the Mandelbrot algorithm has been recorded 50 subsequent times. Every  ten cycles a pause of ten seconds was introduced. This pause should allow the processors to cool down and avoid throttling due to overheat.
In order to test different parallel execution paradigms (threads, SIMD (single instruction multiple data), GPU) and combinations thereof, the following flavors of the algorithm have been implemented for each architecture: pure Java, pure C, C and handwritten scalar assembler, C and handwritten SIMD assembler, C and OpenCL \cite{OpenCL} with scalar instructions and C and OpenCL with vector instructions. Single threaded and multithreaded versions have been used. The multithreaded versions were implemented using the producer-consumer principle \cite{Dijkstra1972} and the workload was divided into 67 chunks of 16 subsequent lines. Assembler instructions were implemented as inline assembler in C. Parts of the program written in C were called from Java using JNI (Java native interface). For all versions that contained C, multithreading was implemented in C with the pthread library. % library in order to avoid multiple switching from Java to C.
On all devices, WLAN was disabled before the program execution was started. For the Java, C and OpenCL versions, care has been taken to avoid implicit type casts. Conditional jump instructions have been eliminated wherever possible in the assembler code.
See Table \ref{tab:execenvir} for the list of executions environments. 

All implementations (independently of the programming language) followed the same scheme: outer loop over the lines, inner loop over pixel of a line and a while loop for the calculations for each pixel (see Listing \ref{lst:mandel1}). The C programs were compiled with Clang 8.0 and the default compiler options set by Android Studio. Additionally, the optimization level was set to -O3.

For all assembler programs except for the x86 SIMD variant no local variables were required. Instead, all intermediate results of the calculations were kept in processor registers. The only RAM (random access memory) access was the storage of the results. %(once per pixel/every two pixels/every four pixels). 
For the implementation of the SIMD x86  assembler three local variables were necessary, as register pressure was too high. The local variables were accessed only in the outer loops.

\begin{lstfloat}
\begin{lstlisting}[basicstyle=\ttfamily,language=Java,caption={Java implementation of the calculation of a window of the Mandelbrot set for one thread and single precision},captionpos=b,label=lst:mandel1]
float ci = starty;
for( int i1=0; i1<HEIGHT; i1++ ){

  float cr = startx;
  for( int i2=0; i2<WIDTH; i2++ ){

    float zr = 0;
    float zi = 0;
    int iter = 0;

    while (((zr*zr + zi*zi)<=4f) 
                 && (iter<MAXITER)){
      float t = zr*zr - zi*zi + cr;
      zi = 2f*zr*zi + ci;
      zr = t;
      iter++;
    }

    result[i1*WIDTH + i2] = iter;
    cr += xstepsize;
  }
  ci -= ystepsize;
}

\end{lstlisting}
\end{lstfloat}

\subsection{Precision}

Not all types of precisions were available for all devices.
Single precision was always available. Many assembler languages support half precision as storage format, but among the CPUs available for this trial only Arm-v8 assembler supports half precision arithmetic as optional extension (Arm-v8.2) which was not present on the Arm-v8 CPU used. Half precision was available only on the GPU of the Arm-v7 tablet. Double precision was not available for the Arm-v7  SIMD instructions and on the GPU. 

\begin{table}[b!]
\centering
\begin{tabular}{ |l|c|c|c| }
\hline
   & Window 1 & Window 2 & Window 3 \\
\hline
$x_1$ & -2 & -0,739 & -0,737 \\
$y_1$ & 1  & 0,1448888 & 0,14667 \\
$x_2$ & 1  & -0,734 & -0,736 \\
$y_2$ & -1 &  0,1415 & 0,146 \\
\hline
iterations & 80 & 1500 & 3000 \\
\hline
\end{tabular}
\caption{Coordinates and maximum number of iterations of the three windows selected. $x_1 + i\cdot y_1$ = left upper corner, $x_2 + i\cdot y_2$ = right lower corner}
\label{tab:wins}
\end{table}

\begin{table*}[bth]
\centering
\begin{tabular}{ |l|c|c|c|c| } 
\hline
   & x86 & x86\_64 & Arm-v7 & Arm-v8 \\
\hline
Android Version & 7 & 7 & 9 & 7 \\
Android API & 24 & 24 & 28 & 24 \\
Manufacturer & Intel & Intel & Samsung & ASUS\\
Model & NUC7i7BNH & NUC7i7BNH & SM-T510 & ZenPad 10 \\
CPU & Core i7-7567U & Core i7-7567U & Exynos 7885 &  Mediatek MT8163\\
CPU-speed (MHz) & 400-3500 & 400-3500 & 449-1768 & 600-1300 \\
Virtualization & qemu/kvm & qemu/kvm & - & - \\
GPU & - & - & Mali-G71 MP 2 & Mali-T720 (a) \\
Precision CPU & s+d & s+d & s+d & s+d \\ 
Precision SIMD & s+d & s+d & s & s+d \\
Precision GPU & - & - & h+s & - \\
Cores       & 4 (b) & 4 (b) & 8 (c) & 4 (d) \\
RAM (MB)    &  2048 & 2048 &  2879 &  1953 \\
32/64 Bit CPU & 64 & 64 & 64 & 64 \\  \hline
\end{tabular}
\caption{Execution environments. h=half, s=single, d=double. (a) was present but could not be used due to Android operating system restrictions. (b) with Intel Hyper-threading technology (c) 2 x Cortex A73 + 6 x Cortex A53.
(d) 4 x Cortex A53}
\label{tab:execenvir}
\end{table*}

\subsection{SIMD}

SIMD instructions are available on CPUs and GPUs. On the CPU, vectors can hold four single or two double precision floating point numbers (Arm NEON and Intel SSE). Using OpenCL \cite{OpenCL} for accessing the GPU, SIMD vector instructions can have 2, 4, 8 or 16 entries even if the ISA (instruction set architecture) of the GPU does not support all of these vector widths. If the ISA does not support a vector width, the OpenCL compiler must split the vector into chunks of appropriate vector sizes or convert the instruction into a loop. 

SIMD instructions apply arithmetical or relational operations 
on all elements of a vector concurrently.
SIMD data parallelism for the calculation of the Mandelbrot set is not straightforward: neighbouring pixels generally will not need the same number of iterations to compute the result. But in the case of the Mandelbrot set, for many parts of the picture the number of iterations for subsequent pixels will be similar. A blending mask was used to exclude vector items that should not be updated any more once their final result has been computed.

\label{expl:blending}
Listing \ref{lst:mandel2} shows a part of the innermost loop with the SIMD blending procedure in x86\_64 assembler. {\tt xmm7} holds four single precision floats that represent the results of the calculation of $(re^2 + im^2)$ of four subsequent pixels. {\tt xmm10} holds four single precision constants (4.0f). In line 21 the SIMD comparison is made. The result of the comparison is written back into the {\tt xmm7} register. After the execution of line 21, all bits of each vector entry of {\tt xmm7} are zero if the comparison for the corresponding vector entry was false, one else.

{\tt xmm5} holds four 32-bit integers and is initialized with 1 before the innermost loop is entered (line 1). This register serves as increment for the iteration counter. In line 25, if the above comparison was false, the corresponding entry in {\tt xmm5} will be set to zero and always remain zero.

\begin{lstfloat}
\begin{lstlisting}[language={[x64]Assembler},caption={x86\_64 Assembler listing with SSE4.1 SIMD extensions of the innermost loop. \texttt{xmm4} holds four integer iteration counters, \texttt{xmm5} holds the iteration step (four times zero or one integer values), \texttt{xmm7} holds the result $re^2 + im^2$ of four consecutive pixels, \texttt{xmm9} has all bits set, \texttt{xmm10} holds four times 4.0f, \texttt{edx0}=0; other explanations see \ref{expl:blending}. }, captionpos=b,label=lst:mandel2,aboveskip=5mm, belowskip=5mm,basicstyle=\ttfamily]
  movdqa xmm5,xmm14   
        ; xmm5 <- (1,1,1,1).U32, 
        ; iteration step
  movdqa xmm4,xmm15   
        ; xmm4 <- (0,0,0,0).U32, 
        ; iteration counter
  mov rcx,r9   ; ecx <- MAXITER

.Label2:

  cmp ecx,0    ; maximum iterations 
               ; reached ?
  je .Label3   ; yes -> quit inner 
               ; loop
  dec ecx      ; ecx--

  (...) ; eleven SIMD floating point
        ; instructions to update 
        ; real and imaginary part

  cmpps xmm7,xmm10,2  
        ; (re^2 + im^2) <= 4.0f ?
        ; xmm10 = 
        ; (4.0f,4.0f,4.0f,4.0f).F32
  pand xmm5,xmm7      
        ; blend iteration offset
  paddd xmm4,xmm5     
        ; step vector iteration 
        ; counter
  ptest xmm5,xmm9     
        ; ZF=1 if xmm5 & 
        ;  xmm9 == (0,0,0,0).U32
        ; xmm9 has all bits set
  cmovz ecx,edx       
        ; quit loop if ZF==1 
        ; (edx=0)
  jmp .Label2
 
.Label3:
\end{lstlisting}
\end{lstfloat}

{\tt xmm4} is the iteration counter and holds four 32 bit integers. They are initialized with zero before the innermost loop is started (line 4). This register counts the number of iterations needed until the condition in line 21 becomes false. Line 27 adds one or zero to each vector entry of the iteration counter, depending on the values of {\tt xmm5}. Line 30 tests if all entries of {\tt xmm5} are zero. If all entries of {\tt xmm5}  were zero, in line 34 {\tt ecx} is loaded with zero. %(which means that the innermost loop will be quit afterwards). 
If not, %all values if {\tt xmm5} are zero, 
{\tt ecx} will remain untouched.
{\tt ecx} can become zero if the comparison in line 21 is false for all vector entries or if the maximum number of iterations has been reached. Similar constructs have been used in all SIMD assembler implementations and show that SIMD can be used even if the vector entries need a similar but unequal  number of arithmetic operations to calculate the result. The instructions executed in the lines 11-37 in listing \ref{lst:mandel2} correspond to the lines 11-16 in listing \ref{lst:mandel1} (executed for four values in parallel).

\subsection{GPU - OpenCL}
\label{refx1}

For deploying calculations on the GPU, the OpenCL \cite{OpenCL} standard has been chosen. 
OpenCL is a framework that uses a C-style programming language and implements task and data based parallelism. Programs can be compiled and deployed on a great variety of architectures like CPUs, GPUs, FPGAs (Field Programmable Gate Array) and many more. 
An advantage of OpenCL is its portability across devices \cite{DU2012391}. %OpenCL has been for accelerating calculations for many problems, e.g. \cite{Rucci2018} and \cite{Cornelis2017}.

%Android devices were selected as applications can be designed, implemented and deployed comfortably using Android studio. 
Many Android tablets are shipped with an OpenCL-capable GPU and the necessary libraries, although 
OpenCL is not officially supported by Android. 
% but many tablets are equipped with a OpenCL shared library and a GPU that supports OpenCL. 
OpenCL can be used from C linking an appropriate shared library.
C compilers require shared libraries to be present at compile-time. This is not the case for the usual Android project build process, which takes place on an external host (e.g., with Android Studio). 

Many manuals recommend building and shipping the device's OpenCL library together with the apk file \cite{Ross2014}. This approach has compatibility issues with other GPU vendors and future GPUs of the same manufacturer.
For this project another approach has been selected: The POSIX standard provides a method to dynamically load shared libraries and resolve the symbols at runtime (with the '{\tt dlopen}' command). The native OpenCL libraries do not need to be present at compiletime.
Once a library has been loaded at runtime, the symbols (names of the methods) in the library must be resolved manually. In the current implementation of the wrapper library this is not done immediately but only on request if a call to a method is really needed.

Unfortunately there is one obstacle to this approach: Since Android 7 Google does not allow vendor provided libraries to be loaded if they do not appear in a special list. If this list is stored on a read-only filesystem, it can be impossible to create that list for older devices (which was the case for the Arm-v8 tablet).

%One approach to exhibit the OpenCL functionality of the shared library to an Android project consists in copying the OpenCL library on the device into the Android project on the host and making it part of it (as external native library). The library would be packed and shipped together with the apk file after compilation. This is the way recommended by most manuals. Beside of copyright issues, this method has several drawbacks: After compilation the project would not be compatible to GPUs of other manufactors or GPUs released by the same manufactor in the future.     

%To avoid this obstacle one can load the library from the device and resolve all symbols at runtime (with the dlopen command). This method is provided by the POSIX standard. Unfortunately there is one difficulty: Since Android 7 google prohibits the use of third party libraries that are not present and packed together with the project at compile-time. But there are some exceptions from this rule: Vendor and manufactor provided libraries on the device can be used even if they are not present at compile time if they appear in a special whitelist. This is the case for newer devices. If the list is not present on older devices, it can be impossible to create the file if the subdirectory where the file has to be created resides on a read-only filesystem (which was the case with the Arm-v8 tablet).

\subsubsection{The wrapper library}

For this project an OpenCL wrapper library has been created. This library implements the entire OpenCL 3.0 standard and forwards the calls to its methods to the corresponding method of the native OpenCL shared library of the device. The wrapper library is compatible with any (lower) OpenCL version of the native OpenCL library. If one tries to call OpenCL methods of a standard that is not provided by the native library, an error is returned (the corresponding symbol can not be resolved).
The wrapper library is linked to the app at compile-time.

At the very beginning of the program, a method has to be called to load the native library from the device. The only parameter for this method is the path to the native library on the device. Figure \ref{fig:sequwrap} in the supplements shows a sequence diagram of the use of the wrapper library in conjunction with the native library.

Unlike desktops or laptops where the programmer is supposed to have permanent access to all resources (especially the GPU) during the whole lifespan of a program, on Android devices a user can choose to stop an app at any time. Programs that use the GPU have to account for that, because otherwise the resources on the GPU might remain blocked and the GPU might not be fully available by other programs (especially if the GPU has its own memory). Ideally the app should also free any memory not used when it looses the focus. To accomplish these issues two mechanisms have been implemented. The first one is part of every call to an OpenCL method of the wrapper library and the programmer does not have to take care of this mechanism. The second one must be integrated into the app because the resources blocked on the GPU should be freed by the programmer and not the wrapper library.

\subsubsection{The unloading mechanism}
\label{expl:unload}

Android devices usually have limited memory resources. If an app loses focus, it is put on the so-called 'backstack'. 
If the dynamically loaded native OpenCL library is temporarily unused, the memory it occupies can be released using a reader/writer lock, a mutex and a flag. A RW lock is needed to avoid unloading the native library while there are still active calls to the native library.
A flag holds the status of the native library (has never been loaded, currently loaded, unloaded) and is protected by the mutex in order to allow  concurrent access to it. 

\begin{figure*}[tb]
\centering
\includegraphics[width=0.9\textwidth]{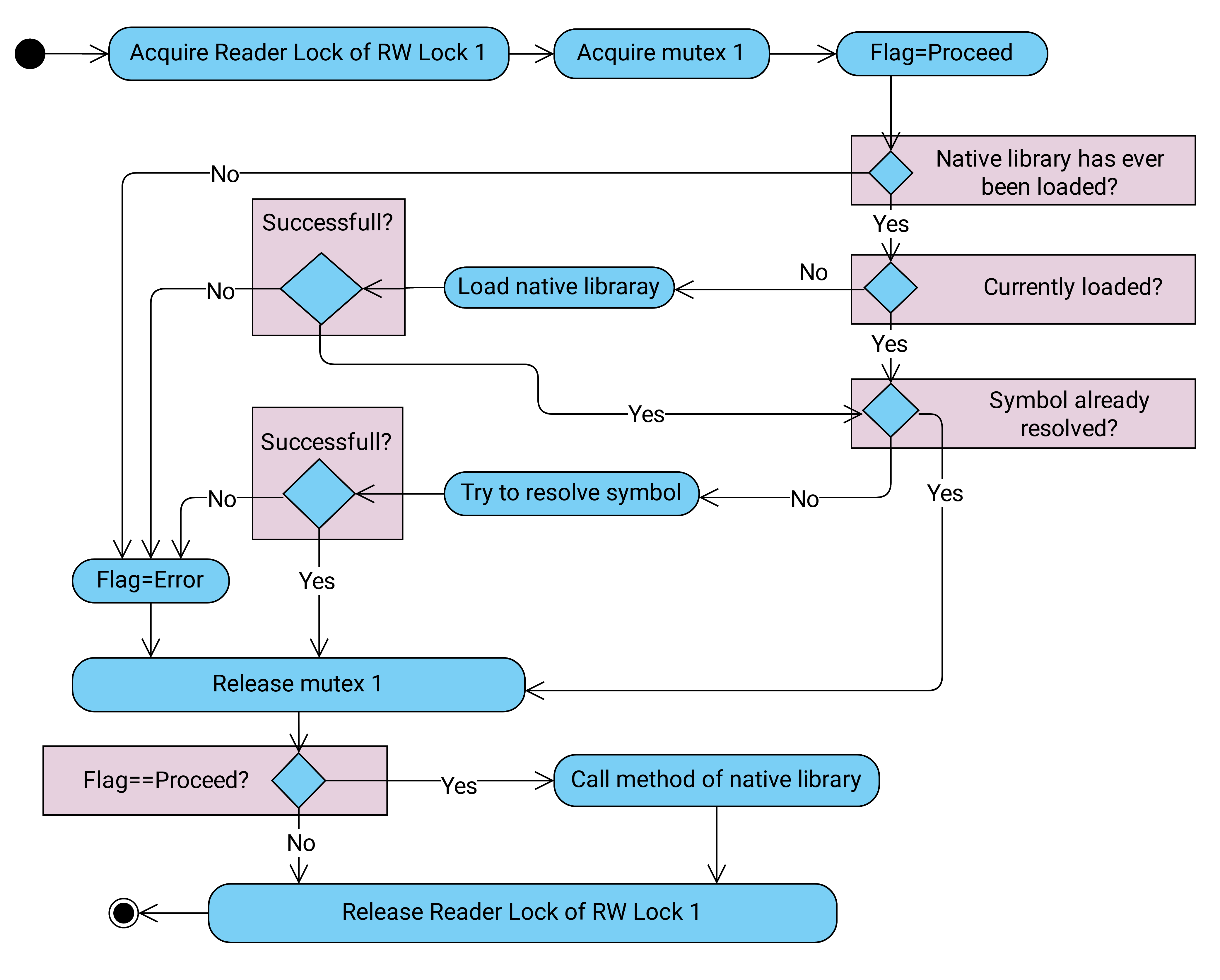}
\caption{Activity diagram of the mechanism that allows to load and unload dynamically the native OpenCL library. This mechanism is part of every OpenCL method in the wrapper library. For further explanations see \ref{expl:unload}.}
\label{fig:lock1}
\end{figure*}

Figure \ref{fig:lock1} depicts the use of the RW locks and the mutex.
At the beginning of every call to a method of the wrapper OpenCL library, the reader lock of the reader/writer lock and the mutex that protects the flag are acquired. Deadlocks cannot occur, as the locks are always blocked  in the same order.  
Once exclusive access to the flag has been gained,
the method of the wrapper library checks if the native library has ever been loaded. If not, the method returns with an error because the path and filename from where to load the native library have not yet been specified. If path and filename are known, the method of the wrapper library tests if the native library is currently loaded and tries to load it, if it has been unloaded. 
Furthermore, the wrapper library checks if the symbol (i.e. the method name) in the dynamically loaded library has already been resolved. If not, the wrapper library tries to resolve the symbol.

The mutex is released before the call to the corresponding method of the native OpenCL library and the reader/writer lock afterwards.
This mechanism allows the use of OpenCL functions of the native library by multiple threads concurrently. 
The lock should be reader-preferred, because otherwise, as long as there are readers waiting, the library would be reloaded immediately after it has been unloaded. 

\begin{figure*}[tb]
\centering
\includegraphics[width=0.8\textwidth]{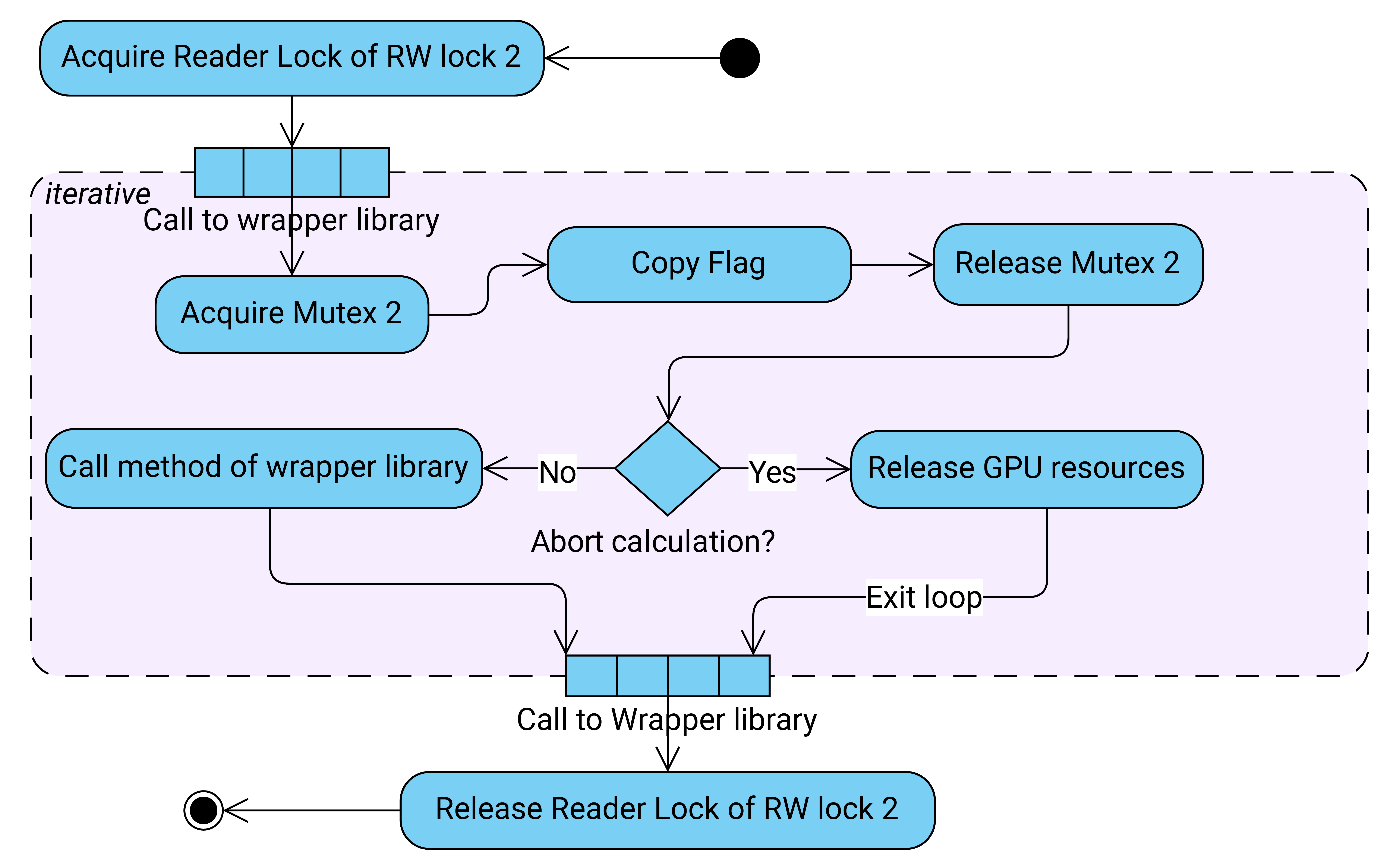}
\caption{Activity diagram of the mechanism that should be executed every time before a call to the wrapper library is made. This diagram should be implemented in the app. For further explanation see \ref{expl:requnload}.}
\label{fig:lock2}
\end{figure*}

\begin{figure*}[bt]
\centering
\includegraphics[width=0.7\textwidth]{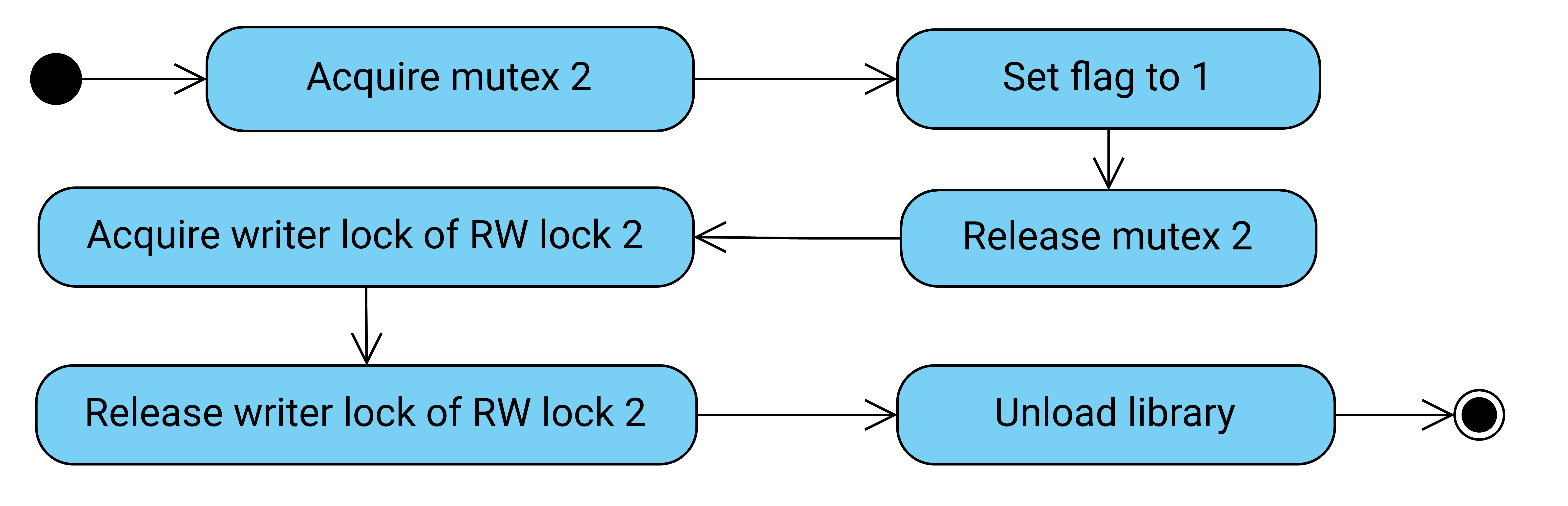}
\caption{Activity diagram of the method the locks the access to the wrapper library. For further explanations see \ref{expl:requnload}.}
\label{fig:lock3}
\end{figure*}

%When a program is stopped and put on the 'backstack' or when the programmer does not need the GPU any more, the native OpenCL library can be unloaded. Memory can be saved if no kernels are executed on the GPU.

The wrapper library has a method that allows one to unload the dynamically loaded native library. %Care must be given to not unload the library while there are still active calls pending.
Before releasing the dynamically loaded OpenCL library, the method responsible for unloading the library acquires the writer part of the reader/writer lock and gains exclusive access to the native OpenCL library. 
There are no pending calls to the native library once the writer lock has been acquired. Next the mutex for the flag is acquired and the flag is set accordingly.
Afterwards the library is unloaded using '{\tt dlclose}'.
Finally the mutex and the lock are released.
%Meanwhile, no other method can call functions of the OpenCL library (the writer lock guarantees exclusive access).
The Android operating system decides if and when the freed shared library is effectively unloaded and the memory freed.
%After the library has been unloaded, all subsequent calls to OpenCL methods of the wrapper library try to reload the library again before it is used.

If calculations on the GPU should be resumed, the shared library does not have to be reloaded explicitly but will be loaded implicitly upon the next call to a wrapper library method by the mechanism described above.
If the library should not be reloaded any more, the programmer has to make sure that no more calls to the wrapper library are made once it has been unloaded. This is the purpose of the second locking mechanism described in the next subsection.

%This mechanism only prevents that the library is unloaded while there are still calls to the library executed. This mechanism is not responsible for freeing the buffers allocated by the programmer on the GPU device.
%The reader/writer lock guarantees only that the library is not unloaded while methods of the native library are executed. %It does not guarantee that the memory allocated on the GPU by the programmer is freed. 

\subsubsection{GPU memory release upon request}
\label{expl:requnload}

If an app is stopped or killed (by the operating system or the user), the app's '{\tt onStop}' and '{\tt onDestroy}' methods are called by the operating system. The execution of these methods cannot be delayed for a longer period for finishing calculations on the GPU, because the user would otherwise be notified that the app is not responding. On the other hand, once these methods have been called, resources on the GPU should be freed correctly.

The OpenCL wrapper library is state-less and therefore the resources a software developer has blocked on the GPU have to be freed by the app and cannot be freed by the
%. Therefore the correct release of the resources has to be implemented by the app and cannot be accomplished by the 
wrapper library.

To address this issue,  a writer-preferred reader/writer lock, a mutex and a flag are sufficient. The flag is protected by the mutex.
Initially the flag is set to zero.
%If a method needs to acquire the reader/writer lock and the semaphore, the lock must be acquired first (cascaded access to avoid dead-locks).  
%The method described now is not part of the wrapper library but has to be implemented by the programmer.

Figure \ref{fig:lock2} illustrates a possible mechanism to stop execution of the app and free resources blocked on the GPU.
If a method of an app wants to use the GPU, the method should acquire the reader lock of the reader/writer lock before the first call to the wrapper library (except the call to the method that loads the library the very first time). This lock is held until the last call to the wrapper library has been made. After the lock has been acquired, the mutex should be acquired and the flag should be checked. If the flag is not zero, the mutex and the lock should be released and the method should return with an error. %after having released all GPU resources.
If the flag is zero, the mutex should be released and the calculations on the GPU can be carried out. After all calculations have been finished, the reader lock should be released. The reader lock can be acquired multiple times concurrently by different threads, allowing the parallel use of the wrapper library.

The approach described here works well if the kernel execution lasts only for a short period.
If calculations on the GPU are more complex and, especially if they involve several subsequent calls to different kernels, the reader lock should be acquired at the very beginning and released once all GPU resources have been freed. The execution of a kernel cannot be easily stopped. But between kernel executions, the flag can be tested (using the mutex) and if the flag meanwhile has been set to one, the resources should be freed, the reader lock should be released and an error should be returned.

%If one has to use OpenCL kernels whose execution takes very long, there is a possibility of stopping kernels while they are executed. One must create a small buffer of shared memory on the GPU. 
%If the execution of the kernel on the GPU should be stopped by another thread, a new kernel is launched concurrently using an independent command queue. The only scope of this kernel is to write a predefined value into the shared buffer. Afterwards, the second kernel quits.
%During the execution, the first kernel occasionally checks the contents of the shared memory and quits if the predefined value value has been stored in the buffer. 

If the '{\tt onStop}' and '{\tt onDestroy}' methods of an app are called by the operating system, they acquire the mutex (without first acquiring the lock) and set the flag to one (see figure \ref{fig:lock3}). The mutex is released immediately afterwards.
Then the writer lock of the reader/writer lock should be acquired.
It is important that the lock is writer-preferred, because otherwise the method could starve.
Once the writer lock has been acquired, no more readers can hold the reader lock and all GPU resources must have been freed. 
The writer lock should be released immediately afterwards and the native OpenCL library can be unloaded. 
No GPU resources will be allocated ever again and no calls to the wrapper library can be made, because the flag has been set to one and is checked before the GPU is used.
The native library can be unloaded at this moment because no readers will be waiting  any more (all resources have been released, no new readers can start to use the GPU).

If the calculations should be resumed ('{\tt onResume}') it is sufficient to acquire the mutex and set the flag again to zero (and release the mutex afterwards). If the native library has been unloaded, it will be loaded automatically when the next call to a wrapper functions is made.

Besides the initial call to load the native OpenCL library and the locking mechanism for the premature termination of the calculations, 
any program using OpenCL should run without any modification using the  methods of the wrapper library instead of the methods of the native OpenCL library (just the linking process of the C compiler has to be modified accordingly). 
Of course neither the mechanism of loading/unloading the native OpenCL libraries at runtime nor the mechanism of premature suspension are  specific to Android and work on all POSIX compliant operating systems.

\subsection{The App}

In order to test the mechanisms described above, a small app with three activities has been written and deployed on the tablets. The first activity allows to enter the path and file name of the native OpenCL library on the device. The second activity is responsible for the proper selection of the parameter for the calculations (method, number of threads, vector size, window and precision). The third activity performs the calculations on the device based on the selections made in the second activity and displays the final result. %The source code of the OpenCL wrapper library can be downloaded from {\tt https://gitlab.\newline com/rfritze/androidopencl.git}. The app is also available for Android devices in the Google Play App Store (only for Android 6.0/API 23 or higher, search "Mandelbrot-OpenCL").

\section{Results}

\begin{figure*}
\centering
\includegraphics[width=0.9\textwidth]{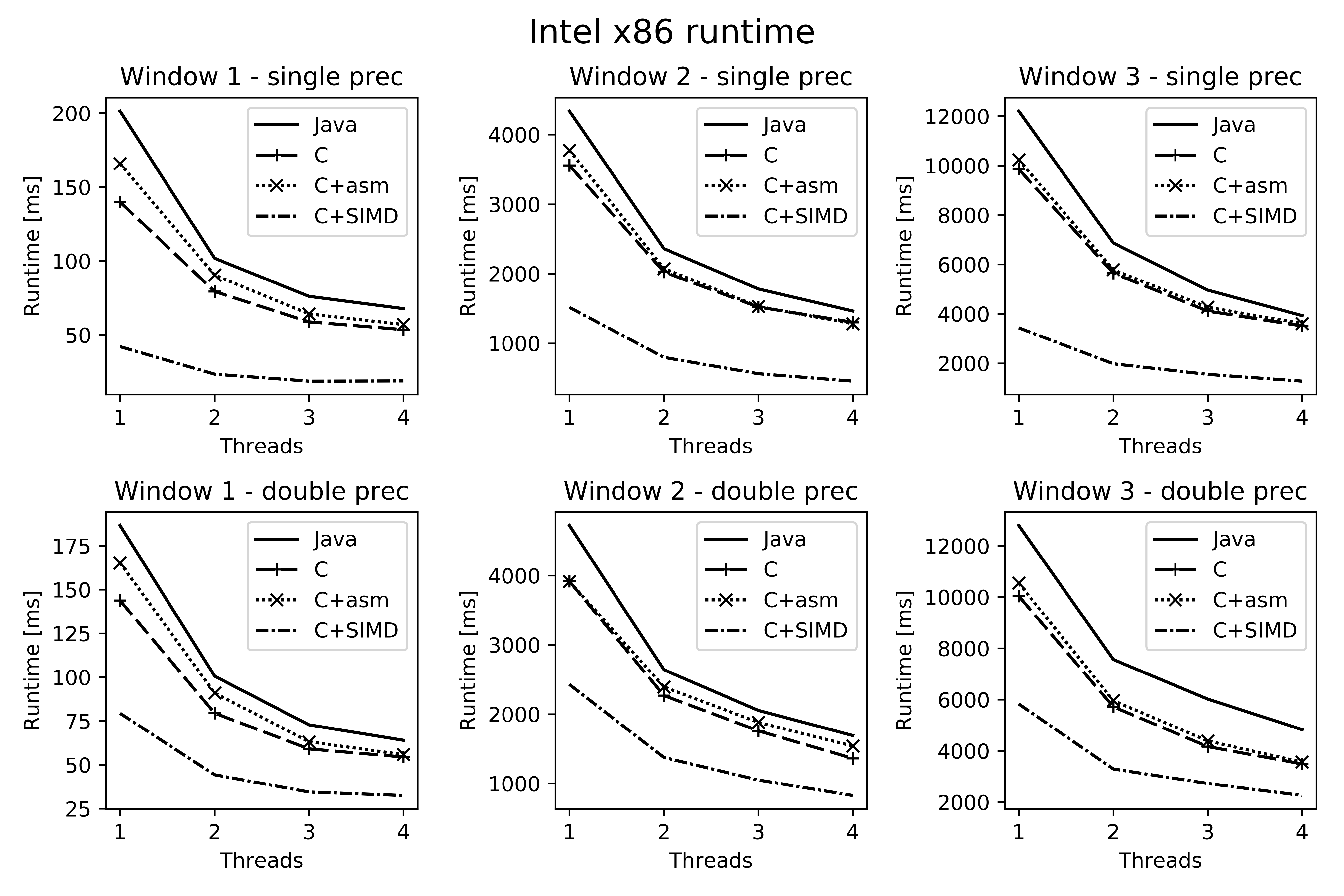}
\caption{Median of the runtime of the Mandelbrot algorithm on the virtual Intel x86 tablet. Further explanations see \ref{result:intel}.}
\label{fig:runtime_x86}
\end{figure*}

\begin{figure*}
\centering
\includegraphics[width=0.9\textwidth]{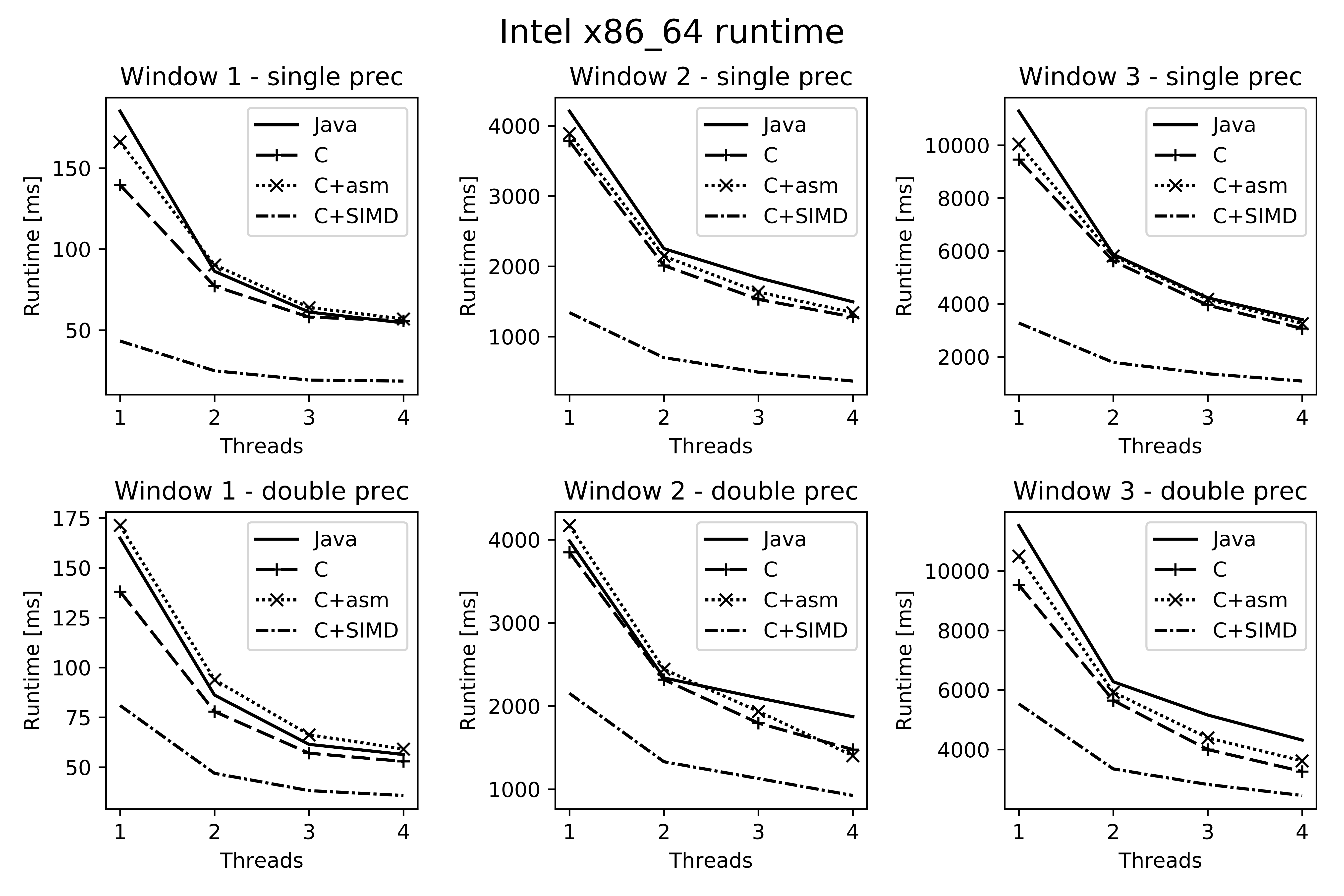}
\caption{Median of the runtime of the Mandelbrot algorithm on the virtual Intel x86\_64 tablet. Further explanations see \ref{result:intel}.}
\label{fig:runtime_x86_64}
\end{figure*}

\begin{figure*}
\centering
\includegraphics[width=0.9\textwidth]{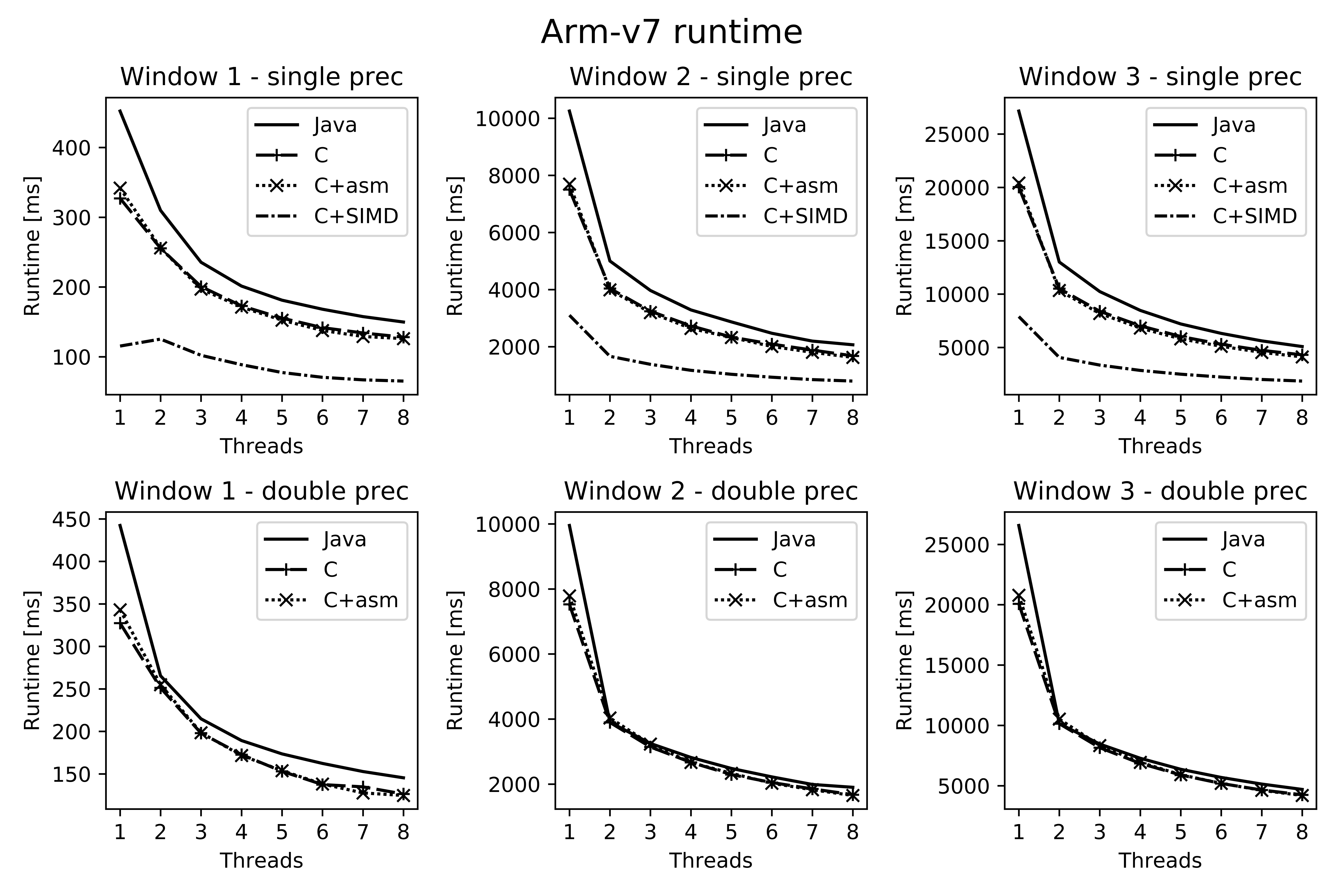}
\caption{Median of the runtime of the Mandelbrot algorithm on the Arm-v7 tablet. Further explanations see \ref{results:armv7}.}
\label{fig:runtime_armv7}
\end{figure*}

\begin{figure*}
\centering
\includegraphics[width=0.9\textwidth]{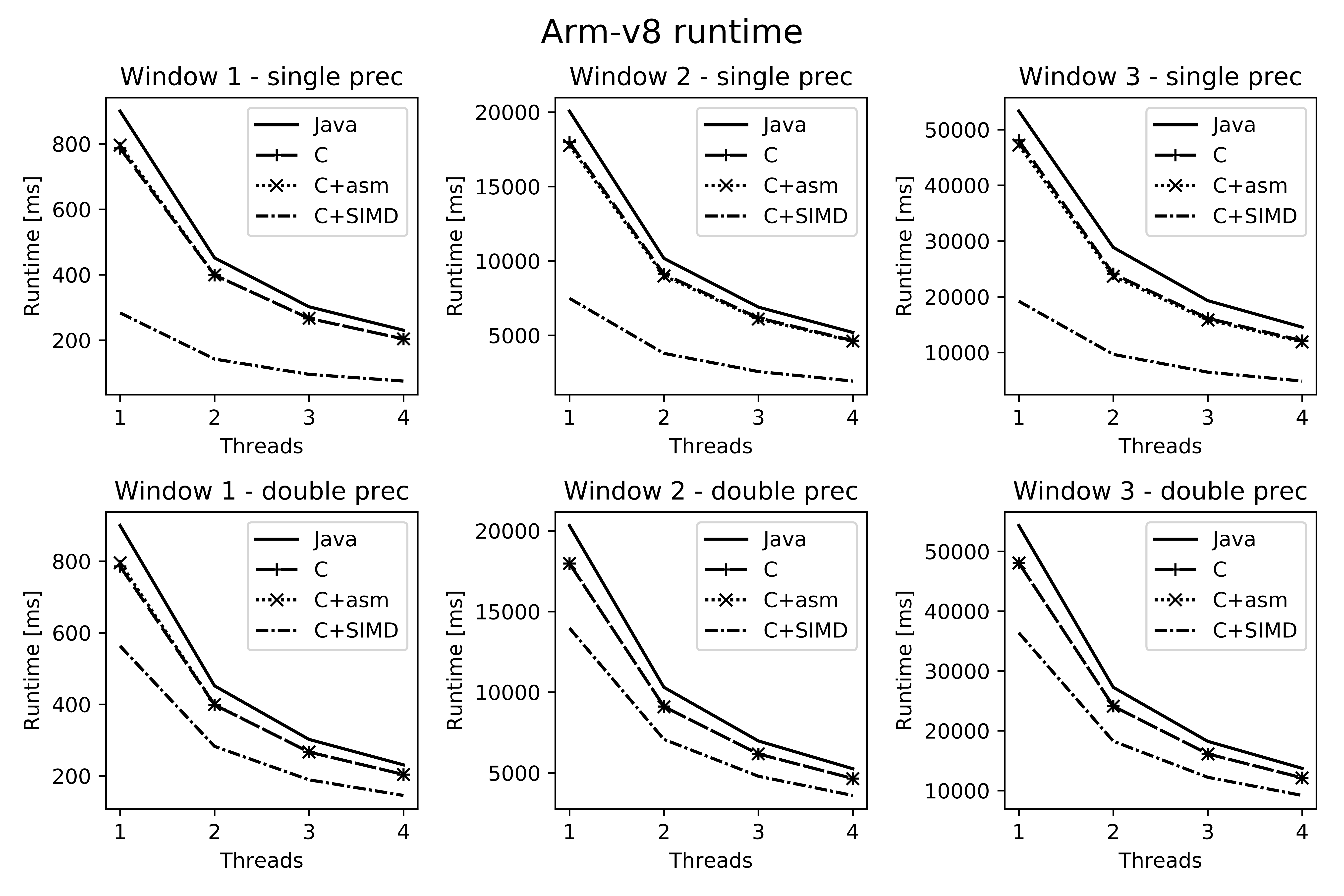}
\caption{Median of the runtime of the Mandelbrot algorithm on the Arm-v8 tablet. Further explanations see \ref{results:armv8}.}
\label{fig:runtime_armv8}
\end{figure*}

Tables \ref{tab:result_x86_1} to
\ref{tab:result_armv8_2}  in the supplements show the results for the calculations on the CPU. 
The first line of each cell shows the mean (or the median if the sample was not normally distributed) of the wall clock time in milliseconds. The second line shows the confidence interval (if the sample was normally distributed). The third line shows the speedups. The first value is the speedup with respect to Java with the same number of threads, the same window and the same precision (first value of the same row). The second value shows the speedup with respect to the same method, same window, same precision but only one thread (first value in each column). The third value shows the speedup with respect to the single threaded Java version with the same precision and same window (value in the left upper corner for each window).
The small letters indicate the method that has been used for the calculation of the significance.
Normal distribution was tested with the Shapiro-Wilk test \cite{Shapiro1965} and homoscedasticity was tested with the Levene test \cite{Levene1960}. If two samples were normally distributed with equal variances, the student t-test \cite{student1908} was used to test significance of the difference of the means. If the two samples were distributed normally but had unequal variances, the Welch test \cite{Welch1947} was used. If the two samples had similar distributions but were not normally distributed, the Wilcoxon-Mann-Whitney test \cite{Wilcoxon1945} \cite{MannWhitney1947} was used. If none of the cases above were true, the Median test \cite{Mood1950} was used. 
The results displayed in the tables are summarized in the Figures \ref{fig:runtime_x86}, \ref{fig:runtime_x86_64},
\ref{fig:runtime_armv7} and \ref{fig:runtime_armv8}.

For all types of architecture and for all methods, for a given number of threads and a given precision there was a %highly
significant (p$<$0.05) difference in the runtime between the windows. The more one zoomed into the Mandelbrot set, the longer took the calculations.
%The more one zoomed into the Mandelbrot set, the more time the calculations required to complete. 
When more threads were used, the execution completed
faster. The difference of runtime was highest between one and two threads and
declined as more threads were used.
%The more threads were used, the faster was the execution. The difference of runtime was highest between one and two threads and declined the more threads were used.
Except for the Intel architectures, runtimes can not be directly compared between the different devices as the execution environments differ. 
Multithreaded SIMD instructions gave had a very good performance.
Independently of the architecture the maximum speedups were achieved with SIMD and with four threads they ranged from 5.23 to 15.38 (median 10.485) for single precision and from 4.35 to 5.98 (median 5.75) for double precision.

\subsection{Intel based instruction sets (x86 and x86\_64)}
\label{result:intel}

The Java version was consistently faster on the x86\_64 than on the x86 device. For all other versions there was no major runtime difference, independently of the precision, the window and the number of threads
(see Tables \ref{tab:result_x86_1},
\ref{tab:result_x86_2},
\ref{tab:result_x86_64_1},
\ref{tab:result_x86_64_2},
and Figures \ref{fig:runtime_x86}, \ref{fig:runtime_x86_64}). 
The code optimization for both Intel architectures and the fact that the x86 SIMD variant needed three local variables had no influence on the runtime.

%Although handwritten assembler code was different for the two architectures and for the x86 SIMD version three local variables had to be used. 

With two exceptions, the single threaded Java version was the slowest for all windows. By trend the single precision versions were a bit faster than the double precision calculations, although sometimes, especially for the single threaded versions, double precision was faster than its single precision counterpart.

Plain C was always a bit faster than Java. The runtimes of handwritten scalar assembler lay somewhere between Java and C with two exceptions were handwritten double precision assembler was slower than Java. SIMD was much faster than all other methods. 

For single precision a maximum speedup with respect to a single threaded Java version of about 15.38 (x86, window 1, SIMD, 4 threads) and for double precision a maximum speedup of about 5.84 (x86, window 1, SIMD, 4 threads) could be achieved. For the x86\_64 architecture the maximum speedups were a little bit lower.
SIMD single precision instructions were almost twice as fast as double precision instructions. 

%The runtime for Java with one thread and single precision was 187.26 ms (95\%CI   184.30-190.23) for window \#1, 3905.16 ms 
%(95\%CI 3899.13-3911.18ms) for window \#2 and 10333.49 ms (95\%CI 10322.33-10344.65 ms) for windoow \#3.    

\subsection{Arm based architectures}

%The programs compiled for Arm-v7 and Arm-v8 architecture were tested on tablets.

\subsubsection{Arm-v7}
\label{results:armv7}

Calculations in single precision with Java were slightly slower than double precision  (see Tables \ref{tab:result_armv7_1}, \ref{tab:result_armv7_2}, \ref{tab:result_armv7_3} and Figure \ref{fig:runtime_armv7}), whereas for C and handwritten scalar assembler there was not such a difference. SIMD instructions with arm-v7 architecture support only single precision. 

A clear explanation for this phenomenon could not be found. Java by default does not convert single precision floating point values to double precision for the standard arithmetic operations (plus, minus, multiplication) if both values have single precision.

As more threads were used, the calculations with Java were performed faster, but the difference in speedup between subsequent numbers of threads became smaller.
%For Java, the more threads were used, the faster were the calculations, but the difference of the speedup between subsequent numbers of threads became lower.
The decline in speedup depended on the precision and the window:  The longer the calculations required, the greater the resulting speedup.
%The longer took the calculations, the higher was the speedup.
The same decline of speedup was observed with the C, scalar assembler and handwritten assembler  implementations. 

For higher workloads (Window 2 and 3) there is a marked loss of speedup gain if three or more threads are used. This phenomenon can be explained by the fact that the Arm-v7 tablet used has an Exynos 7885 processor, that has two fast Cortex A 73 cores (clock rate 2,3 GHz, out of order execution). The other cores are slower Cortex A 53 processors (clocked at 1,6 GHz, in order execution).

C and handwritten assembler were almost equally fast and always faster than Java by a small amount. SIMD assembler was always much faster than any other method.

The maximum speedups achieved (with respect to a single threaded Java implementation) were 14.58 (SIMD with eight threads) for single precision and 6.44 (C with eight threads) for double precision.

\begin{table}[tbh]
\setlength{\tabcolsep}{3pt}
\centering
\begin{tabular}{ |c|c|c|c|c| }
\hline
  &   & \multicolumn{3}{c|}{Window} \\
\hline
V & P & 1 & 2 & 3 \\
\hline
  \multirow{6}{*}{1} &\multirow{3}{*}{h} 
 & 98.44 & {\cellcolor[HTML]{D0D0D0} 519.13 }  & {\cellcolor[HTML]{D0D0D0} 1,418.84 } \\
 &  & {\footnotesize (95.49-101.39)} & {\cellcolor[HTML]{D0D0D0} {\footnotesize (513.93-524.32)} }  & {\cellcolor[HTML]{D0D0D0} {\footnotesize (*)} } \\
 &  & {\footnotesize [\underline{4.65}${}^{a}$, 1.00${}^{a}$]} & {\cellcolor[HTML]{D0D0D0} {\footnotesize [\underline{19.84}${}^{a}$, 1.00${}^{a}$]} }  & {\cellcolor[HTML]{D0D0D0} {\footnotesize [\underline{19.07}${}^{d}$, 1.00${}^{c}$]} } \\
\cline{2-5}
 & \multirow{3}{*}{s} 
 & 100.65 & 617.87 & 1,339.25\\
 &  & {\footnotesize (96.48-104.82)} & {\footnotesize (613.91-621.82)} & {\footnotesize (1,335.02-1,343.49)}\\
 &  & {\footnotesize [\underline{4.54}${}^{a}$, 1.00${}^{a}$]} & {\footnotesize [\underline{16.67}${}^{a}$, 1.00${}^{a}$]} & {\footnotesize [\underline{20.20}${}^{b}$, 1.00${}^{a}$]}\\
\hline
  \multirow{6}{*}{2} &\multirow{3}{*}{h} 
 & 113.34 & {\cellcolor[HTML]{D0D0D0} 568.43 }  & {\cellcolor[HTML]{D0D0D0} 1,520.18 } \\
 &  & {\footnotesize (110.83-115.85)} & {\cellcolor[HTML]{D0D0D0} {\footnotesize (564.60-572.27)} }  & {\cellcolor[HTML]{D0D0D0} {\footnotesize (1,516.58-1,523.79)} } \\
 &  & {\footnotesize [\underline{4.04}${}^{a}$, \underline{0.87}${}^{a}$]} & {\cellcolor[HTML]{D0D0D0} {\footnotesize [\underline{18.12}${}^{a}$, \underline{0.91}${}^{a}$]} }  & {\cellcolor[HTML]{D0D0D0} {\footnotesize [\underline{17.80}${}^{b}$, \underline{0.93}${}^{c}$]} } \\
\cline{2-5}
 & \multirow{3}{*}{s} 
 & 115.00 & 783.42 & 1,703.65\\
 &  & {\footnotesize (112.17-117.83)} & {\footnotesize (779.47-787.36)} & {\footnotesize (1,699.79-1,707.51)}\\
 &  & {\footnotesize [\underline{3.98}${}^{a}$, \underline{0.88}${}^{a}$]} & {\footnotesize [\underline{13.15}${}^{a}$, \underline{0.79}${}^{a}$]} & {\footnotesize [\underline{15.88}${}^{b}$, \underline{0.79}${}^{a}$]}\\
\hline
  \multirow{6}{*}{4} &\multirow{3}{*}{h} 
 & 105.85 & {\cellcolor[HTML]{D0D0D0} 523.52 }  & {\cellcolor[HTML]{D0D0D0} 1,309.81 } \\
 &  & {\footnotesize (102.16-109.53)} & {\cellcolor[HTML]{D0D0D0} {\footnotesize (520.68-526.36)} }  & {\cellcolor[HTML]{D0D0D0} {\footnotesize (*)} } \\
 &  & {\footnotesize [\underline{4.32}${}^{a}$, \underline{0.93}${}^{b}$]} & {\cellcolor[HTML]{D0D0D0} {\footnotesize [\underline{19.68}${}^{a}$, 0.99${}^{b}$]} }  & {\cellcolor[HTML]{D0D0D0} {\footnotesize [\underline{20.66}${}^{d}$, \underline{1.08}${}^{c}$]} } \\
\cline{2-5}
 & \multirow{3}{*}{s} 
 & 110.65 & 793.30 & 1,684.79\\
 &  & {\footnotesize (106.28-115.02)} & {\footnotesize (790.05-796.55)} & {\footnotesize (1,680.93-1,688.66)}\\
 &  & {\footnotesize [\underline{4.13}${}^{a}$, \underline{0.91}${}^{a}$]} & {\footnotesize [\underline{12.99}${}^{a}$, \underline{0.78}${}^{a}$]} & {\footnotesize [\underline{16.06}${}^{b}$, \underline{0.79}${}^{a}$]}\\
\hline
  \multirow{6}{*}{8} &\multirow{3}{*}{h} 
 & 118.15 & {\cellcolor[HTML]{D0D0D0} 557.25 }  & {\cellcolor[HTML]{D0D0D0} 1,329.62 } \\
 &  & {\footnotesize (113.57-122.73)} & {\cellcolor[HTML]{D0D0D0} {\footnotesize (553.50-561.00)} }  & {\cellcolor[HTML]{D0D0D0} {\footnotesize (1,325.56-1,333.69)} } \\
 &  & {\footnotesize [\underline{3.87}${}^{a}$, \underline{0.83}${}^{a}$]} & {\cellcolor[HTML]{D0D0D0} {\footnotesize [\underline{18.49}${}^{a}$, \underline{0.93}${}^{a}$]} }  & {\cellcolor[HTML]{D0D0D0} {\footnotesize [\underline{20.35}${}^{b}$, \underline{1.07}${}^{c}$]} } \\
\cline{2-5}
 & \multirow{3}{*}{s} 
 & 125.35 & 858.20 & 1,830.56\\
 &  & {\footnotesize (121.33-129.37)} & {\footnotesize (854.62-861.77)} & {\footnotesize (1,827.12-1,834.00)}\\
 &  & {\footnotesize [\underline{3.65}${}^{a}$, \underline{0.80}${}^{a}$]} & {\footnotesize [\underline{12.00}${}^{a}$, \underline{0.72}${}^{a}$]} & {\footnotesize [\underline{14.78}${}^{b}$, \underline{0.73}${}^{a}$]}\\
\hline
  \multirow{6}{*}{16} &\multirow{3}{*}{h} 
 & 332.12 & {\cellcolor[HTML]{D0D0D0} 2,812.84 }  & {\cellcolor[HTML]{D0D0D0} 11,917.28 } \\
 &  & {\footnotesize (328.62-335.63)} & {\cellcolor[HTML]{D0D0D0} {\footnotesize (2,797.34-2,828.34)} }  & {\cellcolor[HTML]{D0D0D0} {\footnotesize (*)} } \\
 &  & {\footnotesize [\underline{1.38}${}^{a}$, \underline{0.30}${}^{a}$]} & {\cellcolor[HTML]{D0D0D0} {\footnotesize [\underline{3.66}${}^{a}$, \underline{0.18}${}^{b}$]} }  & {\cellcolor[HTML]{D0D0D0} {\footnotesize [\underline{2.27}${}^{d}$, \underline{0.12}${}^{d}$]} } \\
\cline{2-5}
 & \multirow{3}{*}{s} 
 & 504.37 & 10,091.85 & 27,460.23\\
 &  & {\footnotesize (500.86-507.89)} & {\footnotesize (*)} & {\footnotesize (27,426.08-27,494.38)}\\
 &  & {\footnotesize [\underline{0.91}${}^{a}$, \underline{0.20}${}^{a}$]} & {\footnotesize [\underline{1.02}${}^{c}$, \underline{0.06}${}^{d}$]} & {\footnotesize [\underline{0.99}${}^{b}$, \underline{0.05}${}^{b}$]}\\
\hline
\end{tabular}
\caption{Mean wall clock time with Arm-v7 instruction set, GPU and OpenCL  in ms. W=window, V=number of vector elements, P=precision, h=half, s=single. Values in curved brackets are the values of the 95\% confidence interval. Values in square brackets are the calculated speedups (see text for explanations). (*)=no normal distribution (median instead of mean, no confidence interval).${}^{a}$=t-test, ${}^{b}$=Welch-test, ${}^{c}$=Mann-Whithney-U test, ${}^{d}$=Median test, underline=statistically significant (p$<$0.05). Grey color=result not acceptable due to numerical problems.}
\label{tab:result_armv7_GPU}
\end{table}

\subsubsection{Arm-v8}
\label{results:armv8}

Java was again slower than any other method. C and assembler were approximately equally fast and both were faster than Java. 
For all windows SIMD was much faster than all other methods (maximum speedup 11.00 for single precision and 5.98 for double precision with respect to a single-threaded Java version).
Again using single precision SIMD instructions was almost twice as fast compared to double precision SIMD instructions.

\subsubsection{Arm-v7 GPU with OpenCL}

%The method describe above has been used to access the GPU on the tablet. 

Table \ref{tab:result_armv7_GPU} shows the runtime and speedup for different windows, vector sizes and precisions. The first line of each cell shows the mean wall clock time in milliseconds. The second line shows the 95\% confidence interval of the mean. If the values were not normally distributed, no confidence interval is displayed and the median, instead of the mean, is shown in the first line.
The third line displays the speedup with respect  to a single threaded Java version on the CPU (first value) and to the calculation on the GPU with a scalar version of the same window and the same precision (i.e., topmost value in the same column).
Measurement of the wall lock time was started in Java immediately prior to the call to the native method that performed the calculations on the GPU.

Half precision worked well for Window 1 but failed to return precise results due to numerical issues for Windows 2 and 3 (see Figure \ref{fig:wins}d).
The results that were rejected are marked grey in Tables \ref{tab:result_armv7_GPU} and \ref{tab:result_armv7_GPU2}.

With respect to Java (single precision) a maximum speedup of 4.65 for half precision and 20.20 for single precision could be achieved. The maximum speedup for half precision was 20.66, but the resulting picture had to be discarded. Greater workloads resulted in a greater speedup.
% The higher the workload was, the higher was the speedup.

OpenCL by default offers vector instructions with a length of 2, 4, 8 or 16 elements, even if the GPU natively  does not support vectors of all of these sizes. 
The compiler determines how the vector instructions are implemented.
%It depends on the compiler how the vector instructions are implemented.
For both precisions tested, scalar operations were faster than any vector operation. 
For both (single and half precision) performance degraded notably with 16 vector elements.
While for the scalar version and Window 1 there was almost no runtime difference between half and single precision, the more elements a vector had and the greater the workload was, the faster  the resulting calculations with half instead of single precision.

Table \ref{tab:result_armv7_GPU2} depicts detailed timing of different steps of the algorithm that calculates the Mandelbrot set on the GPU. The same data as in Table \ref{tab:result_armv7_GPU} was used.
The column 'wall clock time' is the median runtime measured in Java. The JNI overhead has been measured taking the time at the very beginning and the end of the C method. The lock overhead was the time needed for acquiring the reader writer lock, the semaphore and testing the flag. 
The OCL overhead is the time needed for setting up and destroying the entire OpenCL environment. This encompasses the compilation of the kernel and the data transfer to and from the GPU.

The JNI overhead is independent of the three variables tested (window, vector size, precision) and is very small (almost always less than 3 milliseconds). The overhead for the locking mechanism is also independent of the three variables and is even smaller (less than 30 microseconds). The OpenCL overhead depends on the workload and the vector size and ranged from 72.99 ms (scalar version, Window 0) to 193.94 ms (vector with 16 elements, Window 2).
For smaller workloads the OpenCL framework overhead can attain two thirds of the entire runtime whereas for high workloads the fraction of runtime for the OpenCL framework overhead can drop to less than 10\%.

\begin{table*}[tbh]
\centering
\begin{tabular}{ |c|c|c|r|r|r|r|r|r|r| }
\hline
W & V & P & wall clock & JNI      & JNI  & lock     & lock & OCL      & OCL \\
  &   &   & time       & overhead &  \%  & overhead &  \%  & overhaed &  \% \\
\hline
  \multirow{10}{*}{1} &   \multirow{2}{*}{1} & h
 & 98.97 & 0.61 & 0.62\% & 0.01 & 0.01\% & 73.40 & 74.17\%\\
 &  & s
 & 98.29 & 0.37 & 0.38\% & 0.01 & 0.01\% & 72.99 & 74.26\%\\
\cline{2-10}
 &   \multirow{2}{*}{2} & h
 & 111.60 & 0.61 & 0.55\% & 0.01 & 0.01\% & 84.70 & 75.89\%\\
 &  & s
 & 114.05 & 0.35 & 0.31\% & 0.01 & 0.01\% & 84.26 & 73.87\%\\
\cline{2-10}
 &   \multirow{2}{*}{4} & h
 & 108.91 & 0.68 & 0.63\% & 0.01 & 0.01\% & 85.10 & 78.14\%\\
 &  & s
 & 105.92 & 0.25 & 0.24\% & 0.01 & 0.01\% & 77.59 & 73.25\%\\
\cline{2-10}
 &   \multirow{2}{*}{8} & h
 & 116.12 & 0.56 & 0.49\% & 0.01 & 0.01\% & 90.04 & 77.54\%\\
 &  & s
 & 126.15 & 0.38 & 0.30\% & 0.01 & 0.01\% & 94.73 & 75.09\%\\
\cline{2-10}
 &   \multirow{2}{*}{16} & h
 & 331.65 & 0.95 & 0.29\% & 0.02 & 0.01\% & 183.20 & 55.24\%\\
 &  & s
 & 504.44 & 0.40 & 0.08\% & 0.02 & 0.00\% & 189.24 & 37.51\%\\
\cline{2-10}
\hline
  \multirow{10}{*}{2} &   \multirow{2}{*}{1} & h
 & {\cellcolor[HTML]{D0D0D0} 520.36} & {\cellcolor[HTML]{D0D0D0} 2.42} & {\cellcolor[HTML]{D0D0D0} 0.46\%} & {\cellcolor[HTML]{D0D0D0} 0.02} & {\cellcolor[HTML]{D0D0D0} 0.00\%} & {\cellcolor[HTML]{D0D0D0} 119.06} & {\cellcolor[HTML]{D0D0D0} 22.88\%}\\
 &  & s
 & 616.49 & 1.00 & 0.16\% & 0.02 & 0.00\% & 121.26 & 19.67\%\\
\cline{2-10}
 &   \multirow{2}{*}{2} & h
 & {\cellcolor[HTML]{D0D0D0} 568.48} & {\cellcolor[HTML]{D0D0D0} 1.57} & {\cellcolor[HTML]{D0D0D0} 0.28\%} & {\cellcolor[HTML]{D0D0D0} 0.02} & {\cellcolor[HTML]{D0D0D0} 0.00\%} & {\cellcolor[HTML]{D0D0D0} 132.31} & {\cellcolor[HTML]{D0D0D0} 23.27\%}\\
 &  & s
 & 781.55 & 0.92 & 0.12\% & 0.02 & 0.00\% & 131.93 & 16.88\%\\
\cline{2-10}
 &   \multirow{2}{*}{4} & h
 & {\cellcolor[HTML]{D0D0D0} 521.65} & {\cellcolor[HTML]{D0D0D0} 1.54} & {\cellcolor[HTML]{D0D0D0} 0.29\%} & {\cellcolor[HTML]{D0D0D0} 0.02} & {\cellcolor[HTML]{D0D0D0} 0.00\%} & {\cellcolor[HTML]{D0D0D0} 138.14} & {\cellcolor[HTML]{D0D0D0} 26.48\%}\\
 &  & s
 & 793.20 & 1.12 & 0.14\% & 0.02 & 0.00\% & 138.13 & 17.41\%\\
\cline{2-10}
 &   \multirow{2}{*}{8} & h
 & {\cellcolor[HTML]{D0D0D0} 558.85} & {\cellcolor[HTML]{D0D0D0} 1.15} & {\cellcolor[HTML]{D0D0D0} 0.21\%} & {\cellcolor[HTML]{D0D0D0} 0.02} & {\cellcolor[HTML]{D0D0D0} 0.00\%} & {\cellcolor[HTML]{D0D0D0} 146.28} & {\cellcolor[HTML]{D0D0D0} 26.17\%}\\
 &  & s
 & 860.08 & 0.98 & 0.11\% & 0.02 & 0.00\% & 148.34 & 17.25\%\\
\cline{2-10}
 &   \multirow{2}{*}{16} & h
 & {\cellcolor[HTML]{D0D0D0} 2795.65} & {\cellcolor[HTML]{D0D0D0} 3.00} & {\cellcolor[HTML]{D0D0D0} 0.11\%} & {\cellcolor[HTML]{D0D0D0} 0.03} & {\cellcolor[HTML]{D0D0D0} 0.00\%} & {\cellcolor[HTML]{D0D0D0} 189.25} & {\cellcolor[HTML]{D0D0D0} 6.77\%}\\
 &  & s
 & 10091.85 & 1.00 & 0.01\% & 0.02 & 0.00\% & 190.92 & 1.89\%\\
\cline{2-10}
\hline
  \multirow{10}{*}{3} &   \multirow{2}{*}{1} & h
 & {\cellcolor[HTML]{D0D0D0} 1418.84} & {\cellcolor[HTML]{D0D0D0} 2.59} & {\cellcolor[HTML]{D0D0D0} 0.18\%} & {\cellcolor[HTML]{D0D0D0} 0.02} & {\cellcolor[HTML]{D0D0D0} 0.00\%} & {\cellcolor[HTML]{D0D0D0} 120.46} & {\cellcolor[HTML]{D0D0D0} 8.49\%}\\
 &  & s
 & 1339.31 & 1.23 & 0.09\% & 0.02 & 0.00\% & 122.82 & 9.17\%\\
\cline{2-10}
 &   \multirow{2}{*}{2} & h
 & {\cellcolor[HTML]{D0D0D0} 1521.70} & {\cellcolor[HTML]{D0D0D0} 1.76} & {\cellcolor[HTML]{D0D0D0} 0.12\%} & {\cellcolor[HTML]{D0D0D0} 0.06} & {\cellcolor[HTML]{D0D0D0} 0.00\%} & {\cellcolor[HTML]{D0D0D0} 132.24} & {\cellcolor[HTML]{D0D0D0} 8.69\%}\\
 &  & s
 & 1703.75 & 0.93 & 0.05\% & 0.02 & 0.00\% & 133.37 & 7.83\%\\
\cline{2-10}
 &   \multirow{2}{*}{4} & h
 & {\cellcolor[HTML]{D0D0D0} 1309.81} & {\cellcolor[HTML]{D0D0D0} 1.48} & {\cellcolor[HTML]{D0D0D0} 0.11\%} & {\cellcolor[HTML]{D0D0D0} 0.02} & {\cellcolor[HTML]{D0D0D0} 0.00\%} & {\cellcolor[HTML]{D0D0D0} 137.22} & {\cellcolor[HTML]{D0D0D0} 10.48\%}\\
 &  & s
 & 1685.47 & 1.10 & 0.07\% & 0.01 & 0.00\% & 138.82 & 8.24\%\\
\cline{2-10}
 &   \multirow{2}{*}{8} & h
 & {\cellcolor[HTML]{D0D0D0} 1326.49} & {\cellcolor[HTML]{D0D0D0} 1.30} & {\cellcolor[HTML]{D0D0D0} 0.10\%} & {\cellcolor[HTML]{D0D0D0} 0.02} & {\cellcolor[HTML]{D0D0D0} 0.00\%} & {\cellcolor[HTML]{D0D0D0} 147.50} & {\cellcolor[HTML]{D0D0D0} 11.12\%}\\
 &  & s
 & 1830.99 & 0.97 & 0.05\% & 0.02 & 0.00\% & 149.06 & 8.14\%\\
\cline{2-10}
 &   \multirow{2}{*}{16} & h
 & {\cellcolor[HTML]{D0D0D0} 11917.28} & {\cellcolor[HTML]{D0D0D0} 1.07} & {\cellcolor[HTML]{D0D0D0} 0.01\%} & {\cellcolor[HTML]{D0D0D0} 0.02} & {\cellcolor[HTML]{D0D0D0} 0.00\%} & {\cellcolor[HTML]{D0D0D0} 191.76} & {\cellcolor[HTML]{D0D0D0} 1.61\%}\\
 &  & s
 & 27448.24 & 1.17 & 0.00\% & 0.02 & 0.00\% & 193.94 & 0.71\%\\
\cline{2-10}
\hline
\end{tabular}
\caption{Median of runtime on Mali G-71 GPU and OpenCL in milliseconds. W=window, V=number of vector elements, P=precision, h=half, s=single. 'wall clock time' = median of wall clock time measured in Java in ms. 'JNI overhead' = median of overhead in ms for JNI interface. 'JNI \%' = percentage of 'JNI overhead' with respect to 'wall clock time'. 'lock overhead' = median of overhead in ms for locking mechanism. 'lock \%' = percentage of 'lock overhead' with respect to 'wall clock time'. 'OCL overhead' = median of overhead in ms for the setup of the OpenCL environment (creation of command queue, building the programm, data transfer and releasing the resources). 'OCL \%' = percentage of 'OCL overhead' with respect to 'wall clock time'. Grey color=result not acceptable due to numerical problems.}
\label{tab:result_armv7_GPU2}
\end{table*}

\subsubsection{Arm-v7 GPU - Test of the locking mechanisms}

Both locking mechanisms described above were implemented and used. The locking events were logged into a file. The app was halted and restarted multiple times. The log showed that the app was working as expected and resources were freed timely after the app had been stopped. The unloading mechanism worked flawlessly.

\subsubsection{Arm-v7 GPU in comparison with other GPUs}

The median wall clock time of the execution of the OpenCL kernel on the Mali G71 GPU has been compared to an integrated Intel Iris Plus 650, to a and dedicated AMD Radeon VII GPU and to a dedicated Nvidia Titan V GPU. Table \ref{tab:result_GPUcomp} in the supplements depicts the results. The  same kernel as on the tablet has been used, but instead of the Android development environment a plain C program (compiled with GCC 8.2) was used to execute the kernels on the GPUs.
Therefore, only the time for the execution of the kernel was tested. On the Intel GPU the program was executed using lightweight virtualization (Docker \cite{Docker}). Lightweight virtualization is known to have only a very small effect on the runtime \cite{Felter2015}.

Depending on the workload the Intel GPU is five to ten times faster than the GPU on the tablet. Performance for half precision is significantly better than for single precision. Performance for vectors with 16 entries is significantly worse than other vector sizes. Depending on the workload, the AMD GPU is 50-100 times faster than the GPU on the tablet. There is no performance degradation with larger vector sizes. On the AMD GPU half precision calculations are as fast as single precision calculations.
Double precision calculations are two to three times slower than single precision calculations.

Calculations on the Titan V GPU were 120 to 265 times faster than on the GPU of the tablet. Half precision calculations were not available with the OpenCL framework (although the ALUs (arithmetic logical units) of the GPU are capable of performing half precision calculations). Single precision calculations were two to three times faster than double precision calculations.

%This should not considered to be a bad result given the comparatively larges dimensions of the AMD GPU.

\section{Discussion}

%The speedup of the most parallel programming algorithms was less than expected. 

C with -O3 optimization and scalar assembler were almost always faster than Java, but the speedup was small.
%(less than 1.6 fold). 
There was no advantage in developing handwritten scalar assembler, especially given the high effort.
Perhaps expert level assembler programmers might be able to rearrange the assembler code and  increase its performance.

In the case of the Mandelbrot algorithm, calculations can be executed fully in parallel and threads do not need to synchronize. According to Amdahl's law this should give a high theoretically possible speedup. On the Arm tablets used in this trial the loss of speedup between two and three cores can - in part - be explained with the fact that the scalar versions of the same program use a shorter and faster code, as they do not rely on the overhead of the producer/consumer principle. Moreover, on processors that have unequal cores (e.g., some faster ones and some slower ones) the theoretically possible speedup using multithreading can be much lower than expected, 
because using more cores results in a higher probability that cores
%because the more cores are used, the higher is the probability that cores 
with a lower computing power are used. %One possibility to deal with this problem could be the division the workload into smaller chunks, where faster cores are able to execute more chunks than the slower cores.

Handwritten SIMD assembler was much faster than any other method (except OpenCL). C compilers might not always be able to automatically extract vector parallelism. SIMD instructions are available for all instruction sets currently supported by Android. 

If one does not want to rely on auto-vectorization, there are two ways to explicitly use SIMD instructions.
The first is the production of assembler code, which may take longer and might be more error prone, but gives full control of the execution of the program. The second is the use of 
SIMD intrinsic instruction extensions for the C programming language which are available for Intel and ARM CPUs and spare the programmer from using assembler.
In both cases program versions for every CPU type and precision needed have to be written. 

The OpenMP framework \cite{OpenMP} allows to implicitly use vectorization with the \texttt{\#pragma omp simd} statements. OpenMP is supported by Android using an appropriate C compiler.

%A more theoretical possibility to implicitly use SIMD instructions would be - if available - the use of an OpenCL library from the pocl-project [] for the CPUs of Android. 

OpenCL kernels are quite easy to write and, contrary to assembler, there is no need to develop versions specific to each architecture. As an extra bonus, the CPU can be used for other purposes while the GPU carries out calculations. OpenCL kernels can be compiled at runtime and therefore can be adapted or programmed at runtime %therefore allow to be adapted or programmed at runtime 
(e.g., the vector size as was done in this trial).
Disadvantages are that OpenCL is not officially supported by Android and third party libraries are needed, whose use can be disallowed in future versions of the Android operating system. Vendor provided OpenCL libraries should be compiled in a way that they do not depend on other shared libraries (i.e., linked with static libraries). This  avoids compatibility issues. This was the second reason why we were not able to use the OpenCL shared library on the Arm-v8 tablet.

During calculations on the GPU the refresh of the screen might become slower. GPU rendering can be disabled on the device (developer options) or in the app (written in the manifest-XML or in the activity-XML) leaving the entire GPU power for OpenCL. For this trial GPU hardware acceleration was disabled. The very bad performance of vectors with 16 elements may be caused by a lack of registers and the need to store intermediate results in memory. This phenomenon might be specific to the implementation of an algorithm, the precision used and the GPU.

For Java, C and scalar assembler, the selection of the precision is not very important. All platforms used are 64 bit CPUs (the Cortex CPUs of the Arm-v7 tablet would also support Arm-v8). % and therefore single and double precision floating point operations can be executed at the same speed. For older 32 bit CPUs there might be a difference. 
Double precision calculations were even slightly faster than single precision arithmetics on the Arm-v7 tablet with C and Java.
SIMD single precision instructions were much faster than double precision, as single precision allows twice as many floating point operations per instruction than double precision.

On the GPU half precision should be preferred over single precision if no numerical issues are expected. Care should be taken regarding the vector size. For half and single precision, scalar operations were fastest. Longer vectors yielded longer runtimes.
%The longer the vectors were, the longer was the runtime. 
For small workloads the increase in runtime was almost exclusively caused by a longer compilation time (subtract "OCL overhead" from "wall clock time" in Table \ref{tab:result_armv7_GPU2} or see table \ref{tab:result_GPUcomp}). For higher workloads this effect was less pronounced. 
%Perhaps multiple tests might help to establish the optimal vector size for a given problem.

Regarding single precision, for small workloads (Window 1), multithreaded SIMD assembler was faster than the fastest OpenCL version (1.46x). For Window 2 and Window 3 (intermediate and high workload) OpenCL was faster than the fastest SIMD version (1.29x and 1.39x). The overhead for JNI and the locking mechanism in the case of OpenCL were negligible.
The overhead for setting up the OpenCL environment can be considerable (up to 79\%), if for every call the entire building process has to be executed. 
%Nevertheless, the Mali-GPU on the Arm-v7 tablet was able to deliver a similar computing power as a multithreaded program on a 7th generation Intel Core-i processor (although one has to account for small overhead of the virtualization technology).

The availability of a pocl (portable computing language) library \cite{pocl} for Android for the execution of OpenCL kernels on the CPU would be helpful. This would allow to execute the same kernels on the CPU if there were no GPU on the tablet. This would enhance portability and would allow the implicit use of SIMD parallelism
%to use implicitly SIMD parallelism 
on the CPU without the need to write specific assembler versions for each architecture. The use of OpenCL on multicore CPUs in comparison to other parallel programming frameworks has been analyzed by Shen et al. \cite{SHEN2013834}.

%If a GPU is used during the whole lifespan of an app and many small kernels must be executed, a separate thread for the OpenCL environment could be created. This thread is woken up by incoming data, processes it asynchronically on the GPU and puts the result into a result queue. If the app is stopped, the thread is woken up and all GPU resources are released.% and releases all GPU resources. 

%The source code for pocl can be downloaded from the project's homepage. %and has to be built for any  architecture.

The results for OpenCL presented here can be applied to other architectures like the Raspberry Pi 3B+, for whose VideoCore IV GPU an (experimental) OpenCL library has been developed.
Unfortunately, no OpenCL library exists for the VideoCore VI GPU of the Raspberry Pi 4b. %actually there exists no OpenCL library.

The use of multiprocessing (\texttt{fork()} in C) has not been investigated as forking new processes on Android is generally not recommended and sharing memory in a multiprocessing environment is more difficult.

\section{Conclusions}

The following factors were most important for speedup (sorted by the effort needed to implement):

\begin{itemize}
    \item Multithreading
    \item SIMD data parallelism
    \item Use of GPU (via OpenCL)
    \item Precision (for SIMD and to a lesser extent for OpenCL)
\end{itemize}

Multithreading is one of the most relevant factors to contribute to a higher speedup on the CPU. Multithreading is well supported by Java and C.

SIMD programming can achieve high speedups, but using data parallelism is not straightforward if one does not want to rely on auto-vectorization (this is valid not only for Android devices).

OpenCL can currently be used on Android devices if the device's GPU supports this framework. The wrapper library developed for this project allows a comfortable way of accessing the GPU's resources. The use of OpenCL on Android devices would be easier if OpenCL was officially supported by Android, or at least the possibility of the use of vendor supplied libraries would be guaranteed in the future. 

Precision has a less important role for the runtime except for SIMD parallelism on the CPU.

This trial has shown that there are efficient mechanisms to allow OpenCL resources to be freed if the program 
%that allow to free the OpenCL resources if the program 
execution is stopped from outside. Memory can potentially be saved if dynamic libraries are  unloaded while an app is not used.

The insights gained here can allow the development of a Java library that implements the locking mechanisms and allows the programmer to use OpenCL without having the need to use C and JNI.

%The method described above could be part of a high level Java-OpenCL API for tablets that leverages the user from using C and JNI.
%The locking mechanisms could be part of an abstract class.

%The OpenMP \cite{OpenMP} framework offers an easy to use multithreading environment, that is often based based on pthreads. 

Interesting future developments would be processors that support SIMD instructions with more elements (like the AVX instructions for many Intel processors) and half precision. This would allow doubling of the current speedups for handwritten SIMD assembler, as
well as for C compilers and Java interpreters that are able to automatically extract
vector parallelism.
%would allow to double the current speedups
%for handwritten SIMD assembler as well as for C compilers and Java interpreters that are able to automatically extract vector parallelism.

On the other hand, more powerfull GPUs (currently already available) would allow faster calculations on the GPU.
%also allow to speed up calculations on the GPU.
Two GPUs on tablets would decouple the screen related workload from the program arithmetic workload. Using precompiled OpenCL kernels could help to save time. 
 
Not all parallel programming paradigms are suited for all types of problems. The selection of the parallelization type might be driven by the nature of the problem.

%Given the speedups AVX vector instruction would potentially allow to double the current speedups. 

%On all device pure C was a little bit faster than pure Java code, but the speedup never exceeded the factor xxx. C with inlined scalar assembler was not faster than pure C. In some cases hand written assembler was significantly slower than pure C, in others slightly faster or there was no significant difference.
%C with hand written SIMD assembler was always faster than pure C or pure J (more than 3 fold for single precision and a little less than 2 fold for double precision with respect to pure C). SIMD assembler for single precisioon was - if available - almost twice as fast as SIMD assembler with double precision. SIMD vectors can hold either 2 double or four single precision floating point values.

%C with OpenCL was much faster than pure Java (up to 20 times). The higher the workload was, the higher was the speedup. For smaller workloads multithreaded SIMD instructions were almost equally fast than GPU versions.
%The use of half precision did not give a great benefit in terms of speedup. The loss of precision did not allow to calculate two of the three zoomed windows.
%For easily parallelizable code (multithreaded+SIMD instructions) and time critical algoithms with low to modest workload the production of handwritten assembler can be taken into consideration. 

\section*{Remarks}

\begin{itemize}
    %\item Android is a trade mark of Google LLC.
    \item This research did not receive any specific grant from funding agencies in the public, commercial, or not-for-profit sectors.
    \item The sequence diagram in the supplements has been created with the online diagram editor available at \newline {\tt https://sequencediagram.org/}.
    \item The activity diagrams have been created with the online available diagram editor \texttt{ https://online.visual-\newline paradigm.com/diagrams/solutions/free- \newline activity-diagram-tool/}.
    \item Tables and figures were created using {\tt python3} and the libraries {\tt scipy}, {\tt numpy} and {\tt matplotlib}.
\end{itemize}

\bibliography{AndroidOCL}

\begin{thebibliography}{25}
\expandafter\ifx\csname natexlab\endcsname\relax\def\natexlab#1{#1}\fi
\providecommand{\url}[1]{\texttt{#1}}
\providecommand{\href}[2]{#2}
\providecommand{\path}[1]{#1}
\providecommand{\DOIprefix}{doi:}
\providecommand{\ArXivprefix}{arXiv:}
\providecommand{\URLprefix}{URL: }
\providecommand{\Pubmedprefix}{pmid:}
\providecommand{\doi}[1]{\href{http://dx.doi.org/#1}{\path{#1}}}
\providecommand{\Pubmed}[1]{\href{pmid:#1}{\path{#1}}}
\providecommand{\bibinfo}[2]{#2}
\ifx\xfnm\relax \def\xfnm[#1]{\unskip,\space#1}\fi
%Type = Misc
\bibitem[{MontBlanc(2020)}]{MontBlanc}
MontBlanc, \bibinfo{title}{https://www.montblanc-project.eu/project, retireved
  on july 11th, 2020}, \bibinfo{year}{2020}.
%Type = Article
\bibitem[{Douady and Hubbard(1985)}]{Douady1984}
\bibinfo{author}{A.~Douady}, \bibinfo{author}{J.~H. Hubbard},
\newblock \bibinfo{title}{Etude dynamique des polynômes complexes},
\newblock \bibinfo{journal}{Prépublications mathémathiques d’Orsay}
  \bibinfo{volume}{2/4} (\bibinfo{year}{1984/1985}).
%Type = Inproceedings
\bibitem[{Brodtkorb and Hagen(2008)}]{Brodtkorb2008}
\bibinfo{author}{A.~R. Brodtkorb}, \bibinfo{author}{T.~R. Hagen},
\newblock \bibinfo{title}{A comparison of three commodity-level parallel
  architectures: Multi-core cpu, cell be and gpu},
\newblock in: \bibinfo{booktitle}{Proceedings of the 7th International
  Conference on Mathematical Methods for Curves and Surfaces}, MMCS’08,
  \bibinfo{year}{2008}, p. \bibinfo{pages}{70–80}.
%Type = Inproceedings
\bibitem[{{Wang} et~al.(2016){Wang}, {Nurmi}, and {Ahonen}}]{Wang2016}
\bibinfo{author}{K.~{Wang}}, \bibinfo{author}{J.~{Nurmi}},
  \bibinfo{author}{T.~{Ahonen}},
\newblock \bibinfo{title}{Accelerating computation on an android phone with
  opencl parallelism and optimizing workload distribution between a phone and a
  cloud service},
\newblock in: \bibinfo{booktitle}{2016 Intl IEEE Conferences on Ubiquitous
  Intelligence Computing, Advanced and Trusted Computing, Scalable Computing
  and Communications, Cloud and Big Data Computing, Internet of People, and
  Smart World Congress (UIC/ATC/ScalCom/CBDCom/IoP/SmartWorld)},
  \bibinfo{year}{2016}, pp. \bibinfo{pages}{636--642}.
%Type = Inproceedings
\bibitem[{Acosta et~al.(2018)Acosta, Merino, and Totz}]{Acosta2018}
\bibinfo{author}{A.~Acosta}, \bibinfo{author}{C.~Merino},
  \bibinfo{author}{J.~Totz},
\newblock \bibinfo{title}{Analysis of opencl support for mobile gpus on
  android},
\newblock in: \bibinfo{booktitle}{Proceedings of the International Workshop on
  OpenCL}, IWOCL ’18, \bibinfo{publisher}{Association for Computing
  Machinery}, \bibinfo{year}{2018}, pp. \bibinfo{pages}{1--6}.
%Type = Inproceedings
\bibitem[{{Ross} et~al.(2014){Ross}, {Richie}, {Park}, {Shires}, and
  {Pollock}}]{Ross2014}
\bibinfo{author}{J.~A. {Ross}}, \bibinfo{author}{D.~A. {Richie}},
  \bibinfo{author}{S.~J. {Park}}, \bibinfo{author}{D.~R. {Shires}},
  \bibinfo{author}{L.~L. {Pollock}},
\newblock \bibinfo{title}{A case study of opencl on an android mobile gpu},
\newblock in: \bibinfo{booktitle}{2014 IEEE High Performance Extreme Computing
  Conference (HPEC)}, \bibinfo{year}{2014}, pp. \bibinfo{pages}{1--6}.
%Type = Article
\bibitem[{Pérez et~al.(2019)Pérez, Stafford, Bosque, Beivide, Mateo, Teruel,
  Martorell, and Ayguade}]{Perez2019}
\bibinfo{author}{B.~Pérez}, \bibinfo{author}{E.~Stafford},
  \bibinfo{author}{J.~Bosque}, \bibinfo{author}{R.~Beivide},
  \bibinfo{author}{S.~Mateo}, \bibinfo{author}{J.~Teruel},
  \bibinfo{author}{X.~Martorell}, \bibinfo{author}{E.~Ayguade},
\newblock \bibinfo{title}{Auto-tuned opencl kernel co-execution in ompss for
  heterogeneous systems},
\newblock \bibinfo{journal}{Journal of parallel and distributed computing}
  \bibinfo{volume}{125} (\bibinfo{year}{2019}) \bibinfo{pages}{45--57}.
%Type = Inproceedings
\bibitem[{{Pérez} et~al.(2017){Pérez}, {Stafford}, {Bosque}, {Beivide},
  {Mateo}, {Teruel}, {Martorell}, and {Ayguadé}}]{Perez2017}
\bibinfo{author}{B.~{Pérez}}, \bibinfo{author}{E.~{Stafford}},
  \bibinfo{author}{J.~L. {Bosque}}, \bibinfo{author}{R.~{Beivide}},
  \bibinfo{author}{S.~{Mateo}}, \bibinfo{author}{X.~{Teruel}},
  \bibinfo{author}{X.~{Martorell}}, \bibinfo{author}{E.~{Ayguadé}},
\newblock \bibinfo{title}{Extending ompss for opencl kernel co-execution in
  heterogeneous systems},
\newblock in: \bibinfo{booktitle}{2017 29th International Symposium on Computer
  Architecture and High Performance Computing (SBAC-PAD)},
  \bibinfo{year}{2017}, pp. \bibinfo{pages}{1--8}.
%Type = Inproceedings
\bibitem[{{Wang} et~al.(2013){Wang}, {Xiong}, {Yun}, and
  {Cavallaro}}]{Wang2013}
\bibinfo{author}{G.~{Wang}}, \bibinfo{author}{Y.~{Xiong}},
  \bibinfo{author}{J.~{Yun}}, \bibinfo{author}{J.~R. {Cavallaro}},
\newblock \bibinfo{title}{Accelerating computer vision algorithms using opencl
  framework on the mobile gpu - a case study},
\newblock in: \bibinfo{booktitle}{2013 IEEE International Conference on
  Acoustics, Speech and Signal Processing}, \bibinfo{year}{2013}, pp.
  \bibinfo{pages}{2629--2633}.
%Type = Article
\bibitem[{Yokoyama et~al.(2019)Yokoyama, Schulze, Borges et~al.}]{Yokoyama2019}
\bibinfo{author}{D.~Yokoyama}, \bibinfo{author}{B.~Schulze},
  \bibinfo{author}{F.~Borges}, et~al.,
\newblock \bibinfo{title}{The survey on arm processors for hpc},
\newblock \bibinfo{journal}{J Supercomput} \bibinfo{volume}{75}
  (\bibinfo{year}{2019}) \bibinfo{pages}{7003–7036}.
%Type = Misc
\bibitem[{OpenCL(2020)}]{OpenCL}
OpenCL, \bibinfo{title}{https://www.khronos.org/opencl/, retrieved on march
  14th, 2020}, \bibinfo{year}{2020}.
%Type = Article
\bibitem[{Dijkstra(1972)}]{Dijkstra1972}
\bibinfo{author}{E.~Dijkstra},
\newblock \bibinfo{title}{Information streams sharing a finite buffer},
\newblock \bibinfo{journal}{Information Processing Letters} \bibinfo{volume}{1}
  (\bibinfo{year}{1972}) \bibinfo{pages}{179--180}.
%Type = Article
\bibitem[{Du et~al.(2012)Du, Weber, Luszczek, Tomov, Peterson, and
  Dongarra}]{DU2012391}
\bibinfo{author}{P.~Du}, \bibinfo{author}{R.~Weber},
  \bibinfo{author}{P.~Luszczek}, \bibinfo{author}{S.~Tomov},
  \bibinfo{author}{G.~Peterson}, \bibinfo{author}{J.~Dongarra},
\newblock \bibinfo{title}{From cuda to opencl: Towards a performance-portable
  solution for multi-platform gpu programming},
\newblock \bibinfo{journal}{Parallel Computing} \bibinfo{volume}{38}
  (\bibinfo{year}{2012}) \bibinfo{pages}{391 -- 407}.
%Type = Article
\bibitem[{Shapiro and Wilk(1965)}]{Shapiro1965}
\bibinfo{author}{S.~S. Shapiro}, \bibinfo{author}{M.~B. Wilk},
\newblock \bibinfo{title}{{An analysis of variance test for normality (complete
  samples)}},
\newblock \bibinfo{journal}{Biometrika} \bibinfo{volume}{52}
  (\bibinfo{year}{1965}) \bibinfo{pages}{591--611}.
%Type = Incollection
\bibitem[{Levene(1960)}]{Levene1960}
\bibinfo{author}{H.~Levene},
\newblock \bibinfo{title}{Robust tests for equality of variances},
\newblock in: \bibinfo{editor}{I.~Olkin}, \bibinfo{editor}{H.~Hotelling},
  et~al. (Eds.), \bibinfo{booktitle}{Contributions to Probability and
  Statistics: Essays in Honor of Harold Hotelling},
  \bibinfo{publisher}{Stanford University Press}, \bibinfo{year}{1960}, p.
  \bibinfo{pages}{278–292}.
%Type = Article
\bibitem[{Student(1908)}]{student1908}
\bibinfo{author}{Student},
\newblock \bibinfo{title}{The probable error of a mean},
\newblock \bibinfo{journal}{Biometrika} \bibinfo{volume}{6}
  (\bibinfo{year}{1908}) \bibinfo{pages}{1--25}.
%Type = Article
\bibitem[{Welch(1947)}]{Welch1947}
\bibinfo{author}{B.~L. Welch},
\newblock \bibinfo{title}{{The generalization of ''Students's'' problem when
  several different population variances are involved}},
\newblock \bibinfo{journal}{Biometrika} \bibinfo{volume}{34}
  (\bibinfo{year}{1947}) \bibinfo{pages}{28--35}.
%Type = Article
\bibitem[{Wilcoxon(1945)}]{Wilcoxon1945}
\bibinfo{author}{F.~Wilcoxon},
\newblock \bibinfo{title}{Individual comparisons by ranking methods},
\newblock \bibinfo{journal}{Biometrics Bulletin} \bibinfo{volume}{1}
  (\bibinfo{year}{1945}) \bibinfo{pages}{80–83}.
%Type = Article
\bibitem[{Mann and Whitney(1947)}]{MannWhitney1947}
\bibinfo{author}{H.~Mann}, \bibinfo{author}{D.~Whitney},
\newblock \bibinfo{title}{On a test of whether one of two random variables is
  stochastically larger than the other},
\newblock \bibinfo{journal}{Annals of mathematical Statistics}
  \bibinfo{volume}{18} (\bibinfo{year}{1947}) \bibinfo{pages}{50–60}.
%Type = Inbook
\bibitem[{Mood(1950)}]{Mood1950}
\bibinfo{author}{A.~M. Mood}, \bibinfo{title}{Introduction to the Theory of
  Statistics}, \bibinfo{publisher}{McGraw-Hill}, \bibinfo{year}{1950}, pp.
  \bibinfo{pages}{394--399}.
%Type = Misc
\bibitem[{Docker(2020)}]{Docker}
Docker, \bibinfo{title}{https://www.docker.com/, retireved on october 1st,
  2020}, \bibinfo{year}{2020}.
%Type = Article
\bibitem[{Felter et~al.(2015)Felter, Ferreira, Rajamony, and
  Rubio}]{Felter2015}
\bibinfo{author}{W.~Felter}, \bibinfo{author}{A.~Ferreira},
  \bibinfo{author}{R.~Rajamony}, \bibinfo{author}{J.~Rubio},
\newblock \bibinfo{title}{An updated performance comparison of virtual machines
  and linux containers},
\newblock \bibinfo{journal}{2015 IEEE International Symposium on Performance
  Analysis of Systems and Software (ISPASS)}  (\bibinfo{year}{2015})
  \bibinfo{pages}{171--172}.
%Type = Misc
\bibitem[{openmp(2020)}]{OpenMP}
openmp, \bibinfo{title}{https://www.openmp.org/, retrieved on august 14th,
  2020}, \bibinfo{year}{2020}.
%Type = Misc
\bibitem[{pocl(2020)}]{pocl}
pocl, \bibinfo{title}{http://portablecl.org/, retrieved on july 12th, 2020},
  \bibinfo{year}{2020}.
%Type = Article
\bibitem[{Shen et~al.(2013)Shen, Fang, Sips, and Varbanescu}]{SHEN2013834}
\bibinfo{author}{J.~Shen}, \bibinfo{author}{J.~Fang},
  \bibinfo{author}{H.~Sips}, \bibinfo{author}{A.~L. Varbanescu},
\newblock \bibinfo{title}{An application-centric evaluation of opencl on
  multi-core cpus},
\newblock \bibinfo{journal}{Parallel Computing} \bibinfo{volume}{39}
  (\bibinfo{year}{2013}) \bibinfo{pages}{834 -- 850}.

\end{thebibliography}

\onecolumn

\setcounter{page}{1}
\renewcommand\thepage{S-\roman{page}}

\section*{Supplementary material}

\begin{suppfigure*}[h!]
\centering
\includegraphics[width=0.7\textwidth]{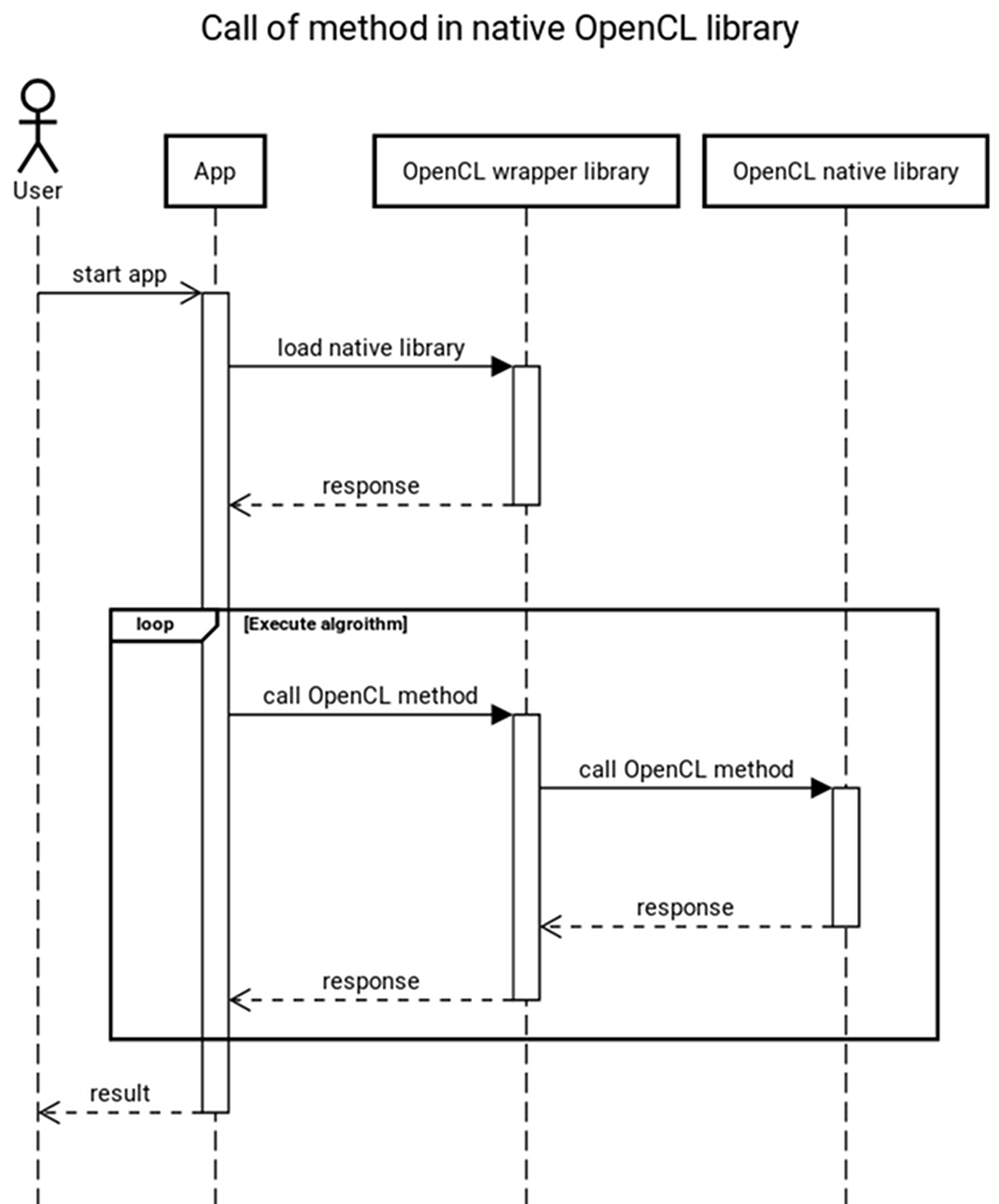}
\caption{Sequence diagram of a call to the native OpenCL library on the Android device. Further explanation see \ref{refx1}.}
\label{fig:sequwrap}
\end{suppfigure*}

\begin{supptable*}
\centering
\setstretch{0.9}
\begin{tabular}{ |c|c|c|c|c|c|c| }
\hline
 W & T & P 
& Java 
& C 
& Assembler 
& SIMD 
\\
\hline
\multirow{24}{*}{1} &   \multirow{6}{*}{1} &\multirow{3}{*}{s} 
 & {\bf 297.78} & 142.01 & 168.67 & 43.48\\
 & &  & {\footnotesize (240.86-354.70)} & {\footnotesize (140.20-143.82)} & {\footnotesize (166.62-170.72)} & {\footnotesize (41.65-45.31)}\\
 & &  & {\footnotesize [1.00${}^{a}$, 1.00${}^{a}$, 1.00${}^{a}$]} & {\footnotesize [\underline{2.10}${}^{b}$, 1.00${}^{a}$, \underline{2.10}${}^{b}$]} & {\footnotesize [\underline{1.77}${}^{b}$, 1.00${}^{a}$, \underline{1.77}${}^{b}$]} & {\footnotesize [\underline{6.85}${}^{b}$, 1.00${}^{a}$, \underline{6.85}${}^{b}$]}\\
\cline{3-7}
 &  & \multirow{3}{*}{d} 
 & {\bf 191.88} & 145.49 & 166.37 & 80.45\\
 & &  & {\footnotesize (185.07-198.69)} & {\footnotesize (143.96-147.01)} & {\footnotesize (164.34-168.41)} & {\footnotesize (78.97-81.93)}\\
 & &  & {\footnotesize [1.00${}^{a}$, 1.00${}^{a}$, 1.00${}^{a}$]} & {\footnotesize [\underline{1.32}${}^{a}$, 1.00${}^{a}$, \underline{1.32}${}^{a}$]} & {\footnotesize [\underline{1.15}${}^{a}$, 1.00${}^{a}$, \underline{1.15}${}^{a}$]} & {\footnotesize [\underline{2.39}${}^{a}$, 1.00${}^{a}$, \underline{2.39}${}^{a}$]}\\
\cline{2-7}
 &   \multirow{6}{*}{2} &\multirow{3}{*}{s} 
 & 104.20 & 80.28 & 91.13 & 24.75\\
 & &  & {\footnotesize (102.59-105.82)} & {\footnotesize (78.84-81.71)} & {\footnotesize (89.80-92.45)} & {\footnotesize (23.43-26.08)}\\
 & &  & {\footnotesize [1.00${}^{a}$, \underline{2.86}${}^{b}$, \underline{2.86}${}^{b}$]} & {\footnotesize [\underline{1.30}${}^{b}$, \underline{1.77}${}^{a}$, \underline{3.71}${}^{b}$]} & {\footnotesize [\underline{1.14}${}^{b}$, \underline{1.85}${}^{a}$, \underline{3.27}${}^{b}$]} & {\footnotesize [\underline{4.21}${}^{b}$, \underline{1.76}${}^{a}$, \underline{12.03}${}^{b}$]}\\
\cline{3-7}
 &  & \multirow{3}{*}{d} 
 & 103.20 & 80.32 & 90.88 & 46.79\\
 & &  & {\footnotesize (99.13-107.28)} & {\footnotesize (79.00-81.63)} & {\footnotesize (89.83-91.93)} & {\footnotesize (45.06-48.53)}\\
 & &  & {\footnotesize [1.00${}^{a}$, \underline{1.86}${}^{a}$, \underline{1.86}${}^{a}$]} & {\footnotesize [\underline{1.28}${}^{a}$, \underline{1.81}${}^{a}$, \underline{2.39}${}^{a}$]} & {\footnotesize [\underline{1.14}${}^{a}$, \underline{1.83}${}^{b}$, \underline{2.11}${}^{a}$]} & {\footnotesize [\underline{2.21}${}^{a}$, \underline{1.72}${}^{a}$, \underline{4.10}${}^{a}$]}\\
\cline{2-7}
 &   \multirow{6}{*}{3} &\multirow{3}{*}{s} 
 & 77.00 & 60.06 & 64.80 & 19.90\\
 & &  & {\footnotesize (75.80-78.21)} & {\footnotesize (58.48-61.64)} & {\footnotesize (63.84-65.76)} & {\footnotesize (18.78-21.02)}\\
 & &  & {\footnotesize [1.00${}^{a}$, \underline{3.87}${}^{b}$, \underline{3.87}${}^{b}$]} & {\footnotesize [\underline{1.28}${}^{a}$, \underline{2.36}${}^{a}$, \underline{4.96}${}^{b}$]} & {\footnotesize [\underline{1.19}${}^{a}$, \underline{2.60}${}^{a}$, \underline{4.60}${}^{b}$]} & {\footnotesize [\underline{3.87}${}^{a}$, \underline{2.18}${}^{a}$, \underline{14.96}${}^{b}$]}\\
\cline{3-7}
 &  & \multirow{3}{*}{d} 
 & 73.54 & 59.54 & 63.65 & 35.20\\
 & &  & {\footnotesize (72.14-74.95)} & {\footnotesize (58.46-60.62)} & {\footnotesize (62.71-64.58)} & {\footnotesize (34.11-36.28)}\\
 & &  & {\footnotesize [1.00${}^{a}$, \underline{2.61}${}^{a}$, \underline{2.61}${}^{a}$]} & {\footnotesize [\underline{1.24}${}^{a}$, \underline{2.44}${}^{a}$, \underline{3.22}${}^{a}$]} & {\footnotesize [\underline{1.16}${}^{a}$, \underline{2.61}${}^{b}$, \underline{3.01}${}^{a}$]} & {\footnotesize [\underline{2.09}${}^{a}$, \underline{2.29}${}^{a}$, \underline{5.45}${}^{a}$]}\\
\cline{2-7}
 &   \multirow{6}{*}{4} &\multirow{3}{*}{s} 
 & 69.72 & 53.97 & 58.67 & {\bf 19.36} \\
 & &  & {\footnotesize (67.72-71.72)} & {\footnotesize (52.69-55.26)} & {\footnotesize (56.51-60.84)} & {\footnotesize (18.14-20.58)}\\
 & &  & {\footnotesize [1.00${}^{a}$, \underline{4.27}${}^{b}$, \underline{4.27}${}^{b}$]} & {\footnotesize [\underline{1.29}${}^{a}$, \underline{2.63}${}^{a}$, \underline{5.52}${}^{b}$]} & {\footnotesize [\underline{1.19}${}^{a}$, \underline{2.87}${}^{a}$, \underline{5.08}${}^{b}$]} & {\footnotesize [\underline{3.60}${}^{a}$, \underline{2.25}${}^{a}$, \underline{{\bf 15.38} }${}^{b}$]}\\
\cline{3-7}
 &  & \multirow{3}{*}{d} 
 & 65.45 & 55.24 & 56.21 & {\bf 32.88} \\
 & &  & {\footnotesize (63.45-67.45)} & {\footnotesize (53.73-56.74)} & {\footnotesize (54.54-57.89)} & {\footnotesize (31.45-34.32)}\\
 & &  & {\footnotesize [1.00${}^{a}$, \underline{2.93}${}^{a}$, \underline{2.93}${}^{a}$]} & {\footnotesize [\underline{1.18}${}^{a}$, \underline{2.63}${}^{a}$, \underline{3.47}${}^{a}$]} & {\footnotesize [\underline{1.16}${}^{a}$, \underline{2.96}${}^{a}$, \underline{3.41}${}^{a}$]} & {\footnotesize [\underline{1.99}${}^{a}$, \underline{2.45}${}^{a}$, \underline{{\bf 5.84} }${}^{a}$]}\\
\Xhline{4\arrayrulewidth}
\multirow{24}{*}{2} &   \multirow{6}{*}{1} &\multirow{3}{*}{s} 
 & {\bf 4,357.41} & 3,582.92 & 3,774.88 & 1,518.02\\
 & &  & {\footnotesize (4,332.78-4,382.04)} & {\footnotesize (3,555.58-3,610.26)} & {\footnotesize (*)} & {\footnotesize (*)}\\
 & &  & {\footnotesize [1.00${}^{a}$, 1.00${}^{a}$, 1.00${}^{a}$]} & {\footnotesize [\underline{1.22}${}^{b}$, 1.00${}^{a}$, \underline{1.22}${}^{b}$]} & {\footnotesize [\underline{1.15}${}^{d}$, 1.00${}^{c}$, \underline{1.15}${}^{d}$]} & {\footnotesize [\underline{2.87}${}^{d}$, 1.00${}^{c}$, \underline{2.87}${}^{d}$]}\\
\cline{3-7}
 &  & \multirow{3}{*}{d} 
 & {\bf 4,773.25} & 3,917.74 & 3,913.92 & 2,429.23\\
 & &  & {\footnotesize (4,646.06-4,900.45)} & {\footnotesize (*)} & {\footnotesize (*)} & {\footnotesize (*)}\\
 & &  & {\footnotesize [1.00${}^{a}$, 1.00${}^{a}$, 1.00${}^{a}$]} & {\footnotesize [\underline{1.22}${}^{c}$, 1.00${}^{c}$, \underline{1.22}${}^{c}$]} & {\footnotesize [\underline{1.22}${}^{c}$, 1.00${}^{c}$, \underline{1.22}${}^{c}$]} & {\footnotesize [\underline{1.96}${}^{c}$, 1.00${}^{c}$, \underline{1.96}${}^{c}$]}\\
\cline{2-7}
 &   \multirow{6}{*}{2} &\multirow{3}{*}{s} 
 & 2,360.89 & 2,026.25 & 2,076.01 & 799.97\\
 & &  & {\footnotesize (*)} & {\footnotesize (*)} & {\footnotesize (*)} & {\footnotesize (*)}\\
 & &  & {\footnotesize [1.00${}^{c}$, \underline{1.85}${}^{d}$, \underline{1.85}${}^{d}$]} & {\footnotesize [\underline{1.17}${}^{c}$, \underline{1.77}${}^{c}$, \underline{2.15}${}^{d}$]} & {\footnotesize [\underline{1.14}${}^{c}$, \underline{1.82}${}^{c}$, \underline{2.10}${}^{d}$]} & {\footnotesize [\underline{2.95}${}^{c}$, \underline{1.90}${}^{d}$, \underline{5.45}${}^{c}$]}\\
\cline{3-7}
 &  & \multirow{3}{*}{d} 
 & 2,656.13 & 2,269.43 & 2,394.58 & 1,376.28\\
 & &  & {\footnotesize (2,588.31-2,723.94)} & {\footnotesize (*)} & {\footnotesize (*)} & {\footnotesize (*)}\\
 & &  & {\footnotesize [1.00${}^{a}$, \underline{1.80}${}^{a}$, \underline{1.80}${}^{a}$]} & {\footnotesize [\underline{1.17}${}^{c}$, \underline{1.73}${}^{c}$, \underline{2.10}${}^{c}$]} & {\footnotesize [\underline{1.11}${}^{c}$, \underline{1.63}${}^{d}$, \underline{1.99}${}^{c}$]} & {\footnotesize [\underline{1.93}${}^{c}$, \underline{1.77}${}^{c}$, \underline{3.47}${}^{c}$]}\\
\cline{2-7}
 &   \multirow{6}{*}{3} &\multirow{3}{*}{s} 
 & 1,768.34 & 1,523.62 & 1,531.48 & 565.26\\
 & &  & {\footnotesize (1,753.45-1,783.22)} & {\footnotesize (*)} & {\footnotesize (*)} & {\footnotesize (*)}\\
 & &  & {\footnotesize [1.00${}^{a}$, \underline{2.46}${}^{a}$, \underline{2.46}${}^{a}$]} & {\footnotesize [\underline{1.16}${}^{d}$, \underline{2.35}${}^{c}$, \underline{2.86}${}^{c}$]} & {\footnotesize [\underline{1.15}${}^{c}$, \underline{2.46}${}^{d}$, \underline{2.85}${}^{c}$]} & {\footnotesize [\underline{3.13}${}^{d}$, \underline{2.69}${}^{d}$, \underline{7.71}${}^{c}$]}\\
\cline{3-7}
 &  & \multirow{3}{*}{d} 
 & 2,054.01 & 1,739.97 & 1,880.89 & 1,050.18\\
 & &  & {\footnotesize (*)} & {\footnotesize (1,699.71-1,780.22)} & {\footnotesize (*)} & {\footnotesize (*)}\\
 & &  & {\footnotesize [1.00${}^{c}$, \underline{2.32}${}^{c}$, \underline{2.32}${}^{c}$]} & {\footnotesize [\underline{1.18}${}^{d}$, \underline{2.25}${}^{c}$, \underline{2.74}${}^{a}$]} & {\footnotesize [\underline{1.09}${}^{d}$, \underline{2.08}${}^{d}$, \underline{2.54}${}^{c}$]} & {\footnotesize [\underline{1.96}${}^{c}$, \underline{2.31}${}^{d}$, \underline{4.55}${}^{c}$]}\\
\cline{2-7}
 &   \multirow{6}{*}{4} &\multirow{3}{*}{s} 
 & 1,465.99 & 1,303.88 & 1,283.36 & {\bf 458.49} \\
 & &  & {\footnotesize (*)} & {\footnotesize (*)} & {\footnotesize (*)} & {\footnotesize (*)}\\
 & &  & {\footnotesize [1.00${}^{c}$, \underline{2.97}${}^{c}$, \underline{2.97}${}^{c}$]} & {\footnotesize [\underline{1.12}${}^{d}$, \underline{2.75}${}^{c}$, \underline{3.34}${}^{d}$]} & {\footnotesize [\underline{1.14}${}^{c}$, \underline{2.94}${}^{d}$, \underline{3.40}${}^{c}$]} & {\footnotesize [\underline{3.20}${}^{d}$, \underline{3.31}${}^{d}$, \underline{ {\bf 9.50} }${}^{c}$]}\\
\cline{3-7}
 &  & \multirow{3}{*}{d} 
 & 1,693.03 & 1,363.14 & 1,542.09 & {\bf 827.88} \\
 & &  & {\footnotesize (*)} & {\footnotesize (*)} & {\footnotesize (*)} & {\footnotesize (*)}\\
 & &  & {\footnotesize [1.00${}^{c}$, \underline{2.82}${}^{c}$, \underline{2.82}${}^{c}$]} & {\footnotesize [\underline{1.24}${}^{c}$, \underline{2.87}${}^{d}$, \underline{3.50}${}^{c}$]} & {\footnotesize [\underline{1.10}${}^{c}$, \underline{2.54}${}^{c}$, \underline{3.10}${}^{c}$]} & {\footnotesize [\underline{2.05}${}^{d}$, \underline{2.93}${}^{d}$, \underline{{\bf 5.77} }${}^{c}$]}\\
 \hline
\end{tabular}
\caption{Mean wall clock time with x86 instruction set and SSE3 extensions in ms. W=window, T=number of threads, P=precision, s=single, d=double. Values in curved brackets are the values of the 95\% confidence interval. Values in square brackets are the calculated speedups (see text for explanations). Minima and maxima in bold face. (*)=no normal distribution (median instead of mean, no confidence interval), ${}^{a}$=t-test, ${}^{b}$=Welch-test, ${}^{c}$=Mann-Whithney-U test, ${}^{d}$=Median test, underline=statistically significant (p$<$0.05)}
\label{tab:result_x86_1}
\end{supptable*}

\begin{supptable*}
\centering
\setstretch{0.9}
\begin{tabular}{ |c|c|c|c|c|c|c| }
\hline
 W & T & P 
& Java 
& C 
& Assembler 
& SIMD 
\\
\hline
\multirow{24}{*}{3} &   \multirow{6}{*}{1} &\multirow{3}{*}{s} 
 & {\bf 12,209.28} & 9,858.81 & 10,238.09 & 3,430.66\\
 & &  & {\footnotesize (*)} & {\footnotesize (*)} & {\footnotesize (*)} & {\footnotesize (*)}\\
 & &  & {\footnotesize [1.00${}^{c}$, 1.00${}^{c}$, 1.00${}^{c}$]} & {\footnotesize [\underline{1.24}${}^{c}$, 1.00${}^{c}$, \underline{1.24}${}^{c}$]} & {\footnotesize [\underline{1.19}${}^{c}$, 1.00${}^{c}$, \underline{1.19}${}^{c}$]} & {\footnotesize [\underline{3.56}${}^{d}$, 1.00${}^{c}$, \underline{3.56}${}^{d}$]}\\
\cline{3-7}
 &  & \multirow{3}{*}{d} 
 & {\bf 12,986.46} & 10,042.16 & 10,544.26 & 5,829.21\\
 & &  & {\footnotesize (12,660.48-13,312.44)} & {\footnotesize (*)} & {\footnotesize (*)} & {\footnotesize (*)}\\
 & &  & {\footnotesize [1.00${}^{a}$, 1.00${}^{a}$, 1.00${}^{a}$]} & {\footnotesize [\underline{1.29}${}^{c}$, 1.00${}^{c}$, \underline{1.29}${}^{c}$]} & {\footnotesize [\underline{1.23}${}^{c}$, 1.00${}^{c}$, \underline{1.23}${}^{c}$]} & {\footnotesize [\underline{2.23}${}^{c}$, 1.00${}^{c}$, \underline{2.23}${}^{c}$]}\\
\cline{2-7}
 &   \multirow{6}{*}{2} &\multirow{3}{*}{s} 
 & 6,789.06 & 5,645.57 & 5,780.18 & 1,963.43\\
 & &  & {\footnotesize (6,721.05-6,857.07)} & {\footnotesize (*)} & {\footnotesize (*)} & {\footnotesize (1,919.17-2,007.69)}\\
 & &  & {\footnotesize [1.00${}^{a}$, \underline{1.80}${}^{d}$, \underline{1.80}${}^{d}$]} & {\footnotesize [\underline{1.20}${}^{c}$, \underline{1.75}${}^{c}$, \underline{2.16}${}^{d}$]} & {\footnotesize [\underline{1.17}${}^{c}$, \underline{1.77}${}^{c}$, \underline{2.11}${}^{c}$]} & {\footnotesize [\underline{3.46}${}^{a}$, \underline{1.75}${}^{c}$, \underline{6.22}${}^{d}$]}\\
\cline{3-7}
 &  & \multirow{3}{*}{d} 
 & 7,536.37 & 5,720.70 & 5,898.06 & 3,292.63\\
 & &  & {\footnotesize (7,366.18-7,706.55)} & {\footnotesize (*)} & {\footnotesize (5,821.74-5,974.38)} & {\footnotesize (*)}\\
 & &  & {\footnotesize [1.00${}^{a}$, \underline{1.72}${}^{a}$, \underline{1.72}${}^{a}$]} & {\footnotesize [\underline{1.32}${}^{c}$, \underline{1.76}${}^{c}$, \underline{2.27}${}^{c}$]} & {\footnotesize [\underline{1.28}${}^{a}$, \underline{1.79}${}^{c}$, \underline{2.20}${}^{a}$]} & {\footnotesize [\underline{2.29}${}^{c}$, \underline{1.77}${}^{c}$, \underline{3.94}${}^{c}$]}\\
\cline{2-7}
 &   \multirow{6}{*}{3} &\multirow{3}{*}{s} 
 & 4,960.31 & 4,114.10 & 4,267.82 & 1,550.22\\
 & &  & {\footnotesize (*)} & {\footnotesize (*)} & {\footnotesize (*)} & {\footnotesize (*)}\\
 & &  & {\footnotesize [1.00${}^{c}$, \underline{2.46}${}^{d}$, \underline{2.46}${}^{d}$]} & {\footnotesize [\underline{1.21}${}^{c}$, \underline{2.40}${}^{d}$, \underline{2.97}${}^{d}$]} & {\footnotesize [\underline{1.16}${}^{c}$, \underline{2.40}${}^{c}$, \underline{2.86}${}^{d}$]} & {\footnotesize [\underline{3.20}${}^{d}$, \underline{2.21}${}^{c}$, \underline{7.88}${}^{d}$]}\\
\cline{3-7}
 &  & \multirow{3}{*}{d} 
 & 5,910.54 & 4,169.11 & 4,390.41 & 2,727.32\\
 & &  & {\footnotesize (5,820.75-6,000.34)} & {\footnotesize (*)} & {\footnotesize (*)} & {\footnotesize (*)}\\
 & &  & {\footnotesize [1.00${}^{a}$, \underline{2.20}${}^{a}$, \underline{2.20}${}^{a}$]} & {\footnotesize [\underline{1.42}${}^{d}$, \underline{2.41}${}^{d}$, \underline{3.11}${}^{c}$]} & {\footnotesize [\underline{1.35}${}^{d}$, \underline{2.40}${}^{d}$, \underline{2.96}${}^{c}$]} & {\footnotesize [\underline{2.17}${}^{d}$, \underline{2.14}${}^{c}$, \underline{4.76}${}^{c}$]}\\
\cline{2-7}
 &   \multirow{6}{*}{4} &\multirow{3}{*}{s} 
 & 3,911.83 & 3,507.68 & 3,604.62 & {\bf 1,276.68} \\
 & &  & {\footnotesize (3,877.04-3,946.63)} & {\footnotesize (*)} & {\footnotesize (*)} & {\footnotesize (*)}\\
 & &  & {\footnotesize [1.00${}^{a}$, \underline{3.12}${}^{d}$, \underline{3.12}${}^{d}$]} & {\footnotesize [\underline{1.12}${}^{d}$, \underline{2.81}${}^{d}$, \underline{3.48}${}^{d}$]} & {\footnotesize [\underline{1.09}${}^{d}$, \underline{2.84}${}^{d}$, \underline{3.39}${}^{d}$]} & {\footnotesize [\underline{3.06}${}^{c}$, \underline{2.69}${}^{c}$, \underline{{\bf 9.56} }${}^{d}$]}\\
\cline{3-7}
 &  & \multirow{3}{*}{d} 
 & 4,829.50 & 3,458.77 & 3,540.65 & {\bf 2,259.53} \\
 & &  & {\footnotesize (*)} & {\footnotesize (3,406.63-3,510.91)} & {\footnotesize (3,486.53-3,594.78)} & {\footnotesize (*)}\\
 & &  & {\footnotesize [1.00${}^{c}$, \underline{2.69}${}^{c}$, \underline{2.69}${}^{c}$]} & {\footnotesize [\underline{1.40}${}^{c}$, \underline{2.90}${}^{d}$, \underline{3.75}${}^{a}$]} & {\footnotesize [\underline{1.36}${}^{c}$, \underline{2.98}${}^{d}$, \underline{3.67}${}^{a}$]} & {\footnotesize [\underline{2.14}${}^{c}$, \underline{2.58}${}^{c}$, \underline{ {\bf 5.75} }${}^{c}$]}\\
\hline
\end{tabular}
\caption{Mean wall clock time with x86 instruction set and SSE3 extensions in ms. W=window, T=number of threads, P=precision, s=single, d=double. Values in curved brackets are the values of the 95\% confidence interval. Values in square brackets are the calculated speedups (see text for explanations). Minima and maxima in bold face. (*)=no normal distribution (median instead of mean, no confidence interval), ${}^{a}$=t-test, ${}^{b}$=Welch-test, ${}^{c}$=Mann-Whithney-U test, ${}^{d}$=Median test, underline=statistically significant (p$<$0.05)}
\label{tab:result_x86_2}
\end{supptable*}

\begin{supptable*}
\centering
\begin{tabular}{ |c|c|c|c|c|c|c| }
\hline
 W & T & P 
& Java 
& C 
& Assembler 
& SIMD 
\\
\hline
\multirow{24}{*}{1} &   \multirow{6}{*}{1} &\multirow{3}{*}{s} 
 & {\bf 187.56} & 141.91 & 166.88 & 45.11\\
 & &  & {\footnotesize (182.52-192.60)} & {\footnotesize (140.09-143.73)} & {\footnotesize (166.00-167.76)} & {\footnotesize (43.23-46.98)}\\
 & &  & {\footnotesize [1.00${}^{a}$, 1.00${}^{a}$, 1.00${}^{a}$]} & {\footnotesize [\underline{1.32}${}^{a}$, 1.00${}^{a}$, \underline{1.32}${}^{a}$]} & {\footnotesize [\underline{1.12}${}^{a}$, 1.00${}^{a}$, \underline{1.12}${}^{a}$]} & {\footnotesize [\underline{4.16}${}^{a}$, 1.00${}^{a}$, \underline{4.16}${}^{a}$]}\\
\cline{3-7}
 &  & \multirow{3}{*}{d} 
 & 168.26 & 138.94 & {\bf 172.24} & 83.03\\
 & &  & {\footnotesize (165.73-170.78)} & {\footnotesize (137.22-140.66)} & {\footnotesize (171.44-173.03)} & {\footnotesize (81.37-84.68)}\\
 & &  & {\footnotesize [1.00${}^{a}$, 1.00${}^{a}$, 1.00${}^{a}$]} & {\footnotesize [\underline{1.21}${}^{a}$, 1.00${}^{a}$, \underline{1.21}${}^{a}$]} & {\footnotesize [\underline{0.98}${}^{a}$, 1.00${}^{a}$, \underline{0.98}${}^{a}$]} & {\footnotesize [\underline{2.03}${}^{a}$, 1.00${}^{a}$, \underline{2.03}${}^{a}$]}\\
\cline{2-7}
 &   \multirow{6}{*}{2} &\multirow{3}{*}{s} 
 & 86.85 & 78.14 & 90.77 & 25.63\\
 & &  & {\footnotesize (85.63-88.07)} & {\footnotesize (77.01-79.27)} & {\footnotesize (89.48-92.07)} & {\footnotesize (24.41-26.84)}\\
 & &  & {\footnotesize [1.00${}^{a}$, \underline{2.16}${}^{a}$, \underline{2.16}${}^{a}$]} & {\footnotesize [\underline{1.11}${}^{a}$, \underline{1.82}${}^{a}$, \underline{2.40}${}^{a}$]} & {\footnotesize [\underline{0.96}${}^{a}$, \underline{1.84}${}^{a}$, \underline{2.07}${}^{a}$]} & {\footnotesize [\underline{3.39}${}^{a}$, \underline{1.76}${}^{a}$, \underline{7.32}${}^{a}$]}\\
\cline{3-7}
 &  & \multirow{3}{*}{d} 
 & 87.00 & 79.15 & 93.75 & 49.36\\
 & &  & {\footnotesize (85.63-88.37)} & {\footnotesize (77.63-80.67)} & {\footnotesize (92.69-94.82)} & {\footnotesize (47.75-50.96)}\\
 & &  & {\footnotesize [1.00${}^{a}$, \underline{1.93}${}^{a}$, \underline{1.93}${}^{a}$]} & {\footnotesize [\underline{1.10}${}^{a}$, \underline{1.76}${}^{a}$, \underline{2.13}${}^{a}$]} & {\footnotesize [\underline{0.93}${}^{a}$, \underline{1.84}${}^{a}$, \underline{1.79}${}^{a}$]} & {\footnotesize [\underline{1.76}${}^{b}$, \underline{1.68}${}^{b}$, \underline{3.41}${}^{a}$]}\\
\cline{2-7}
 &   \multirow{6}{*}{3} &\multirow{3}{*}{s} 
 & 62.52 & 58.77 & 64.72 & 20.55\\
 & &  & {\footnotesize (60.86-64.18)} & {\footnotesize (57.54-60.00)} & {\footnotesize (63.71-65.73)} & {\footnotesize (19.24-21.86)}\\
 & &  & {\footnotesize [1.00${}^{a}$, \underline{3.00}${}^{a}$, \underline{3.00}${}^{a}$]} & {\footnotesize [\underline{1.06}${}^{a}$, \underline{2.41}${}^{a}$, \underline{3.19}${}^{a}$]} & {\footnotesize [\underline{0.97}${}^{a}$, \underline{2.58}${}^{a}$, \underline{2.90}${}^{a}$]} & {\footnotesize [\underline{3.04}${}^{a}$, \underline{2.19}${}^{a}$, \underline{9.13}${}^{a}$]}\\
\cline{3-7}
 &  & \multirow{3}{*}{d} 
 & 62.32 & 58.18 & 66.72 & 39.62\\
 & &  & {\footnotesize (60.94-63.70)} & {\footnotesize (56.72-59.65)} & {\footnotesize (65.67-67.76)} & {\footnotesize (38.09-41.15)}\\
 & &  & {\footnotesize [1.00${}^{a}$, \underline{2.70}${}^{a}$, \underline{2.70}${}^{a}$]} & {\footnotesize [\underline{1.07}${}^{a}$, \underline{2.39}${}^{a}$, \underline{2.89}${}^{a}$]} & {\footnotesize [\underline{0.93}${}^{a}$, \underline{2.58}${}^{a}$, \underline{2.52}${}^{a}$]} & {\footnotesize [\underline{1.57}${}^{a}$, \underline{2.10}${}^{a}$, \underline{4.25}${}^{a}$]}\\
\cline{2-7}
 &   \multirow{6}{*}{4} &\multirow{3}{*}{s} 
 & 55.41 & 55.73 & 58.17 & {\bf 19.11} \\
 & &  & {\footnotesize (53.88-56.93)} & {\footnotesize (*)} & {\footnotesize (56.23-60.11)} & {\footnotesize (17.95-20.26)}\\
 & &  & {\footnotesize [1.00${}^{a}$, \underline{3.39}${}^{a}$, \underline{3.39}${}^{a}$]} & {\footnotesize [0.99${}^{c}$, \underline{2.55}${}^{c}$, \underline{3.37}${}^{c}$]} & {\footnotesize [\underline{0.95}${}^{a}$, \underline{2.87}${}^{b}$, \underline{3.22}${}^{a}$]} & {\footnotesize [\underline{2.90}${}^{a}$, \underline{2.36}${}^{a}$, \underline{ {\bf 9.82} }${}^{a}$]}\\
\cline{3-7}
 &  & \multirow{3}{*}{d} 
 & 57.34 & 53.90 & 60.03 & {\bf 36.53} \\
 & &  & {\footnotesize (55.39-59.29)} & {\footnotesize (52.07-55.73)} & {\footnotesize (58.18-61.88)} & {\footnotesize (34.85-38.21)}\\
 & &  & {\footnotesize [1.00${}^{a}$, \underline{2.93}${}^{a}$, \underline{2.93}${}^{a}$]} & {\footnotesize [\underline{1.06}${}^{a}$, \underline{2.58}${}^{a}$, \underline{3.12}${}^{a}$]} & {\footnotesize [\underline{0.96}${}^{a}$, \underline{2.87}${}^{b}$, \underline{2.80}${}^{a}$]} & {\footnotesize [\underline{1.57}${}^{a}$, \underline{2.27}${}^{a}$, \underline{ {\bf 4.61} }${}^{a}$]}\\
\hline
\end{tabular}
\caption{Mean wall clock time with x86\_64 instruction set and SSE4.2 extensions in ms. W=window, T=number of threads, P=precision, s=single, d=double. Values in curved brackets are the values of the 95\% confidence interval. Values in square brackets are the calculated speedups (see text for explanations). Minima and maxima in bold face. (*)=no normal distribution (median instead of mean, no confidence interval), ${}^{a}$=t-test, ${}^{b}$=Welch-test, ${}^{c}$=Mann-Whithney-U test, ${}^{d}$=Median test, underline=statistically significant (p$<$0.05)}
\label{tab:result_x86_64_1}
\end{supptable*}

\begin{supptable*}
\centering
\begin{tabular}{ |c|c|c|c|c|c|c| }
\hline
 W & T & P 
& Java 
& C 
& Assembler 
& SIMD 
\\
\hline
\multirow{24}{*}{2} &   \multirow{6}{*}{1} &\multirow{3}{*}{s} 
 & {\bf 4,236.19} & 3,821.24 & 3,977.30 & 1,341.01\\
 & &  & {\footnotesize (4,207.62-4,264.75)} & {\footnotesize (3,765.17-3,877.32)} & {\footnotesize (3,914.39-4,040.22)} & {\footnotesize (*)}\\
 & &  & {\footnotesize [1.00${}^{a}$, 1.00${}^{a}$, 1.00${}^{a}$]} & {\footnotesize [\underline{1.11}${}^{b}$, 1.00${}^{a}$, \underline{1.11}${}^{b}$]} & {\footnotesize [\underline{1.07}${}^{b}$, 1.00${}^{a}$, \underline{1.07}${}^{b}$]} & {\footnotesize [\underline{3.16}${}^{c}$, 1.00${}^{c}$, \underline{3.16}${}^{c}$]}\\
\cline{3-7}
 &  & \multirow{3}{*}{d} 
 & 4,021.92 & 3,848.30 & {\bf 4,170.62} & 2,152.90\\
 & &  & {\footnotesize (3,949.05-4,094.78)} & {\footnotesize (*)} & {\footnotesize (*)} & {\footnotesize (*)}\\
 & &  & {\footnotesize [1.00${}^{a}$, 1.00${}^{a}$, 1.00${}^{a}$]} & {\footnotesize [\underline{1.05}${}^{c}$, 1.00${}^{c}$, \underline{1.05}${}^{c}$]} & {\footnotesize [\underline{0.96}${}^{c}$, 1.00${}^{c}$, \underline{0.96}${}^{c}$]} & {\footnotesize [\underline{1.87}${}^{c}$, 1.00${}^{c}$, \underline{1.87}${}^{c}$]}\\
\cline{2-7}
 &   \multirow{6}{*}{2} &\multirow{3}{*}{s} 
 & 2,252.54 & 2,012.16 & 2,145.86 & 699.99\\
 & &  & {\footnotesize (*)} & {\footnotesize (*)} & {\footnotesize (*)} & {\footnotesize (*)}\\
 & &  & {\footnotesize [1.00${}^{c}$, \underline{1.88}${}^{d}$, \underline{1.88}${}^{d}$]} & {\footnotesize [\underline{1.12}${}^{d}$, \underline{1.90}${}^{d}$, \underline{2.11}${}^{c}$]} & {\footnotesize [\underline{1.05}${}^{d}$, \underline{1.85}${}^{d}$, \underline{1.97}${}^{d}$]} & {\footnotesize [\underline{3.22}${}^{d}$, \underline{1.92}${}^{d}$, \underline{6.05}${}^{c}$]}\\
\cline{3-7}
 &  & \multirow{3}{*}{d} 
 & 2,338.51 & 2,318.18 & 2,441.74 & 1,330.98\\
 & &  & {\footnotesize (*)} & {\footnotesize (*)} & {\footnotesize (*)} & {\footnotesize (*)}\\
 & &  & {\footnotesize [1.00${}^{c}$, \underline{1.72}${}^{c}$, \underline{1.72}${}^{c}$]} & {\footnotesize [1.01${}^{c}$, \underline{1.66}${}^{c}$, \underline{1.73}${}^{c}$]} & {\footnotesize [0.96${}^{d}$, \underline{1.71}${}^{d}$, \underline{1.65}${}^{c}$]} & {\footnotesize [\underline{1.76}${}^{c}$, \underline{1.62}${}^{d}$, \underline{3.02}${}^{c}$]}\\
\cline{2-7}
 &   \multirow{6}{*}{3} &\multirow{3}{*}{s} 
 & 1,812.41 & 1,530.58 & 1,635.36 & 493.53\\
 & &  & {\footnotesize (1,790.54-1,834.28)} & {\footnotesize (*)} & {\footnotesize (*)} & {\footnotesize (*)}\\
 & &  & {\footnotesize [1.00${}^{a}$, \underline{2.34}${}^{a}$, \underline{2.34}${}^{a}$]} & {\footnotesize [\underline{1.18}${}^{c}$, \underline{2.50}${}^{d}$, \underline{2.77}${}^{c}$]} & {\footnotesize [\underline{1.11}${}^{c}$, \underline{2.43}${}^{d}$, \underline{2.59}${}^{c}$]} & {\footnotesize [\underline{3.67}${}^{d}$, \underline{2.72}${}^{d}$, \underline{8.58}${}^{d}$]}\\
\cline{3-7}
 &  & \multirow{3}{*}{d} 
 & 2,098.70 & 1,795.00 & 1,935.61 & 1,128.83\\
 & &  & {\footnotesize (*)} & {\footnotesize (*)} & {\footnotesize (*)} & {\footnotesize (*)}\\
 & &  & {\footnotesize [1.00${}^{c}$, \underline{1.92}${}^{c}$, \underline{1.92}${}^{c}$]} & {\footnotesize [\underline{1.17}${}^{c}$, \underline{2.14}${}^{c}$, \underline{2.24}${}^{c}$]} & {\footnotesize [\underline{1.08}${}^{c}$, \underline{2.15}${}^{c}$, \underline{2.08}${}^{c}$]} & {\footnotesize [\underline{1.86}${}^{d}$, \underline{1.91}${}^{c}$, \underline{3.56}${}^{c}$]}\\
\cline{2-7}
 &   \multirow{6}{*}{4} &\multirow{3}{*}{s} 
 & 1,495.02 & 1,279.85 & 1,328.15 & {\bf 366.85}\\
 & &  & {\footnotesize (*)} & {\footnotesize (*)} & {\footnotesize (1,307.94-1,348.36)} & {\footnotesize (*)}\\
 & &  & {\footnotesize [1.00${}^{c}$, \underline{2.83}${}^{c}$, \underline{2.83}${}^{c}$]} & {\footnotesize [\underline{1.17}${}^{c}$, \underline{2.99}${}^{d}$, \underline{3.31}${}^{c}$]} & {\footnotesize [\underline{1.13}${}^{c}$, \underline{2.99}${}^{b}$, \underline{3.19}${}^{a}$]} & {\footnotesize [\underline{4.08}${}^{d}$, \underline{3.66}${}^{d}$, \underline{ {\bf 11.55} }${}^{d}$]}\\
\cline{3-7}
 &  & \multirow{3}{*}{d} 
 & 1,872.47 & 1,476.92 & 1,403.56 & {\bf 925.22} \\
 & &  & {\footnotesize (*)} & {\footnotesize (*)} & {\footnotesize (*)} & {\footnotesize (*)}\\
 & &  & {\footnotesize [1.00${}^{c}$, \underline{2.15}${}^{c}$, \underline{2.15}${}^{c}$]} & {\footnotesize [\underline{1.27}${}^{d}$, \underline{2.61}${}^{c}$, \underline{2.72}${}^{c}$]} & {\footnotesize [\underline{1.33}${}^{d}$, \underline{2.97}${}^{c}$, \underline{2.87}${}^{c}$]} & {\footnotesize [\underline{2.02}${}^{d}$, \underline{2.33}${}^{c}$, \underline{ {\bf 4.35} }${}^{c}$]}\\
\Xhline{4\arrayrulewidth}
\multirow{24}{*}{3} &   \multirow{6}{*}{1} &\multirow{3}{*}{s} 
 & {\bf 11,326.66} & 9,512.57 & 10,037.57 & 3,276.98\\
 & &  & {\footnotesize (11,245.88-11,407.43)} & {\footnotesize (9,458.47-9,566.66)} & {\footnotesize (*)} & {\footnotesize (*)}\\
 & &  & {\footnotesize [1.00${}^{a}$, 1.00${}^{a}$, 1.00${}^{a}$]} & {\footnotesize [\underline{1.19}${}^{a}$, 1.00${}^{a}$, \underline{1.19}${}^{a}$]} & {\footnotesize [\underline{1.13}${}^{c}$, 1.00${}^{c}$, \underline{1.13}${}^{c}$]} & {\footnotesize [\underline{3.46}${}^{c}$, 1.00${}^{c}$, \underline{3.46}${}^{c}$]}\\
\cline{3-7}
 &  & \multirow{3}{*}{d} 
 & {\bf 11,577.91} & 9,524.42 & 10,492.35 & 5,550.86\\
 & &  & {\footnotesize (11,396.59-11,759.23)} & {\footnotesize (*)} & {\footnotesize (*)} & {\footnotesize (5,488.93-5,612.78)}\\
 & &  & {\footnotesize [1.00${}^{a}$, 1.00${}^{a}$, 1.00${}^{a}$]} & {\footnotesize [\underline{1.22}${}^{c}$, 1.00${}^{c}$, \underline{1.22}${}^{c}$]} & {\footnotesize [\underline{1.10}${}^{c}$, 1.00${}^{c}$, \underline{1.10}${}^{c}$]} & {\footnotesize [\underline{2.09}${}^{a}$, 1.00${}^{a}$, \underline{2.09}${}^{a}$]}\\
\cline{2-7}
 &   \multirow{6}{*}{2} &\multirow{3}{*}{s} 
 & 5,828.26 & 5,553.59 & 5,762.30 & 1,786.84\\
 & &  & {\footnotesize (5,745.91-5,910.62)} & {\footnotesize (5,460.63-5,646.54)} & {\footnotesize (5,693.84-5,830.76)} & {\footnotesize (*)}\\
 & &  & {\footnotesize [1.00${}^{a}$, \underline{1.94}${}^{a}$, \underline{1.94}${}^{a}$]} & {\footnotesize [\underline{1.05}${}^{a}$, \underline{1.71}${}^{b}$, \underline{2.04}${}^{a}$]} & {\footnotesize [1.01${}^{a}$, \underline{1.74}${}^{c}$, \underline{1.97}${}^{a}$]} & {\footnotesize [\underline{3.26}${}^{d}$, \underline{1.83}${}^{c}$, \underline{6.34}${}^{d}$]}\\
\cline{3-7}
 &  & \multirow{3}{*}{d} 
 & 6,275.84 & 5,648.08 & 5,924.89 & 3,349.58\\
 & &  & {\footnotesize (*)} & {\footnotesize (*)} & {\footnotesize (*)} & {\footnotesize (*)}\\
 & &  & {\footnotesize [1.00${}^{c}$, \underline{1.84}${}^{c}$, \underline{1.84}${}^{c}$]} & {\footnotesize [\underline{1.11}${}^{c}$, \underline{1.69}${}^{d}$, \underline{2.05}${}^{c}$]} & {\footnotesize [\underline{1.06}${}^{c}$, \underline{1.77}${}^{c}$, \underline{1.95}${}^{c}$]} & {\footnotesize [\underline{1.87}${}^{c}$, \underline{1.66}${}^{c}$, \underline{3.46}${}^{c}$]}\\
\cline{2-7}
 &   \multirow{6}{*}{3} &\multirow{3}{*}{s} 
 & 4,229.38 & 3,951.35 & 4,176.13 & 1,356.72\\
 & &  & {\footnotesize (*)} & {\footnotesize (*)} & {\footnotesize (*)} & {\footnotesize (*)}\\
 & &  & {\footnotesize [1.00${}^{c}$, \underline{2.68}${}^{c}$, \underline{2.68}${}^{c}$]} & {\footnotesize [\underline{1.07}${}^{c}$, \underline{2.41}${}^{c}$, \underline{2.87}${}^{c}$]} & {\footnotesize [1.01${}^{c}$, \underline{2.40}${}^{d}$, \underline{2.71}${}^{c}$]} & {\footnotesize [\underline{3.12}${}^{d}$, \underline{2.42}${}^{d}$, \underline{8.35}${}^{d}$]}\\
\cline{3-7}
 &  & \multirow{3}{*}{d} 
 & 5,158.15 & 3,955.32 & 4,392.99 & 2,826.95\\
 & &  & {\footnotesize (*)} & {\footnotesize (3,911.35-3,999.29)} & {\footnotesize (*)} & {\footnotesize (*)}\\
 & &  & {\footnotesize [1.00${}^{c}$, \underline{2.24}${}^{c}$, \underline{2.24}${}^{c}$]} & {\footnotesize [\underline{1.30}${}^{d}$, \underline{2.41}${}^{c}$, \underline{2.93}${}^{b}$]} & {\footnotesize [\underline{1.17}${}^{c}$, \underline{2.39}${}^{c}$, \underline{2.64}${}^{c}$]} & {\footnotesize [\underline{1.82}${}^{c}$, \underline{1.96}${}^{c}$, \underline{4.10}${}^{c}$]}\\
\cline{2-7}
 &   \multirow{6}{*}{4} &\multirow{3}{*}{s} 
 & 3,403.69 & 3,069.36 & 3,254.17 & {\bf 1,087.24} \\
 & &  & {\footnotesize (*)} & {\footnotesize (*)} & {\footnotesize (3,209.43-3,298.90)} & {\footnotesize (1,058.40-1,116.09)}\\
 & &  & {\footnotesize [1.00${}^{c}$, \underline{3.33}${}^{c}$, \underline{3.33}${}^{c}$]} & {\footnotesize [\underline{1.11}${}^{d}$, \underline{3.10}${}^{d}$, \underline{3.69}${}^{d}$]} & {\footnotesize [\underline{1.05}${}^{c}$, \underline{3.08}${}^{d}$, \underline{3.48}${}^{b}$]} & {\footnotesize [\underline{3.13}${}^{d}$, \underline{3.01}${}^{d}$, \underline{ {\bf 10.42} }${}^{b}$]}\\
\cline{3-7}
 &  & \multirow{3}{*}{d} 
 & 4,316.69 & 3,250.89 & 3,619.40 & {\bf 2,457.99} \\
 & &  & {\footnotesize (*)} & {\footnotesize (3,208.97-3,292.81)} & {\footnotesize (*)} & {\footnotesize (*)}\\
 & &  & {\footnotesize [1.00${}^{c}$, \underline{2.68}${}^{c}$, \underline{2.68}${}^{c}$]} & {\footnotesize [\underline{1.33}${}^{c}$, \underline{2.93}${}^{c}$, \underline{3.56}${}^{b}$]} & {\footnotesize [\underline{1.19}${}^{c}$, \underline{2.90}${}^{d}$, \underline{3.20}${}^{c}$]} & {\footnotesize [\underline{1.76}${}^{c}$, \underline{2.26}${}^{d}$, \underline{ {\bf 4.71} }${}^{c}$]}\\
\hline
\end{tabular}
\caption{Mean wall clock time with x86\_64 instruction set and SSE4.2 extensions in ms. W=window, T=number of threads, P=precision, s=single, d=double. Values in curved brackets are the values of the 95\% confidence interval. Values in square brackets are the calculated speedups (see text for explanations). Minima and maxima in bold face. (*)=no normal distribution (median instead of mean, no confidence interval), ${}^{a}$=t-test, ${}^{b}$=Welch-test, ${}^{c}$=Mann-Whithney-U test, ${}^{d}$=Median test, underline=statistically significant (p$<$0.05)}
\label{tab:result_x86_64_2}
\end{supptable*}

\begin{supptable*}
\centering
\setstretch{0.9}
\begin{tabular}{ |c|c|c|c|c|c|c| }
\hline
 W & T & P 
& Java 
& C 
& Assembler 
& SIMD 
\\
\hline
\multirow{48}{*}{1} &   \multirow{6}{*}{1} &\multirow{3}{*}{s} 
 & {\bf 457.35} & 330.48 & 345.01 & 118.71\\
 & &  & {\footnotesize (450.39-464.31)} & {\footnotesize (326.90-334.07)} & {\footnotesize (341.74-348.28)} & {\footnotesize (115.04-122.38)}\\
 & &  & {\footnotesize [1.00${}^{a}$, 1.00${}^{a}$, 1.00${}^{a}$]} & {\footnotesize [\underline{1.38}${}^{a}$, 1.00${}^{a}$, \underline{1.38}${}^{a}$]} & {\footnotesize [\underline{1.33}${}^{a}$, 1.00${}^{a}$, \underline{1.33}${}^{a}$]} & {\footnotesize [\underline{3.85}${}^{a}$, 1.00${}^{a}$, \underline{3.85}${}^{a}$]}\\
\cline{3-7}
 &  & \multirow{3}{*}{d} 
 & {\bf 454.16} & 330.83 & 346.04 & -\\
 & &  & {\footnotesize (433.97-474.34)} & {\footnotesize (327.26-334.40)} & {\footnotesize (342.51-349.57)} & \\
 & &  & {\footnotesize [1.00${}^{a}$, 1.00${}^{a}$, 1.00${}^{a}$]} & {\footnotesize [\underline{1.37}${}^{a}$, 1.00${}^{a}$, \underline{1.37}${}^{a}$]} & {\footnotesize [\underline{1.31}${}^{a}$, 1.00${}^{a}$, \underline{1.31}${}^{a}$]} & \\
\cline{2-7}
 &   \multirow{6}{*}{2} &\multirow{3}{*}{s} 
 & 353.69 & 273.86 & 274.06 & 130.55\\
 & &  & {\footnotesize (329.51-377.87)} & {\footnotesize (258.48-289.23)} & {\footnotesize (258.90-289.21)} & {\footnotesize (125.50-135.61)}\\
 & &  & {\footnotesize [1.00${}^{a}$, \underline{1.29}${}^{b}$, \underline{1.29}${}^{b}$]} & {\footnotesize [\underline{1.29}${}^{a}$, \underline{1.21}${}^{b}$, \underline{1.67}${}^{b}$]} & {\footnotesize [\underline{1.29}${}^{b}$, \underline{1.26}${}^{b}$, \underline{1.67}${}^{b}$]} & {\footnotesize [\underline{2.71}${}^{b}$, \underline{0.91}${}^{b}$, \underline{3.50}${}^{a}$]}\\
\cline{3-7}
 &  & \multirow{3}{*}{d} 
 & 281.21 & 275.44 & 279.92 & -\\
 & &  & {\footnotesize (267.31-295.12)} & {\footnotesize (259.30-291.58)} & {\footnotesize (263.22-296.62)} & \\
 & &  & {\footnotesize [1.00${}^{a}$, \underline{1.61}${}^{a}$, \underline{1.61}${}^{a}$]} & {\footnotesize [1.02${}^{a}$, \underline{1.20}${}^{b}$, \underline{1.65}${}^{a}$]} & {\footnotesize [1.00${}^{a}$, \underline{1.24}${}^{b}$, \underline{1.62}${}^{a}$]} & \\
\cline{2-7}
 &   \multirow{6}{*}{3} &\multirow{3}{*}{s} 
 & 235.47 & 200.68 & 197.73 & 103.09\\
 & &  & {\footnotesize (*)} & {\footnotesize (*)} & {\footnotesize (195.59-199.87)} & {\footnotesize (99.39-106.79)}\\
 & &  & {\footnotesize [1.00${}^{c}$, \underline{1.94}${}^{c}$, \underline{1.94}${}^{c}$]} & {\footnotesize [\underline{1.17}${}^{c}$, \underline{1.65}${}^{c}$, \underline{2.28}${}^{c}$]} & {\footnotesize [\underline{1.19}${}^{c}$, \underline{1.74}${}^{a}$, \underline{2.31}${}^{a}$]} & {\footnotesize [\underline{2.28}${}^{d}$, \underline{1.15}${}^{b}$, \underline{4.44}${}^{a}$]}\\
\cline{3-7}
 &  & \multirow{3}{*}{d} 
 & 215.14 & 196.90 & 198.46 & -\\
 & &  & {\footnotesize (213.45-216.83)} & {\footnotesize (195.11-198.69)} & {\footnotesize (*)} & \\
 & &  & {\footnotesize [1.00${}^{a}$, \underline{2.11}${}^{a}$, \underline{2.11}${}^{a}$]} & {\footnotesize [\underline{1.09}${}^{a}$, \underline{1.68}${}^{a}$, \underline{2.31}${}^{a}$]} & {\footnotesize [\underline{1.08}${}^{c}$, \underline{1.74}${}^{c}$, \underline{2.29}${}^{c}$]} & \\
\cline{2-7}
 &   \multirow{6}{*}{4} &\multirow{3}{*}{s} 
 & 201.35 & 172.98 & 171.12 & 87.49\\
 & &  & {\footnotesize (*)} & {\footnotesize (*)} & {\footnotesize (*)} & {\footnotesize (84.30-90.67)}\\
 & &  & {\footnotesize [1.00${}^{c}$, \underline{2.27}${}^{c}$, \underline{2.27}${}^{c}$]} & {\footnotesize [\underline{1.16}${}^{c}$, \underline{1.91}${}^{c}$, \underline{2.64}${}^{c}$]} & {\footnotesize [\underline{1.18}${}^{c}$, \underline{2.02}${}^{c}$, \underline{2.67}${}^{c}$]} & {\footnotesize [\underline{2.30}${}^{d}$, \underline{1.36}${}^{b}$, \underline{5.23}${}^{a}$]}\\
\cline{3-7}
 &  & \multirow{3}{*}{d} 
 & 189.25 & 172.75 & 171.70 & -\\
 & &  & {\footnotesize (*)} & {\footnotesize (*)} & {\footnotesize (*)} & \\
 & &  & {\footnotesize [1.00${}^{c}$, \underline{2.40}${}^{c}$, \underline{2.40}${}^{c}$]} & {\footnotesize [\underline{1.10}${}^{c}$, \underline{1.92}${}^{c}$, \underline{2.63}${}^{c}$]} & {\footnotesize [\underline{1.10}${}^{c}$, \underline{2.02}${}^{c}$, \underline{2.65}${}^{c}$]} & \\
\cline{2-7}
 &   \multirow{6}{*}{5} &\multirow{3}{*}{s} 
 & 181.12 & 155.41 & 152.47 & 77.46\\
 & &  & {\footnotesize (*)} & {\footnotesize (*)} & {\footnotesize (*)} & {\footnotesize (*)}\\
 & &  & {\footnotesize [1.00${}^{c}$, \underline{2.53}${}^{c}$, \underline{2.53}${}^{c}$]} & {\footnotesize [\underline{1.17}${}^{c}$, \underline{2.13}${}^{c}$, \underline{2.94}${}^{c}$]} & {\footnotesize [\underline{1.19}${}^{c}$, \underline{2.26}${}^{c}$, \underline{3.00}${}^{c}$]} & {\footnotesize [\underline{2.34}${}^{c}$, \underline{1.53}${}^{d}$, \underline{5.90}${}^{c}$]}\\
\cline{3-7}
 &  & \multirow{3}{*}{d} 
 & 173.55 & 153.03 & 153.84 & -\\
 & &  & {\footnotesize (*)} & {\footnotesize (*)} & {\footnotesize (*)} & \\
 & &  & {\footnotesize [1.00${}^{c}$, \underline{2.62}${}^{c}$, \underline{2.62}${}^{c}$]} & {\footnotesize [\underline{1.13}${}^{c}$, \underline{2.16}${}^{c}$, \underline{2.97}${}^{c}$]} & {\footnotesize [\underline{1.13}${}^{c}$, \underline{2.25}${}^{c}$, \underline{2.95}${}^{c}$]} & \\
\cline{2-7}
 &   \multirow{6}{*}{6} &\multirow{3}{*}{s} 
 & 166.11 & 141.51 & 137.70 & 72.68\\
 & &  & {\footnotesize (163.82-168.40)} & {\footnotesize (139.33-143.69)} & {\footnotesize (*)} & {\footnotesize (69.48-75.88)}\\
 & &  & {\footnotesize [1.00${}^{a}$, \underline{2.75}${}^{a}$, \underline{2.75}${}^{a}$]} & {\footnotesize [\underline{1.17}${}^{a}$, \underline{2.34}${}^{a}$, \underline{3.23}${}^{a}$]} & {\footnotesize [\underline{1.21}${}^{c}$, \underline{2.51}${}^{c}$, \underline{3.32}${}^{c}$]} & {\footnotesize [\underline{2.29}${}^{b}$, \underline{1.63}${}^{b}$, \underline{6.29}${}^{a}$]}\\
\cline{3-7}
 &  & \multirow{3}{*}{d} 
 & 162.33 & 137.45 & 138.15 & -\\
 & &  & {\footnotesize (*)} & {\footnotesize (*)} & {\footnotesize (*)} & \\
 & &  & {\footnotesize [1.00${}^{c}$, \underline{2.80}${}^{c}$, \underline{2.80}${}^{c}$]} & {\footnotesize [\underline{1.18}${}^{c}$, \underline{2.41}${}^{c}$, \underline{3.30}${}^{c}$]} & {\footnotesize [\underline{1.18}${}^{c}$, \underline{2.50}${}^{c}$, \underline{3.29}${}^{c}$]} & \\
\cline{2-7}
 &   \multirow{6}{*}{7} &\multirow{3}{*}{s} 
 & 156.59 & 134.14 & 129.09 & {\bf 68.86}\\
 & &  & {\footnotesize (154.17-159.01)} & {\footnotesize (*)} & {\footnotesize (*)} & {\footnotesize (66.26-71.45)}\\
 & &  & {\footnotesize [1.00${}^{a}$, \underline{2.92}${}^{a}$, \underline{2.92}${}^{a}$]} & {\footnotesize [\underline{1.17}${}^{c}$, \underline{2.46}${}^{c}$, \underline{3.41}${}^{c}$]} & {\footnotesize [\underline{1.21}${}^{c}$, \underline{2.67}${}^{d}$, \underline{3.54}${}^{c}$]} & {\footnotesize [\underline{2.27}${}^{a}$, \underline{1.72}${}^{a}$, \underline{{\bf 6.64} }${}^{a}$]}\\
\cline{3-7}
 &  & \multirow{3}{*}{d} 
 & 152.73 & 134.78 & 130.42 & -\\
 & &  & {\footnotesize (150.84-154.62)} & {\footnotesize (*)} & {\footnotesize (127.96-132.87)} & \\
 & &  & {\footnotesize [1.00${}^{a}$, \underline{2.97}${}^{a}$, \underline{2.97}${}^{a}$]} & {\footnotesize [\underline{1.13}${}^{c}$, \underline{2.45}${}^{c}$, \underline{3.37}${}^{c}$]} & {\footnotesize [\underline{1.17}${}^{b}$, \underline{2.65}${}^{b}$, \underline{3.48}${}^{a}$]} & \\
\cline{2-7}
 &   \multirow{6}{*}{8} &\multirow{3}{*}{s} 
 & 149.80 & 128.17 & 126.58 & 69.42\\
 & &  & {\footnotesize (147.29-152.31)} & {\footnotesize (*)} & {\footnotesize (124.63-128.53)} & {\footnotesize (66.17-72.67)}\\
 & &  & {\footnotesize [1.00${}^{a}$, \underline{3.05}${}^{a}$, \underline{3.05}${}^{a}$]} & {\footnotesize [\underline{1.17}${}^{c}$, \underline{2.58}${}^{c}$, \underline{3.57}${}^{c}$]} & {\footnotesize [\underline{1.18}${}^{a}$, \underline{2.73}${}^{a}$, \underline{3.61}${}^{a}$]} & {\footnotesize [\underline{2.16}${}^{a}$, \underline{1.71}${}^{a}$, \underline{6.59}${}^{a}$]}\\
\cline{3-7}
 &  & \multirow{3}{*}{d} 
 & 145.43 & 129.05 & {\bf 126.21} & -\\
 & &  & {\footnotesize (*)} & {\footnotesize (126.80-131.29)} & {\footnotesize (124.28-128.15)} & \\
 & &  & {\footnotesize [1.00${}^{c}$, \underline{3.12}${}^{c}$, \underline{3.12}${}^{c}$]} & {\footnotesize [\underline{1.13}${}^{c}$, \underline{2.56}${}^{a}$, \underline{3.52}${}^{a}$]} & {\footnotesize [\underline{1.15}${}^{c}$, \underline{2.74}${}^{a}$, \underline{{\bf 3.60} }${}^{a}$]} & \\
\hline
\end{tabular}
\caption{Mean wall clock time with Arm-v7 instruction set and NEON extensions in ms. W=window, T=number of threads, P=precision, s=single, d=double. Values in curved brackets are the values of the 95\% confidence interval. Values in square brackets are the calculated speedups (see text for explanations). Minima and maxima in bold face. (*)=no normal distribution (median instead of mean, no confidence interval), ${}^{a}$=t-test, ${}^{b}$=Welch-test, ${}^{c}$=Mann-Whithney-U test, ${}^{d}$=Median test, underline=statistically significant (p$<$0.05). For armv7 there are no double precision SIMD instructions and support for single precision SIMD instructions is optional.}
\label{tab:result_armv7_1}
\end{supptable*}

\begin{supptable*}
\centering
\setstretch{0.9}
\begin{tabular}{ |c|c|c|c|c|c|c| }
\hline
 W & T & P 
& Java 
& C 
& Assembler 
& SIMD 
\\
\hline
\multirow{48}{*}{2} &   \multirow{6}{*}{1} &\multirow{3}{*}{s} 
 & {\bf 10,301.97} & 7,506.50 & 7,697.42 & 3,106.00\\
 & &  & {\footnotesize (10,203.80-10,400.13)} & {\footnotesize (7,502.93-7,510.08)} & {\footnotesize (7,693.94-7,700.89)} & {\footnotesize (3,102.07-3,109.93)}\\
 & &  & {\footnotesize [1.00${}^{a}$, 1.00${}^{a}$, 1.00${}^{a}$]} & {\footnotesize [\underline{1.37}${}^{a}$, 1.00${}^{a}$, \underline{1.37}${}^{a}$]} & {\footnotesize [\underline{1.34}${}^{a}$, 1.00${}^{a}$, \underline{1.34}${}^{a}$]} & {\footnotesize [\underline{3.32}${}^{a}$, 1.00${}^{a}$, \underline{3.32}${}^{a}$]}\\
\cline{3-7}
 &  & \multirow{3}{*}{d} 
 & {\bf 10,125.76} & 7,530.26 & 7,793.73 & -\\
 & &  & {\footnotesize (9,782.54-10,468.99)} & {\footnotesize (7,526.72-7,533.80)} & {\footnotesize (7,790.32-7,797.14)} & \\
 & &  & {\footnotesize [1.00${}^{a}$, 1.00${}^{a}$, 1.00${}^{a}$]} & {\footnotesize [\underline{1.34}${}^{a}$, 1.00${}^{a}$, \underline{1.34}${}^{a}$]} & {\footnotesize [\underline{1.30}${}^{a}$, 1.00${}^{a}$, \underline{1.30}${}^{a}$]} & \\
\cline{2-7}
 &   \multirow{6}{*}{2} &\multirow{3}{*}{s} 
 & 5,185.94 & 4,347.59 & 4,740.35 & 1,856.78\\
 & &  & {\footnotesize (4,995.67-5,376.21)} & {\footnotesize (4,119.84-4,575.35)} & {\footnotesize (4,347.56-5,133.14)} & {\footnotesize (1,716.11-1,997.44)}\\
 & &  & {\footnotesize [1.00${}^{a}$, \underline{1.99}${}^{a}$, \underline{1.99}${}^{a}$]} & {\footnotesize [\underline{1.19}${}^{a}$, \underline{1.73}${}^{b}$, \underline{2.37}${}^{b}$]} & {\footnotesize [\underline{1.09}${}^{b}$, \underline{1.62}${}^{b}$, \underline{2.17}${}^{b}$]} & {\footnotesize [\underline{2.79}${}^{a}$, \underline{1.67}${}^{b}$, \underline{5.55}${}^{a}$]}\\
\cline{3-7}
 &  & \multirow{3}{*}{d} 
 & 4,292.88 & 4,314.44 & 4,197.26 & -\\
 & &  & {\footnotesize (4,014.28-4,571.48)} & {\footnotesize (4,009.02-4,619.85)} & {\footnotesize (4,016.75-4,377.76)} & \\
 & &  & {\footnotesize [1.00${}^{a}$, \underline{2.36}${}^{a}$, \underline{2.36}${}^{a}$]} & {\footnotesize [1.00${}^{a}$, \underline{1.75}${}^{b}$, \underline{2.35}${}^{a}$]} & {\footnotesize [1.02${}^{a}$, \underline{1.86}${}^{a}$, \underline{2.41}${}^{a}$]} & \\
\cline{2-7}
 &   \multirow{6}{*}{3} &\multirow{3}{*}{s} 
 & 3,968.83 & 3,242.95 & 3,193.00 & 1,389.74\\
 & &  & {\footnotesize (*)} & {\footnotesize (*)} & {\footnotesize (3,186.17-3,199.82)} & {\footnotesize (1,374.47-1,405.01)}\\
 & &  & {\footnotesize [1.00${}^{c}$, \underline{2.60}${}^{c}$, \underline{2.60}${}^{c}$]} & {\footnotesize [\underline{1.22}${}^{d}$, \underline{2.31}${}^{d}$, \underline{3.18}${}^{c}$]} & {\footnotesize [\underline{1.24}${}^{c}$, \underline{2.41}${}^{b}$, \underline{3.23}${}^{a}$]} & {\footnotesize [\underline{2.86}${}^{c}$, \underline{2.23}${}^{a}$, \underline{7.41}${}^{a}$]}\\
\cline{3-7}
 &  & \multirow{3}{*}{d} 
 & 3,259.14 & 3,154.53 & 3,227.43 & -\\
 & &  & {\footnotesize (3,249.96-3,268.31)} & {\footnotesize (3,130.69-3,178.37)} & {\footnotesize (3,221.52-3,233.33)} & \\
 & &  & {\footnotesize [1.00${}^{a}$, \underline{3.11}${}^{a}$, \underline{3.11}${}^{a}$]} & {\footnotesize [\underline{1.03}${}^{a}$, \underline{2.39}${}^{a}$, \underline{3.21}${}^{a}$]} & {\footnotesize [\underline{1.01}${}^{a}$, \underline{2.41}${}^{b}$, \underline{3.14}${}^{a}$]} & \\
\cline{2-7}
 &   \multirow{6}{*}{4} &\multirow{3}{*}{s} 
 & 3,288.26 & 2,730.45 & 2,641.34 & 1,177.51\\
 & &  & {\footnotesize (*)} & {\footnotesize (*)} & {\footnotesize (2,636.17-2,646.51)} & {\footnotesize (1,172.03-1,182.98)}\\
 & &  & {\footnotesize [1.00${}^{c}$, \underline{3.13}${}^{c}$, \underline{3.13}${}^{c}$]} & {\footnotesize [\underline{1.20}${}^{d}$, \underline{2.75}${}^{d}$, \underline{3.77}${}^{c}$]} & {\footnotesize [\underline{1.24}${}^{d}$, \underline{2.91}${}^{b}$, \underline{3.90}${}^{a}$]} & {\footnotesize [\underline{2.79}${}^{d}$, \underline{2.64}${}^{a}$, \underline{8.75}${}^{a}$]}\\
\cline{3-7}
 &  & \multirow{3}{*}{d} 
 & 2,819.69 & 2,667.35 & 2,669.91 & -\\
 & &  & {\footnotesize (*)} & {\footnotesize (2,662.29-2,672.41)} & {\footnotesize (2,663.23-2,676.58)} & \\
 & &  & {\footnotesize [1.00${}^{c}$, \underline{3.59}${}^{c}$, \underline{3.59}${}^{c}$]} & {\footnotesize [\underline{1.06}${}^{c}$, \underline{2.82}${}^{b}$, \underline{3.80}${}^{a}$]} & {\footnotesize [\underline{1.06}${}^{c}$, \underline{2.92}${}^{b}$, \underline{3.79}${}^{a}$]} & \\
\cline{2-7}
 &   \multirow{6}{*}{5} &\multirow{3}{*}{s} 
 & 2,868.80 & 2,341.53 & 2,317.27 & 1,034.20\\
 & &  & {\footnotesize (*)} & {\footnotesize (2,335.63-2,347.43)} & {\footnotesize (2,310.17-2,324.36)} & {\footnotesize (*)}\\
 & &  & {\footnotesize [1.00${}^{c}$, \underline{3.59}${}^{c}$, \underline{3.59}${}^{c}$]} & {\footnotesize [\underline{1.23}${}^{c}$, \underline{3.21}${}^{b}$, \underline{4.40}${}^{a}$]} & {\footnotesize [\underline{1.24}${}^{c}$, \underline{3.32}${}^{b}$, \underline{4.45}${}^{a}$]} & {\footnotesize [\underline{2.77}${}^{d}$, \underline{3.00}${}^{c}$, \underline{9.96}${}^{c}$]}\\
\cline{3-7}
 &  & \multirow{3}{*}{d} 
 & 2,481.74 & 2,294.46 & 2,331.88 & -\\
 & &  & {\footnotesize (*)} & {\footnotesize (*)} & {\footnotesize (*)} & \\
 & &  & {\footnotesize [1.00${}^{c}$, \underline{4.08}${}^{c}$, \underline{4.08}${}^{c}$]} & {\footnotesize [\underline{1.08}${}^{d}$, \underline{3.28}${}^{d}$, \underline{4.41}${}^{c}$]} & {\footnotesize [\underline{1.06}${}^{c}$, \underline{3.34}${}^{d}$, \underline{4.34}${}^{c}$]} & \\
\cline{2-7}
 &   \multirow{6}{*}{6} &\multirow{3}{*}{s} 
 & 2,470.37 & 2,090.37 & 2,008.40 & 932.93\\
 & &  & {\footnotesize (*)} & {\footnotesize (*)} & {\footnotesize (*)} & {\footnotesize (928.65-937.20)}\\
 & &  & {\footnotesize [1.00${}^{c}$, \underline{4.17}${}^{c}$, \underline{4.17}${}^{c}$]} & {\footnotesize [\underline{1.18}${}^{d}$, \underline{3.59}${}^{d}$, \underline{4.93}${}^{c}$]} & {\footnotesize [\underline{1.23}${}^{d}$, \underline{3.83}${}^{d}$, \underline{5.13}${}^{c}$]} & {\footnotesize [\underline{2.65}${}^{d}$, \underline{3.33}${}^{a}$, \underline{11.04}${}^{a}$]}\\
\cline{3-7}
 &  & \multirow{3}{*}{d} 
 & 2,222.04 & 2,047.90 & 2,027.07 & -\\
 & &  & {\footnotesize (*)} & {\footnotesize (*)} & {\footnotesize (*)} & \\
 & &  & {\footnotesize [1.00${}^{c}$, \underline{4.56}${}^{c}$, \underline{4.56}${}^{c}$]} & {\footnotesize [\underline{1.09}${}^{c}$, \underline{3.68}${}^{d}$, \underline{4.94}${}^{c}$]} & {\footnotesize [\underline{1.10}${}^{c}$, \underline{3.84}${}^{d}$, \underline{5.00}${}^{c}$]} & \\
\cline{2-7}
 &   \multirow{6}{*}{7} &\multirow{3}{*}{s} 
 & 2,199.47 & 1,881.40 & 1,803.05 & 847.77\\
 & &  & {\footnotesize (2,194.74-2,204.19)} & {\footnotesize (*)} & {\footnotesize (*)} & {\footnotesize (*)}\\
 & &  & {\footnotesize [1.00${}^{a}$, \underline{4.68}${}^{a}$, \underline{4.68}${}^{a}$]} & {\footnotesize [\underline{1.17}${}^{c}$, \underline{3.99}${}^{d}$, \underline{5.48}${}^{c}$]} & {\footnotesize [\underline{1.22}${}^{c}$, \underline{4.27}${}^{d}$, \underline{5.71}${}^{c}$]} & {\footnotesize [\underline{2.59}${}^{d}$, \underline{3.66}${}^{c}$, \underline{12.15}${}^{c}$]}\\
\cline{3-7}
 &  & \multirow{3}{*}{d} 
 & 1,996.40 & 1,852.41 & 1,824.95 & -\\
 & &  & {\footnotesize (1,990.50-2,002.29)} & {\footnotesize (*)} & {\footnotesize (*)} & \\
 & &  & {\footnotesize [1.00${}^{a}$, \underline{5.07}${}^{a}$, \underline{5.07}${}^{a}$]} & {\footnotesize [\underline{1.08}${}^{c}$, \underline{4.07}${}^{d}$, \underline{5.47}${}^{c}$]} & {\footnotesize [\underline{1.09}${}^{c}$, \underline{4.27}${}^{d}$, \underline{5.55}${}^{c}$]} & \\
\cline{2-7}
 &   \multirow{6}{*}{8} &\multirow{3}{*}{s} 
 & 2,065.79 & 1,688.52 & 1,638.14 & {\bf 797.54} \\
 & &  & {\footnotesize (2,057.37-2,074.21)} & {\footnotesize (1,682.68-1,694.36)} & {\footnotesize (1,626.65-1,649.63)} & {\footnotesize (793.82-801.27)}\\
 & &  & {\footnotesize [1.00${}^{a}$, \underline{4.99}${}^{a}$, \underline{4.99}${}^{a}$]} & {\footnotesize [\underline{1.22}${}^{a}$, \underline{4.45}${}^{b}$, \underline{6.10}${}^{a}$]} & {\footnotesize [\underline{1.26}${}^{a}$, \underline{4.70}${}^{b}$, \underline{6.29}${}^{a}$]} & {\footnotesize [\underline{2.59}${}^{b}$, \underline{3.89}${}^{a}$, \underline{{\bf 12.92} }${}^{a}$]}\\
\cline{3-7}
 &  & \multirow{3}{*}{d} 
 & 1,920.05 & 1,686.80 & {\bf 1,660.85} & -\\
 & &  & {\footnotesize (1,905.01-1,935.10)} & {\footnotesize (1,676.47-1,697.13)} & {\footnotesize (1,647.10-1,674.61)} & \\
 & &  & {\footnotesize [1.00${}^{a}$, \underline{5.27}${}^{a}$, \underline{5.27}${}^{a}$]} & {\footnotesize [\underline{1.14}${}^{a}$, \underline{4.46}${}^{b}$, \underline{6.00}${}^{a}$]} & {\footnotesize [\underline{1.16}${}^{a}$, \underline{4.69}${}^{b}$, \underline{{\bf 6.10} }${}^{a}$]} & \\
\hline
\end{tabular}
\caption{Mean wall clock time with Arm-v7 instruction set and NEON extensions in ms. W=window, T=number of threads, P=precision, s=single, d=double. Values in curved brackets are the values of the 95\% confidence interval. Values in square brackets are the calculated speedups (see text for explanations). Minima and maxima in bold face. (*)=no normal distribution (median instead of mean, no confidence interval), ${}^{a}$=t-test, ${}^{b}$=Welch-test, ${}^{c}$=Mann-Whithney-U test, ${}^{d}$=Median test, underline=statistically significant (p$<$0.05). For armv7 there are no double precision SIMD instructions and support for single precision SIMD instructions is optional.}
\label{tab:result_armv7_2}
\end{supptable*}

\begin{supptable*}
\centering
\setstretch{0.9}
\begin{tabular}{ |c|c|c|c|c|c|c| }
\hline
 W & T & P 
& Java 
& C 
& Assembler 
& SIMD 
\\
\hline
\multirow{48}{*}{3} &   \multirow{6}{*}{1} &\multirow{3}{*}{s} 
 & {\bf 27,054.64} & 20,032.31 & 20,410.31 & 7,891.85\\
 & &  & {\footnotesize (26,763.87-27,345.41)} & {\footnotesize (20,028.56-20,036.05)} & {\footnotesize (20,406.90-20,413.72)} & {\footnotesize (*)}\\
 & &  & {\footnotesize [1.00${}^{a}$, 1.00${}^{a}$, 1.00${}^{a}$]} & {\footnotesize [\underline{1.35}${}^{b}$, 1.00${}^{a}$, \underline{1.35}${}^{b}$]} & {\footnotesize [\underline{1.33}${}^{b}$, 1.00${}^{a}$, \underline{1.33}${}^{b}$]} & {\footnotesize [\underline{3.43}${}^{d}$, 1.00${}^{c}$, \underline{3.43}${}^{d}$]}\\
\cline{3-7}
 &  & \multirow{3}{*}{d} 
 & {\bf 27,019.56} & 20,080.50 & 20,793.25 & -\\
 & &  & {\footnotesize (26,116.21-27,922.91)} & {\footnotesize (20,076.93-20,084.07)} & {\footnotesize (20,789.83-20,796.68)} & \\
 & &  & {\footnotesize [1.00${}^{a}$, 1.00${}^{a}$, 1.00${}^{a}$]} & {\footnotesize [\underline{1.35}${}^{a}$, 1.00${}^{a}$, \underline{1.35}${}^{a}$]} & {\footnotesize [\underline{1.30}${}^{a}$, 1.00${}^{a}$, \underline{1.30}${}^{a}$]} & \\
\cline{2-7}
 &   \multirow{6}{*}{2} &\multirow{3}{*}{s} 
 & 13,450.19 & 10,810.30 & 10,553.15 & 4,277.06\\
 & &  & {\footnotesize (13,119.79-13,780.60)} & {\footnotesize (10,554.82-11,065.77)} & {\footnotesize (10,336.72-10,769.58)} & {\footnotesize (4,071.48-4,482.65)}\\
 & &  & {\footnotesize [1.00${}^{a}$, \underline{2.01}${}^{a}$, \underline{2.01}${}^{a}$]} & {\footnotesize [\underline{1.24}${}^{a}$, \underline{1.85}${}^{b}$, \underline{2.50}${}^{a}$]} & {\footnotesize [\underline{1.27}${}^{a}$, \underline{1.93}${}^{b}$, \underline{2.56}${}^{a}$]} & {\footnotesize [\underline{3.14}${}^{a}$, \underline{1.85}${}^{d}$, \underline{6.33}${}^{a}$]}\\
\cline{3-7}
 &  & \multirow{3}{*}{d} 
 & 10,469.37 & 10,558.60 & 11,267.43 & -\\
 & &  & {\footnotesize (10,116.66-10,822.09)} & {\footnotesize (10,281.00-10,836.20)} & {\footnotesize (10,724.64-11,810.21)} & \\
 & &  & {\footnotesize [1.00${}^{a}$, \underline{2.58}${}^{a}$, \underline{2.58}${}^{a}$]} & {\footnotesize [0.99${}^{a}$, \underline{1.90}${}^{b}$, \underline{2.56}${}^{a}$]} & {\footnotesize [\underline{0.93}${}^{a}$, \underline{1.85}${}^{b}$, \underline{2.40}${}^{a}$]} & \\
\cline{2-7}
 &   \multirow{6}{*}{3} &\multirow{3}{*}{s} 
 & 10,235.33 & 8,395.36 & 8,183.07 & 3,344.23\\
 & &  & {\footnotesize (*)} & {\footnotesize (8,383.07-8,407.66)} & {\footnotesize (*)} & {\footnotesize (3,338.11-3,350.35)}\\
 & &  & {\footnotesize [1.00${}^{c}$, \underline{2.64}${}^{d}$, \underline{2.64}${}^{d}$]} & {\footnotesize [\underline{1.22}${}^{c}$, \underline{2.39}${}^{b}$, \underline{3.22}${}^{b}$]} & {\footnotesize [\underline{1.25}${}^{c}$, \underline{2.49}${}^{d}$, \underline{3.31}${}^{d}$]} & {\footnotesize [\underline{3.06}${}^{c}$, \underline{2.36}${}^{d}$, \underline{8.09}${}^{b}$]}\\
\cline{3-7}
 &  & \multirow{3}{*}{d} 
 & 8,449.32 & 8,161.82 & 8,333.35 & -\\
 & &  & {\footnotesize (*)} & {\footnotesize (8,140.54-8,183.10)} & {\footnotesize (8,327.67-8,339.03)} & \\
 & &  & {\footnotesize [1.00${}^{c}$, \underline{3.20}${}^{c}$, \underline{3.20}${}^{c}$]} & {\footnotesize [\underline{1.04}${}^{c}$, \underline{2.46}${}^{a}$, \underline{3.31}${}^{a}$]} & {\footnotesize [\underline{1.01}${}^{c}$, \underline{2.50}${}^{b}$, \underline{3.24}${}^{a}$]} & \\
\cline{2-7}
 &   \multirow{6}{*}{4} &\multirow{3}{*}{s} 
 & 8,469.14 & 7,025.81 & 6,810.72 & 2,845.26\\
 & &  & {\footnotesize (*)} & {\footnotesize (7,018.29-7,033.34)} & {\footnotesize (*)} & {\footnotesize (2,838.71-2,851.80)}\\
 & &  & {\footnotesize [1.00${}^{c}$, \underline{3.19}${}^{d}$, \underline{3.19}${}^{d}$]} & {\footnotesize [\underline{1.21}${}^{d}$, \underline{2.85}${}^{b}$, \underline{3.85}${}^{b}$]} & {\footnotesize [\underline{1.24}${}^{c}$, \underline{3.00}${}^{d}$, \underline{3.97}${}^{d}$]} & {\footnotesize [\underline{2.98}${}^{c}$, \underline{2.77}${}^{c}$, \underline{9.51}${}^{b}$]}\\
\cline{3-7}
 &  & \multirow{3}{*}{d} 
 & 7,270.45 & 6,850.59 & 6,935.36 & -\\
 & &  & {\footnotesize (7,266.33-7,274.56)} & {\footnotesize (*)} & {\footnotesize (*)} & \\
 & &  & {\footnotesize [1.00${}^{a}$, \underline{3.72}${}^{a}$, \underline{3.72}${}^{a}$]} & {\footnotesize [\underline{1.06}${}^{c}$, \underline{2.93}${}^{d}$, \underline{3.94}${}^{c}$]} & {\footnotesize [\underline{1.05}${}^{c}$, \underline{3.00}${}^{d}$, \underline{3.90}${}^{c}$]} & \\
\cline{2-7}
 &   \multirow{6}{*}{5} &\multirow{3}{*}{s} 
 & 7,201.13 & 6,025.95 & 5,808.30 & 2,492.77\\
 & &  & {\footnotesize (7,196.28-7,205.99)} & {\footnotesize (6,021.90-6,030.01)} & {\footnotesize (5,803.33-5,813.28)} & {\footnotesize (2,486.56-2,498.98)}\\
 & &  & {\footnotesize [1.00${}^{a}$, \underline{3.76}${}^{b}$, \underline{3.76}${}^{b}$]} & {\footnotesize [\underline{1.20}${}^{a}$, \underline{3.32}${}^{a}$, \underline{4.49}${}^{b}$]} & {\footnotesize [\underline{1.24}${}^{a}$, \underline{3.51}${}^{b}$, \underline{4.66}${}^{b}$]} & {\footnotesize [\underline{2.89}${}^{a}$, \underline{3.17}${}^{d}$, \underline{10.85}${}^{b}$]}\\
\cline{3-7}
 &  & \multirow{3}{*}{d} 
 & 6,359.34 & 5,886.54 & 5,909.65 & -\\
 & &  & {\footnotesize (*)} & {\footnotesize (*)} & {\footnotesize (5,905.49-5,913.80)} & \\
 & &  & {\footnotesize [1.00${}^{c}$, \underline{4.25}${}^{c}$, \underline{4.25}${}^{c}$]} & {\footnotesize [\underline{1.08}${}^{c}$, \underline{3.41}${}^{d}$, \underline{4.59}${}^{c}$]} & {\footnotesize [\underline{1.08}${}^{c}$, \underline{3.52}${}^{b}$, \underline{4.57}${}^{a}$]} & \\
\cline{2-7}
 &   \multirow{6}{*}{6} &\multirow{3}{*}{s} 
 & 6,320.64 & 5,299.29 & 5,103.03 & 2,220.69\\
 & &  & {\footnotesize (6,305.06-6,336.22)} & {\footnotesize (5,293.86-5,304.71)} & {\footnotesize (5,099.79-5,106.27)} & {\footnotesize (*)}\\
 & &  & {\footnotesize [1.00${}^{a}$, \underline{4.28}${}^{b}$, \underline{4.28}${}^{b}$]} & {\footnotesize [\underline{1.19}${}^{a}$, \underline{3.78}${}^{b}$, \underline{5.11}${}^{b}$]} & {\footnotesize [\underline{1.24}${}^{a}$, \underline{4.00}${}^{b}$, \underline{5.30}${}^{b}$]} & {\footnotesize [\underline{2.85}${}^{c}$, \underline{3.55}${}^{d}$, \underline{12.18}${}^{d}$]}\\
\cline{3-7}
 &  & \multirow{3}{*}{d} 
 & 5,683.66 & 5,188.35 & 5,207.69 & -\\
 & &  & {\footnotesize (*)} & {\footnotesize (5,182.96-5,193.75)} & {\footnotesize (5,180.04-5,235.34)} & \\
 & &  & {\footnotesize [1.00${}^{c}$, \underline{4.75}${}^{c}$, \underline{4.75}${}^{c}$]} & {\footnotesize [\underline{1.10}${}^{c}$, \underline{3.87}${}^{b}$, \underline{5.21}${}^{a}$]} & {\footnotesize [\underline{1.09}${}^{c}$, \underline{3.99}${}^{a}$, \underline{5.19}${}^{a}$]} & \\
\cline{2-7}
 &   \multirow{6}{*}{7} &\multirow{3}{*}{s} 
 & 5,615.93 & 4,727.66 & 4,535.33 & 1,993.89\\
 & &  & {\footnotesize (5,610.35-5,621.51)} & {\footnotesize (4,724.28-4,731.04)} & {\footnotesize (4,533.03-4,537.63)} & {\footnotesize (*)}\\
 & &  & {\footnotesize [1.00${}^{a}$, \underline{4.82}${}^{b}$, \underline{4.82}${}^{b}$]} & {\footnotesize [\underline{1.19}${}^{a}$, \underline{4.24}${}^{a}$, \underline{5.72}${}^{b}$]} & {\footnotesize [\underline{1.24}${}^{b}$, \underline{4.50}${}^{a}$, \underline{5.97}${}^{b}$]} & {\footnotesize [\underline{2.82}${}^{d}$, \underline{3.96}${}^{d}$, \underline{13.57}${}^{d}$]}\\
\cline{3-7}
 &  & \multirow{3}{*}{d} 
 & 5,141.18 & 4,642.51 & 4,616.64 & -\\
 & &  & {\footnotesize (5,133.89-5,148.46)} & {\footnotesize (*)} & {\footnotesize (*)} & \\
 & &  & {\footnotesize [1.00${}^{a}$, \underline{5.26}${}^{a}$, \underline{5.26}${}^{a}$]} & {\footnotesize [\underline{1.11}${}^{c}$, \underline{4.33}${}^{c}$, \underline{5.82}${}^{c}$]} & {\footnotesize [\underline{1.11}${}^{c}$, \underline{4.50}${}^{c}$, \underline{5.85}${}^{c}$]} & \\
\cline{2-7}
 &   \multirow{6}{*}{8} &\multirow{3}{*}{s} 
 & 5,101.61 & 4,321.30 & 4,117.84 & {\bf 1,855.60} \\
 & &  & {\footnotesize (5,080.93-5,122.30)} & {\footnotesize (4,302.95-4,339.65)} & {\footnotesize (4,108.16-4,127.52)} & {\footnotesize (1,844.41-1,866.80)}\\
 & &  & {\footnotesize [1.00${}^{a}$, \underline{5.30}${}^{b}$, \underline{5.30}${}^{b}$]} & {\footnotesize [\underline{1.18}${}^{a}$, \underline{4.64}${}^{b}$, \underline{6.26}${}^{b}$]} & {\footnotesize [\underline{1.24}${}^{a}$, \underline{4.96}${}^{b}$, \underline{6.57}${}^{b}$]} & {\footnotesize [\underline{2.75}${}^{a}$, \underline{4.25}${}^{c}$, \underline{{\bf 14.58} }${}^{b}$]}\\
\cline{3-7}
 &  & \multirow{3}{*}{d} 
 & 4,698.09 & 4,281.58 & {\bf 4,193.66} & -\\
 & &  & {\footnotesize (4,690.38-4,705.81)} & {\footnotesize (4,254.14-4,309.02)} & {\footnotesize (4,183.41-4,203.91)} & \\
 & &  & {\footnotesize [1.00${}^{a}$, \underline{5.75}${}^{a}$, \underline{5.75}${}^{a}$]} & {\footnotesize [\underline{1.10}${}^{b}$, \underline{4.69}${}^{b}$, \underline{6.31}${}^{a}$]} & {\footnotesize [\underline{1.12}${}^{a}$, \underline{4.96}${}^{b}$, \underline{{\bf 6.44} }${}^{a}$]} & \\
\hline
\end{tabular}
\caption{Mean wall clock time with Arm-v7 instruction set and NEON extensions in ms. W=window, T=number of threads, P=precision, s=single, d=double. Values in curved brackets are the values of the 95\% confidence interval. Values in square brackets are the calculated speedups (see text for explanations). Minima and maxima in bold face. (*)=no normal distribution (median instead of mean, no confidence interval), ${}^{a}$=t-test, ${}^{b}$=Welch-test, ${}^{c}$=Mann-Whithney-U test, ${}^{d}$=Median test, underline=statistically significant (p$<$0.05). For armv7 there are no double precision SIMD instructions and support for single precision SIMD instructions is optional. }
\label{tab:result_armv7_3}
\end{supptable*}

\begin{supptable*}
\centering
\setstretch{0.9}
\begin{tabular}{ |c|c|c|c|c|c|c| }
\hline
 W & T & P 
& Java 
& C 
& Assembler 
& SIMD 
\\
\hline
\multirow{24}{*}{0} &   \multirow{6}{*}{1} &\multirow{3}{*}{s} 
 & {\bf 912.26} & 788.26 & 799.85 & 285.15\\
 & &  & {\footnotesize (891.92-932.60)} & {\footnotesize (786.74-789.78)} & {\footnotesize (797.02-802.68)} & {\footnotesize (283.63-286.67)}\\
 & &  & {\footnotesize [1.00${}^{a}$, 1.00${}^{a}$, 1.00${}^{a}$]} & {\footnotesize [\underline{1.16}${}^{a}$, 1.00${}^{a}$, \underline{1.16}${}^{a}$]} & {\footnotesize [\underline{1.14}${}^{a}$, 1.00${}^{a}$, \underline{1.14}${}^{a}$]} & {\footnotesize [\underline{3.20}${}^{a}$, 1.00${}^{a}$, \underline{3.20}${}^{a}$]}\\
\cline{3-7}
 &  & \multirow{3}{*}{d} 
 & {\bf 916.20} & 788.59 & 800.82 & 567.00\\
 & &  & {\footnotesize (887.49-944.91)} & {\footnotesize (786.51-790.67)} & {\footnotesize (796.84-804.81)} & {\footnotesize (564.33-569.67)}\\
 & &  & {\footnotesize [1.00${}^{a}$, 1.00${}^{a}$, 1.00${}^{a}$]} & {\footnotesize [\underline{1.16}${}^{a}$, 1.00${}^{a}$, \underline{1.16}${}^{a}$]} & {\footnotesize [\underline{1.14}${}^{a}$, 1.00${}^{a}$, \underline{1.14}${}^{a}$]} & {\footnotesize [\underline{1.62}${}^{a}$, 1.00${}^{a}$, \underline{1.62}${}^{a}$]}\\
\cline{2-7}
 &   \multirow{6}{*}{2} &\multirow{3}{*}{s} 
 & 465.39 & 406.36 & 408.50 & 151.76\\
 & &  & {\footnotesize (449.52-481.26)} & {\footnotesize (399.36-413.36)} & {\footnotesize (400.41-416.58)} & {\footnotesize (143.96-159.56)}\\
 & &  & {\footnotesize [1.00${}^{a}$, \underline{1.96}${}^{a}$, \underline{1.96}${}^{a}$]} & {\footnotesize [\underline{1.15}${}^{a}$, \underline{1.94}${}^{a}$, \underline{2.24}${}^{a}$]} & {\footnotesize [\underline{1.14}${}^{a}$, \underline{1.96}${}^{a}$, \underline{2.23}${}^{a}$]} & {\footnotesize [\underline{3.07}${}^{a}$, \underline{1.88}${}^{a}$, \underline{6.01}${}^{a}$]}\\
\cline{3-7}
 &  & \multirow{3}{*}{d} 
 & 468.27 & 406.95 & 409.02 & 290.05\\
 & &  & {\footnotesize (448.81-487.73)} & {\footnotesize (400.20-413.69)} & {\footnotesize (400.82-417.21)} & {\footnotesize (283.79-296.31)}\\
 & &  & {\footnotesize [1.00${}^{a}$, \underline{1.96}${}^{a}$, \underline{1.96}${}^{a}$]} & {\footnotesize [\underline{1.15}${}^{a}$, \underline{1.94}${}^{a}$, \underline{2.25}${}^{a}$]} & {\footnotesize [\underline{1.14}${}^{a}$, \underline{1.96}${}^{a}$, \underline{2.24}${}^{a}$]} & {\footnotesize [\underline{1.61}${}^{a}$, \underline{1.95}${}^{a}$, \underline{3.16}${}^{a}$]}\\
\cline{2-7}
 &   \multirow{6}{*}{3} &\multirow{3}{*}{s} 
 & 314.66 & 278.34 & 277.90 & 107.54\\
 & &  & {\footnotesize (304.09-325.24)} & {\footnotesize (268.74-287.94)} & {\footnotesize (267.68-288.12)} & {\footnotesize (97.83-117.24)}\\
 & &  & {\footnotesize [1.00${}^{a}$, \underline{2.90}${}^{a}$, \underline{2.90}${}^{a}$]} & {\footnotesize [\underline{1.13}${}^{a}$, \underline{2.83}${}^{a}$, \underline{3.28}${}^{a}$]} & {\footnotesize [\underline{1.13}${}^{a}$, \underline{2.88}${}^{a}$, \underline{3.28}${}^{a}$]} & {\footnotesize [\underline{2.93}${}^{a}$, \underline{2.65}${}^{b}$, \underline{8.48}${}^{a}$]}\\
\cline{3-7}
 &  & \multirow{3}{*}{d} 
 & 313.31 & 279.42 & 278.21 & 199.47\\
 & &  & {\footnotesize (303.87-322.74)} & {\footnotesize (268.80-290.03)} & {\footnotesize (268.66-287.76)} & {\footnotesize (190.76-208.19)}\\
 & &  & {\footnotesize [1.00${}^{a}$, \underline{2.92}${}^{a}$, \underline{2.92}${}^{a}$]} & {\footnotesize [\underline{1.12}${}^{a}$, \underline{2.82}${}^{a}$, \underline{3.28}${}^{a}$]} & {\footnotesize [\underline{1.13}${}^{a}$, \underline{2.88}${}^{a}$, \underline{3.29}${}^{a}$]} & {\footnotesize [\underline{1.57}${}^{a}$, \underline{2.84}${}^{a}$, \underline{4.59}${}^{a}$]}\\
\cline{2-7}
 &   \multirow{6}{*}{4} &\multirow{3}{*}{s} 
 & 243.80 & 218.95 & 216.42 & {\bf 86.51} \\
 & &  & {\footnotesize (232.54-255.06)} & {\footnotesize (206.83-231.06)} & {\footnotesize (204.82-228.02)} & {\footnotesize (77.14-95.88)}\\
 & &  & {\footnotesize [1.00${}^{a}$, \underline{3.74}${}^{a}$, \underline{3.74}${}^{a}$]} & {\footnotesize [\underline{1.11}${}^{a}$, \underline{3.60}${}^{b}$, \underline{4.17}${}^{a}$]} & {\footnotesize [\underline{1.13}${}^{a}$, \underline{3.70}${}^{a}$, \underline{4.22}${}^{a}$]} & {\footnotesize [\underline{2.82}${}^{a}$, \underline{3.30}${}^{b}$, \underline{\bf{10.55}}${}^{a}$]}\\
\cline{3-7}
 &  & \multirow{3}{*}{d} 
 & 245.75 & 218.22 & 221.16 & {\bf 158.92} \\
 & &  & {\footnotesize (233.89-257.61)} & {\footnotesize (206.22-230.23)} & {\footnotesize (208.29-234.03)} & {\footnotesize (147.96-169.89)}\\
 & &  & {\footnotesize [1.00${}^{a}$, \underline{3.73}${}^{a}$, \underline{3.73}${}^{a}$]} & {\footnotesize [\underline{1.13}${}^{a}$, \underline{3.61}${}^{a}$, \underline{4.20}${}^{a}$]} & {\footnotesize [\underline{1.11}${}^{a}$, \underline{3.62}${}^{a}$, \underline{4.14}${}^{a}$]} & {\footnotesize [\underline{1.55}${}^{a}$, \underline{3.57}${}^{a}$, \underline{{\bf 5.77}}${}^{a}$]}\\
\Xhline{4\arrayrulewidth}
\multirow{24}{*}{1} &   \multirow{6}{*}{1} &\multirow{3}{*}{s} 
 & {\bf 20,083.78} & 17,986.80 & 17,772.32 & 7,492.27\\
 & &  & {\footnotesize (20,076.60-20,090.96)} & {\footnotesize (17,980.20-17,993.41)} & {\footnotesize (17,764.97-17,779.68)} & {\footnotesize (7,488.99-7,495.55)}\\
 & &  & {\footnotesize [1.00${}^{a}$, 1.00${}^{a}$, 1.00${}^{a}$]} & {\footnotesize [\underline{1.12}${}^{a}$, 1.00${}^{a}$, \underline{1.12}${}^{a}$]} & {\footnotesize [\underline{1.13}${}^{a}$, 1.00${}^{a}$, \underline{1.13}${}^{a}$]} & {\footnotesize [\underline{2.68}${}^{b}$, 1.00${}^{a}$, \underline{2.68}${}^{b}$]}\\
\cline{3-7}
 &  & \multirow{3}{*}{d} 
 & {\bf 20,339.45} & 17,977.88 & 17,991.78 & 13,982.91\\
 & &  & {\footnotesize (20,329.14-20,349.76)} & {\footnotesize (17,973.54-17,982.22)} & {\footnotesize (17,986.73-17,996.82)} & {\footnotesize (13,978.10-13,987.72)}\\
 & &  & {\footnotesize [1.00${}^{a}$, 1.00${}^{a}$, 1.00${}^{a}$]} & {\footnotesize [\underline{1.13}${}^{a}$, 1.00${}^{a}$, \underline{1.13}${}^{a}$]} & {\footnotesize [\underline{1.13}${}^{a}$, 1.00${}^{a}$, \underline{1.13}${}^{a}$]} & {\footnotesize [\underline{1.45}${}^{a}$, 1.00${}^{a}$, \underline{1.45}${}^{a}$]}\\
\cline{2-7}
 &   \multirow{6}{*}{2} &\multirow{3}{*}{s} 
 & 10,228.35 & 9,131.35 & 9,013.93 & 3,797.94\\
 & &  & {\footnotesize (10,134.54-10,322.16)} & {\footnotesize (9,124.06-9,138.64)} & {\footnotesize (9,006.62-9,021.23)} & {\footnotesize (3,791.48-3,804.40)}\\
 & &  & {\footnotesize [1.00${}^{a}$, \underline{1.96}${}^{a}$, \underline{1.96}${}^{a}$]} & {\footnotesize [\underline{1.12}${}^{a}$, \underline{1.97}${}^{a}$, \underline{2.20}${}^{a}$]} & {\footnotesize [\underline{1.13}${}^{a}$, \underline{1.97}${}^{a}$, \underline{2.23}${}^{a}$]} & {\footnotesize [\underline{2.69}${}^{a}$, \underline{1.97}${}^{a}$, \underline{5.29}${}^{a}$]}\\
\cline{3-7}
 &  & \multirow{3}{*}{d} 
 & 10,371.80 & 9,126.61 & 9,124.87 & 7,090.98\\
 & &  & {\footnotesize (10,253.93-10,489.68)} & {\footnotesize (9,118.58-9,134.64)} & {\footnotesize (9,116.97-9,132.78)} & {\footnotesize (7,083.76-7,098.20)}\\
 & &  & {\footnotesize [1.00${}^{a}$, \underline{1.96}${}^{a}$, \underline{1.96}${}^{a}$]} & {\footnotesize [\underline{1.14}${}^{a}$, \underline{1.97}${}^{a}$, \underline{2.23}${}^{a}$]} & {\footnotesize [\underline{1.14}${}^{a}$, \underline{1.97}${}^{a}$, \underline{2.23}${}^{a}$]} & {\footnotesize [\underline{1.46}${}^{a}$, \underline{1.97}${}^{a}$, \underline{2.87}${}^{a}$]}\\
\cline{2-7}
 &   \multirow{6}{*}{3} &\multirow{3}{*}{s} 
 & 6,910.37 & 6,190.24 & 6,109.78 & 2,575.74\\
 & &  & {\footnotesize (6,900.66-6,920.07)} & {\footnotesize (6,179.74-6,200.75)} & {\footnotesize (6,100.84-6,118.72)} & {\footnotesize (2,566.37-2,585.10)}\\
 & &  & {\footnotesize [1.00${}^{a}$, \underline{2.91}${}^{a}$, \underline{2.91}${}^{a}$]} & {\footnotesize [\underline{1.12}${}^{a}$, \underline{2.91}${}^{a}$, \underline{3.24}${}^{a}$]} & {\footnotesize [\underline{1.13}${}^{a}$, \underline{2.91}${}^{a}$, \underline{3.29}${}^{a}$]} & {\footnotesize [\underline{2.68}${}^{a}$, \underline{2.91}${}^{a}$, \underline{7.80}${}^{a}$]}\\
\cline{3-7}
 &  & \multirow{3}{*}{d} 
 & 6,995.74 & 6,190.74 & 6,190.32 & 4,810.10\\
 & &  & {\footnotesize (6,985.64-7,005.85)} & {\footnotesize (6,181.74-6,199.75)} & {\footnotesize (6,180.72-6,199.92)} & {\footnotesize (4,799.27-4,820.94)}\\
 & &  & {\footnotesize [1.00${}^{a}$, \underline{2.91}${}^{a}$, \underline{2.91}${}^{a}$]} & {\footnotesize [\underline{1.13}${}^{a}$, \underline{2.90}${}^{a}$, \underline{3.29}${}^{a}$]} & {\footnotesize [\underline{1.13}${}^{a}$, \underline{2.91}${}^{a}$, \underline{3.29}${}^{a}$]} & {\footnotesize [\underline{1.45}${}^{a}$, \underline{2.91}${}^{a}$, \underline{4.23}${}^{a}$]}\\
\cline{2-7}
 &   \multirow{6}{*}{4} &\multirow{3}{*}{s} 
 & 5,219.61 & 4,671.89 & 4,628.28 & {\bf 1,945.72} \\
 & &  & {\footnotesize (5,204.55-5,234.68)} & {\footnotesize (4,657.56-4,686.23)} & {\footnotesize (4,613.20-4,643.37)} & {\footnotesize (1,932.66-1,958.78)}\\
 & &  & {\footnotesize [1.00${}^{a}$, \underline{3.85}${}^{a}$, \underline{3.85}${}^{a}$]} & {\footnotesize [\underline{1.12}${}^{a}$, \underline{3.85}${}^{a}$, \underline{4.30}${}^{a}$]} & {\footnotesize [\underline{1.13}${}^{a}$, \underline{3.84}${}^{b}$, \underline{4.34}${}^{b}$]} & {\footnotesize [\underline{2.68}${}^{a}$, \underline{3.85}${}^{b}$, \underline{{\bf 10.32}}${}^{a}$]}\\
\cline{3-7}
 &  & \multirow{3}{*}{d} 
 & 5,280.36 & 4,685.29 & 4,680.73 & {\bf 3,633.11} \\
 & &  & {\footnotesize (5,268.03-5,292.69)} & {\footnotesize (4,668.10-4,702.48)} & {\footnotesize (4,663.40-4,698.06)} & {\footnotesize (3,616.93-3,649.30)}\\
 & &  & {\footnotesize [1.00${}^{a}$, \underline{3.85}${}^{a}$, \underline{3.85}${}^{a}$]} & {\footnotesize [\underline{1.13}${}^{a}$, \underline{3.84}${}^{b}$, \underline{4.34}${}^{a}$]} & {\footnotesize [\underline{1.13}${}^{a}$, \underline{3.84}${}^{b}$, \underline{4.35}${}^{a}$]} & {\footnotesize [\underline{1.45}${}^{a}$, \underline{3.85}${}^{b}$, \underline{{\bf 5.60}}${}^{a}$]}\\
\hline
\end{tabular}
\caption{Mean wall clock time with Arm-v8 instruction set and NEON extensions in ms. W=window, T=number of threads, P=precision, s=single, d=double. Values in curved brackets are the values of the 95\% confidence interval. Values in square brackets are the calculated speedups (see text for explanations). Minima and maxima in bold face. (*)=no normal distribution (median instead of mean, no confidence interval), ${}^{a}$=t-test, ${}^{b}$=Welch-test, ${}^{c}$=Mann-Whithney-U test, ${}^{d}$=Median test, underline=statistically significant (p$<$0.05)}
\label{tab:result_armv8_1}
\end{supptable*}

\begin{supptable*}
\centering
\setstretch{0.9}
\begin{tabular}{ |c|c|c|c|c|c|c| }
\hline
 W & T & P 
& Java 
& C 
& Assembler 
& SIMD 
\\
\hline
\multirow{24}{*}{2} &   \multirow{6}{*}{1} &\multirow{3}{*}{s} 
 & {\bf 53,794.44} & 48,078.46 & 47,207.33 & 19,202.21\\
 & &  & {\footnotesize (52,879.59-54,709.29)} & {\footnotesize (48,070.87-48,086.05)} & {\footnotesize (47,196.50-47,218.16)} & {\footnotesize (19,197.98-19,206.45)}\\
 & &  & {\footnotesize [1.00${}^{a}$, 1.00${}^{a}$, 1.00${}^{a}$]} & {\footnotesize [\underline{1.12}${}^{a}$, 1.00${}^{a}$, \underline{1.12}${}^{a}$]} & {\footnotesize [\underline{1.14}${}^{a}$, 1.00${}^{a}$, \underline{1.14}${}^{a}$]} & {\footnotesize [\underline{2.80}${}^{a}$, 1.00${}^{a}$, \underline{2.80}${}^{a}$]}\\
\cline{3-7}
 &  & \multirow{3}{*}{d} 
 & {\bf 55,057.05} & 48,072.52 & 48,075.06 & 36,392.40\\
 & &  & {\footnotesize (53,622.46-56,491.64)} & {\footnotesize (*)} & {\footnotesize (48,066.64-48,083.48)} & {\footnotesize (36,383.03-36,401.77)}\\
 & &  & {\footnotesize [1.00${}^{a}$, 1.00${}^{a}$, 1.00${}^{a}$]} & {\footnotesize [\underline{1.15}${}^{c}$, 1.00${}^{c}$, \underline{1.15}${}^{c}$]} & {\footnotesize [\underline{1.15}${}^{a}$, 1.00${}^{a}$, \underline{1.15}${}^{a}$]} & {\footnotesize [\underline{1.51}${}^{a}$, 1.00${}^{a}$, \underline{1.51}${}^{a}$]}\\
\cline{2-7}
 &   \multirow{6}{*}{2} &\multirow{3}{*}{s} 
 & 28,865.83 & 24,125.08 & 23,705.75 & 9,657.86\\
 & &  & {\footnotesize (28,859.39-28,872.26)} & {\footnotesize (24,116.87-24,133.28)} & {\footnotesize (23,695.49-23,716.01)} & {\footnotesize (9,651.47-9,664.24)}\\
 & &  & {\footnotesize [1.00${}^{a}$, \underline{1.86}${}^{a}$, \underline{1.86}${}^{a}$]} & {\footnotesize [\underline{1.20}${}^{a}$, \underline{1.99}${}^{a}$, \underline{2.23}${}^{a}$]} & {\footnotesize [\underline{1.22}${}^{a}$, \underline{1.99}${}^{a}$, \underline{2.27}${}^{a}$]} & {\footnotesize [\underline{2.99}${}^{a}$, \underline{1.99}${}^{a}$, \underline{5.57}${}^{a}$]}\\
\cline{3-7}
 &  & \multirow{3}{*}{d} 
 & 27,693.69 & 24,128.64 & 24,131.75 & 18,273.23\\
 & &  & {\footnotesize (26,844.37-28,543.01)} & {\footnotesize (24,120.35-24,136.93)} & {\footnotesize (24,124.87-24,138.64)} & {\footnotesize (18,264.60-18,281.86)}\\
 & &  & {\footnotesize [1.00${}^{a}$, \underline{1.99}${}^{a}$, \underline{1.99}${}^{a}$]} & {\footnotesize [\underline{1.15}${}^{a}$, \underline{1.99}${}^{c}$, \underline{2.28}${}^{a}$]} & {\footnotesize [\underline{1.15}${}^{a}$, \underline{1.99}${}^{a}$, \underline{2.28}${}^{a}$]} & {\footnotesize [\underline{1.52}${}^{a}$, \underline{1.99}${}^{a}$, \underline{3.01}${}^{a}$]}\\
\cline{2-7}
 &   \multirow{6}{*}{3} &\multirow{3}{*}{s} 
 & 19,316.16 & 16,146.52 & 15,860.33 & 6,470.98\\
 & &  & {\footnotesize (19,305.75-19,326.57)} & {\footnotesize (16,136.21-16,156.83)} & {\footnotesize (15,851.05-15,869.61)} & {\footnotesize (6,462.17-6,479.78)}\\
 & &  & {\footnotesize [1.00${}^{a}$, \underline{2.78}${}^{a}$, \underline{2.78}${}^{a}$]} & {\footnotesize [\underline{1.20}${}^{a}$, \underline{2.98}${}^{a}$, \underline{3.33}${}^{a}$]} & {\footnotesize [\underline{1.22}${}^{a}$, \underline{2.98}${}^{b}$, \underline{3.39}${}^{a}$]} & {\footnotesize [\underline{2.99}${}^{a}$, \underline{2.97}${}^{a}$, \underline{8.31}${}^{a}$]}\\
\cline{3-7}
 &  & \multirow{3}{*}{d} 
 & 18,249.66 & 16,144.60 & 16,147.18 & 12,253.26\\
 & &  & {\footnotesize (18,239.49-18,259.83)} & {\footnotesize (16,135.95-16,153.24)} & {\footnotesize (16,136.78-16,157.58)} & {\footnotesize (12,217.39-12,289.12)}\\
 & &  & {\footnotesize [1.00${}^{a}$, \underline{3.02}${}^{a}$, \underline{3.02}${}^{a}$]} & {\footnotesize [\underline{1.13}${}^{a}$, \underline{2.98}${}^{c}$, \underline{3.41}${}^{a}$]} & {\footnotesize [\underline{1.13}${}^{a}$, \underline{2.98}${}^{a}$, \underline{3.41}${}^{a}$]} & {\footnotesize [\underline{1.49}${}^{a}$, \underline{2.97}${}^{a}$, \underline{4.49}${}^{a}$]}\\
\cline{2-7}
 &   \multirow{6}{*}{4} &\multirow{3}{*}{s} 
 & 14,569.34 & 12,152.71 & 11,951.54 & {\bf 4,890.87} \\
 & &  & {\footnotesize (14,549.93-14,588.76)} & {\footnotesize (12,135.98-12,169.44)} & {\footnotesize (11,928.73-11,974.34)} & {\footnotesize (4,874.14-4,907.61)}\\
 & &  & {\footnotesize [1.00${}^{a}$, \underline{3.69}${}^{a}$, \underline{3.69}${}^{a}$]} & {\footnotesize [\underline{1.20}${}^{a}$, \underline{3.96}${}^{b}$, \underline{4.43}${}^{a}$]} & {\footnotesize [\underline{1.22}${}^{a}$, \underline{3.95}${}^{b}$, \underline{4.50}${}^{a}$]} & {\footnotesize [\underline{2.98}${}^{a}$, \underline{3.93}${}^{b}$, \underline{{\bf 11.00}}${}^{a}$]}\\
\cline{3-7}
 &  & \multirow{3}{*}{d} 
 & 13,752.50 & 12,164.01 & 12,150.13 & {\bf 9,203.69} \\
 & &  & {\footnotesize (13,735.03-13,769.97)} & {\footnotesize (12,140.61-12,187.42)} & {\footnotesize (12,133.91-12,166.35)} & {\footnotesize (9,190.81-9,216.57)}\\
 & &  & {\footnotesize [1.00${}^{a}$, \underline{4.00}${}^{a}$, \underline{4.00}${}^{a}$]} & {\footnotesize [\underline{1.13}${}^{a}$, \underline{3.95}${}^{d}$, \underline{4.53}${}^{a}$]} & {\footnotesize [\underline{1.13}${}^{a}$, \underline{3.96}${}^{b}$, \underline{4.53}${}^{a}$]} & {\footnotesize [\underline{1.49}${}^{a}$, \underline{3.95}${}^{a}$, \underline{{\bf 5.98}}${}^{a}$]}\\
\hline
\end{tabular}
\caption{Mean wall clock time with Arm-v8 instruction set and NEON extensions in ms. W=window, T=number of threads, P=precision, s=single, d=double. Values in curved brackets are the values of the 95\% confidence interval. Values in square brackets are the calculated speedups (see text for explanations). Minima and maxima in bold face. (*)=no normal distribution (median instead of mean, no confidence interval), ${}^{a}$=t-test, ${}^{b}$=Welch-test, ${}^{c}$=Mann-Whithney-U test, ${}^{d}$=Median test, underline=statistically significant (p$<$0.05)}
\label{tab:result_armv8_2}
\end{supptable*}

\begin{supptable*}
\centering
\begin{tabular}{ |c|c|c|r|r|r|r|r|r|r| }
\hline
W & V & P & wall clock time & wall clock time     & speedup & wall clock time & speedup & wall clock time & speedup \\
  &   &   & Mali-G71 MP 2   & Intel Iris Plus 650 &         & AMD Radeon VII  &         & Nvidia Titan V  &   \\
\hline
  \multirow{15}{*}{1} &   \multirow{3}{*}{1} & h
 & 24.94 & 4.90 & 5.09 & 0.49 & 51.26 & n/a & \\
 &  & s
 & 24.92 & 5.84 & 4.27 & 0.40 & 62.51 & 0.20 & 124.28\\
 &  & d
 & n/a & 15.61 &  & 1.19 &  & 0.27 & \\
\cline{2-10}
 &   \multirow{3}{*}{2} & h
 & 26.27 & 4.20 & 6.26 & 0.50 & 52.12 & n/a & \\
 &  & s
 & 29.43 & 5.87 & 5.01 & 0.67 & 43.65 & 0.21 & 137.30\\
 &  & d
 & n/a & 16.53 &  & 1.26 &  & 0.29 & \\
\cline{2-10}
 &   \multirow{3}{*}{4} & h
 & 23.11 & 3.98 & 5.81 & 0.55 & 42.09 & n/a & \\
 &  & s
 & 28.07 & 5.87 & 4.78 & 0.66 & 42.74 & 0.21 & 131.75\\
 &  & d
 & n/a & 16.67 &  & 1.67 &  & 0.30 & \\
\cline{2-10}
 &   \multirow{3}{*}{8} & h
 & 25.51 & 4.17 & 6.11 & 0.65 & 39.46 & n/a & \\
 &  & s
 & 31.03 & 5.72 & 5.42 & 0.57 & 54.26 & 0.22 & 143.91\\
 &  & d
 & n/a & 15.73 &  & 1.66 &  & 0.34 & \\
\cline{2-10}
 &   \multirow{3}{*}{16} & h
 & 147.47 & 6.59 & 22.37 & 0.79 & 186.74 & n/a & \\
 &  & s
 & 314.79 & 6.33 & 49.70 & 0.83 & 381.55 & 0.27 & 1161.23\\
 &  & d
 & n/a & 119.49 &  & 2.14 &  & 0.60 & \\
\cline{2-10}
\hline
  \multirow{15}{*}{2} &   \multirow{3}{*}{1} & h
 & {\cellcolor[HTML]{D0D0D0} 398.87} & {\cellcolor[HTML]{D0D0D0} 80.93} & {\cellcolor[HTML]{D0D0D0} 4.93} & {\cellcolor[HTML]{D0D0D0} 5.89} & {\cellcolor[HTML]{D0D0D0} 67.74} & {\cellcolor[HTML]{D0D0D0} n/a} & {\cellcolor[HTML]{D0D0D0} }\\
 &  & s
 & 494.20 & 118.81 & 4.16 & 6.55 & 75.46 & 1.99 & 247.87\\
 &  & d
 & n/a & 150.19 &  & 19.93 &  & 4.42 & \\
\cline{2-10}
 &   \multirow{3}{*}{2} & h
 & {\cellcolor[HTML]{D0D0D0} 434.58} & {\cellcolor[HTML]{D0D0D0} 74.53} & {\cellcolor[HTML]{D0D0D0} 5.83} & {\cellcolor[HTML]{D0D0D0} 7.32} & {\cellcolor[HTML]{D0D0D0} 59.36} & {\cellcolor[HTML]{D0D0D0} n/a} & {\cellcolor[HTML]{D0D0D0} }\\
 &  & s
 & 648.68 & 120.15 & 5.40 & 8.14 & 79.72 & 2.58 & 251.22\\
 &  & d
 & n/a & 157.57 &  & 20.26 &  & 5.17 & \\
\cline{2-10}
 &   \multirow{3}{*}{4} & h
 & {\cellcolor[HTML]{D0D0D0} 381.96} & {\cellcolor[HTML]{D0D0D0} 70.26} & {\cellcolor[HTML]{D0D0D0} 5.44} & {\cellcolor[HTML]{D0D0D0} 8.39} & {\cellcolor[HTML]{D0D0D0} 45.53} & {\cellcolor[HTML]{D0D0D0} n/a} & {\cellcolor[HTML]{D0D0D0} }\\
 &  & s
 & 653.93 & 125.87 & 5.20 & 8.68 & 75.35 & 2.68 & 244.06\\
 &  & d
 & n/a & 161.88 &  & 21.49 &  & 5.66 & \\
\cline{2-10}
 &   \multirow{3}{*}{8} & h
 & {\cellcolor[HTML]{D0D0D0} 411.40} & {\cellcolor[HTML]{D0D0D0} 76.08} & {\cellcolor[HTML]{D0D0D0} 5.41} & {\cellcolor[HTML]{D0D0D0} 10.19} & {\cellcolor[HTML]{D0D0D0} 40.36} & {\cellcolor[HTML]{D0D0D0} n/a} & {\cellcolor[HTML]{D0D0D0} }\\
 &  & s
 & 710.74 & 125.70 & 5.65 & 9.16 & 77.56 & 3.16 & 225.08\\
 &  & d
 & n/a & 158.04 &  & 22.22 &  & 7.33 & \\
\cline{2-10}
 &   \multirow{3}{*}{16} & h
 & {\cellcolor[HTML]{D0D0D0} 2603.37} & {\cellcolor[HTML]{D0D0D0} 117.84} & {\cellcolor[HTML]{D0D0D0} 22.09} & {\cellcolor[HTML]{D0D0D0} 11.85} & {\cellcolor[HTML]{D0D0D0} 219.71} & {\cellcolor[HTML]{D0D0D0} n/a} & {\cellcolor[HTML]{D0D0D0} }\\
 &  & s
 & 9899.92 & 159.01 & 62.26 & 9.30 & 1064.67 & 3.40 & 2913.67\\
 &  & d
 & n/a & 1402.76 &  & 23.06 &  & 8.80 & \\
\cline{2-10}
\hline
  \multirow{15}{*}{3} &   \multirow{3}{*}{1} & h
 & {\cellcolor[HTML]{D0D0D0} 1295.78} & {\cellcolor[HTML]{D0D0D0} 130.21} & {\cellcolor[HTML]{D0D0D0} 9.95} & {\cellcolor[HTML]{D0D0D0} 15.48} & {\cellcolor[HTML]{D0D0D0} 83.73} & {\cellcolor[HTML]{D0D0D0} n/a} & {\cellcolor[HTML]{D0D0D0} }\\
 &  & s
 & 1215.24 & 144.37 & 8.42 & 14.77 & 82.30 & 4.70 & 258.73\\
 &  & d
 & n/a & 331.79 &  & 37.00 &  & 10.57 & \\
\cline{2-10}
 &   \multirow{3}{*}{2} & h
 & {\cellcolor[HTML]{D0D0D0} 1387.64} & {\cellcolor[HTML]{D0D0D0} 128.49} & {\cellcolor[HTML]{D0D0D0} 10.80} & {\cellcolor[HTML]{D0D0D0} 16.72} & {\cellcolor[HTML]{D0D0D0} 82.99} & {\cellcolor[HTML]{D0D0D0} n/a} & {\cellcolor[HTML]{D0D0D0} }\\
 &  & s
 & 1569.44 & 149.78 & 10.48 & 16.19 & 96.94 & 6.01 & 261.14\\
 &  & d
 & n/a & 376.49 &  & 35.74 &  & 12.20 & \\
\cline{2-10}
 &   \multirow{3}{*}{4} & h
 & {\cellcolor[HTML]{D0D0D0} 1171.09} & {\cellcolor[HTML]{D0D0D0} 125.39} & {\cellcolor[HTML]{D0D0D0} 9.34} & {\cellcolor[HTML]{D0D0D0} 17.10} & {\cellcolor[HTML]{D0D0D0} 68.49} & {\cellcolor[HTML]{D0D0D0} n/a} & {\cellcolor[HTML]{D0D0D0} }\\
 &  & s
 & 1545.54 & 152.94 & 10.11 & 15.92 & 97.07 & 5.90 & 262.06\\
 &  & d
 & n/a & 398.80 &  & 36.50 &  & 12.85 & \\
\cline{2-10}
 &   \multirow{3}{*}{8} & h
 & {\cellcolor[HTML]{D0D0D0} 1177.66} & {\cellcolor[HTML]{D0D0D0} 130.73} & {\cellcolor[HTML]{D0D0D0} 9.01} & {\cellcolor[HTML]{D0D0D0} 17.33} & {\cellcolor[HTML]{D0D0D0} 67.95} & {\cellcolor[HTML]{D0D0D0} n/a} & {\cellcolor[HTML]{D0D0D0} }\\
 &  & s
 & 1680.94 & 153.75 & 10.93 & 15.95 & 105.41 & 6.48 & 259.49\\
 &  & d
 & n/a & 382.03 &  & 36.54 &  & 16.27 & \\
\cline{2-10}
 &   \multirow{3}{*}{16} & h
 & {\cellcolor[HTML]{D0D0D0} 11724.42} & {\cellcolor[HTML]{D0D0D0} 182.80} & {\cellcolor[HTML]{D0D0D0} 64.14} & {\cellcolor[HTML]{D0D0D0} 18.53} & {\cellcolor[HTML]{D0D0D0} 632.87} & {\cellcolor[HTML]{D0D0D0} n/a} & {\cellcolor[HTML]{D0D0D0} }\\
 &  & s
 & 27253.11 & 192.10 & 141.87 & 16.09 & 1694.22 & 7.90 & 3450.04\\
 &  & d
 & n/a & 3684.83 &  & 37.94 &  & 17.45 & \\
\cline{2-10}
\hline
\end{tabular}
\caption{Comparison of the wall clock time in milliseconds of the GPU on the Arm-v7 tablet, an integrated Intel Iris Plus 650 GPU, a dedicated AMD Radeon VII GPU and a dedicated Nvidia Titan V GPU.W=window, V=number of vector elements, P=precision, h=half, s=single, d=double. n/a = not available. 'wall clock time' = median of the wall clock time for the execution of the kernel ({\tt clEnqueueNDRangeKernel}) on the specified GPU in ms. 'speedup' = speedup with respect to the leftmost column. For the execution on the Intel and the AMD GPUs the same OpenCL kernel as for the tablet was used. Instead of the Android development environment a program written in C (GCC 8.2, -O3) was chosen to prepare, enqueue and execute  the kernel. For the Intel GPU lightweight virtualization (Docker) has been used. Due to the differenct execution environments, only the time for the execution of the kernel can be compared. Although the Titan V GPU is able to perform half precision calculations, this capability is not available with the OpenCL framework. Grey color=result not acceptable due to numerical problems.}
\label{tab:result_GPUcomp}
\end{supptable*}

\end{document}